\documentclass[structabstract, longauth]{aa}  

\usepackage[colorlinks = true,
            linkcolor = blue,
            urlcolor  = blue,
            citecolor = blue,
            anchorcolor = blue]{hyperref}
\usepackage{graphicx}
\usepackage[utf8]{inputenc}
\usepackage{amsmath}	
\usepackage{amssymb}	
\usepackage[dvipsnames]{xcolor}
\usepackage{ulem}
\usepackage{longtable}

\usepackage{threeparttable}

\usepackage{colortbl}

\usepackage{subcaption}
\usepackage{mwe}

\usepackage{enumitem}

\usepackage{pifont}

\usepackage{natbib} 
\usepackage{url} 
\usepackage{txfonts}
\begin{document} 

   \title{An elliptical accretion disk following the tidal disruption event AT~2020zso}
    \titlerunning{A newly formed, highly elliptical accretion disk in a TDE}
    \authorrunning{Wevers et al.}

   \author{T. Wevers\inst{1}\thanks{Email: twevers@eso.org}\href{https://orcid.org/0000-0002-4043-9400}
          \and 
          M.~Nicholl\inst{2} 
          \and 
          M.~Guolo\inst{1,3}\href{https://orcid.org/0000-0002-5063-0751} 
          \and 
          P.~Charalampopoulos\inst{4}\href{https://orcid.org/0000-0002-0326-6715}
          \and 
          M.~Gromadzki\inst{5} 
          \and
          T.M.~Reynolds\inst{6}
          \and 
          E.~Kankare\inst{6}\href{https://orcid.org/0000-0001-8257-3512} 
          \and 
          G.~Leloudas\inst{4} 
        \and
          J.P.~Anderson\inst{1}\href{https://orcid.org/0000-0003-0227-3451} 
          \and 
          I.~Arcavi\inst{7,8}\href{https://orcid.org/0000-0001-7090-4898} 
          \and 
          G.~Cannizzaro\inst{9,10} 
          \and 
          T.-W.~Chen\inst{11}\href{https://orcid.org/0000-0002-1066-6098} 
          \and
          N. Ihanec\inst{5} 
          \and
          C.~Inserra\inst{12}\href{https://orcid.org/0000-0002-3968-4409}
          \and 
          C.P.~Guti\'errez\inst{13,14}\href{https://orcid.org/0000-0003-2375-2064}
            \and
          P.G.~Jonker\inst{9,10}
          \and
          A.~Lawrence\inst{15}\href{https://orcid.org/0000-0002-3134-6093} 
          \and 
          M.R.~Magee\inst{16}\href{https://orcid.org/0000-0002-0629-8931}
          \and 
        T.E.~M{\"u}ller-Bravo\inst{17}\href{https://orcid.org/0000-0003-3939-7167}
          \and 
          F.~Onori\inst{18}
          \and 
          E.~Ridley\inst{3} 
          \and 
          S.~Schulze\inst{19} 
          \and 
          P.~Short\inst{15} 
          \and 
          D.~Hiramatsu\inst{20, 21, 22}\href{https://orcid.org/0000-0002-1125-9187}
          \and 
          M.~Newsome\inst{20, 21}
          \and
          J.H.~Terwel\inst{23}\href{https://orcid.org/0000-0001-9834-3439}
          \and 
          S.~Yang\inst{11}\href{https://orcid.org/0000-0002-2898-6532} 
          \and 
          and D.~Young\inst{24}
          }

   \institute{European Southern Observatory, Alonso de C\'ordova 3107, Casilla 19, Santiago, Chile\\ \email{twevers@eso.org}
    \and 
    Birmingham Institute for Gravitational Wave Astronomy and School of Physics and Astronomy, University of Birmingham,\\ Birmingham B15 2TT, UK 
    \and Department of Physics and Astronomy, Johns Hopkins University, 3400 N. Charles St., Baltimore, MD 21218, USA
    \and
    DTU Space, National Space Institute, Technical University of Denmark, Elektrovej 327, 2800 Kgs.  Lyngby, Denmark
    \and
    Warsaw University Astronomical Observatory, Al. Ujazdowskie 4, 00-478 Warszawa, Poland
    \and
    Department of Physics and Astronomy, University of Turku, 20014 Turku, Finland
    \and
    The School of Physics and Astronomy, Tel Aviv University, Tel Aviv 69978, Israel
    \and 
    CIFAR Azrieli Global Scholars program, CIFAR, Toronto, Canada
    \and
    Department of Astrophysics/IMAPP, Radboud University, P.O. Box 9010, 6500 GL Nijmegen, The Netherlands
    \and 
    SRON, Netherlands Institute for Space Research, Sorbonnelaan 2, 3584 CA, Utrecht, The Netherlands
    \and 
    The Oskar Klein Centre, Department of Astronomy, Stockholm University, AlbaNova, SE-10691 Stockholm, Sweden
    \and 
    School of Physics \& Astronomy, Cardiff University, Queens Buildings, The Parade, Cardiff, CF24 3AA, UK
    \and 
    Finnish Centre for Astronomy with ESO (FINCA), FI-20014 University of Turku, Finland
    \and 
    Tuorla Observatory, Department of Physics and Astronomy, FI-20014 University of Turku, Finland
    \and
    Institute for Astronomy, University of Edinburgh, Royal Observatory, Blackford Hill, EH9 3HJ, UK
    \and 
    Institute of Cosmology and Gravitation, University of Portsmouth, Portsmouth, PO1 3FX, UK 
    \and 
    School of Physics and Astronomy, University of Southampton, Southampton, Hampshire, SO17 1BJ, UK
    \and
    Instituto di Astrofisica e Planetologia Spaziali (INAF), Via Fosso del Cavaliere 100, Roma, I-00133, Italy
    \and 
    The Oskar Klein Centre, Physics Department of Physics, Stockholm University, Albanova University Center, SE 106 91 Stockholm, Sweden
    \and
    Las Cumbres Observatory, 6740 Cortona Drive, Suite 102, Goleta, CA 93117-5575, USA
    \and 
    Department of Physics, University of California, Santa Barbara, CA 93106-9530, USA
    \and Center for Astrophysics, Harvard \& Smithsonian, 60 Garden Street, Cambridge, MA 02138-1516, USA
    \and 
    School of Physics, Trinity College Dublin, The University of Dublin, Dublin 2, Ireland 
    \and 
    Astrophysics Research Centre, School of Mathematics and Physics, Queens University Belfast, Belfast BT7 1NN, UK
        }
   \date{Received 08/11/2021; revised 16/02/2022 ; accepted 08/06/2022}
 
  \abstract
   {}
   {Modeling spectroscopic observations of tidal disruption events (TDEs) to date suggests that the newly-formed accretion disks are mostly quasi-circular. In this work we study the transient event AT~2020zso, hosted by an active galactic nucleus (AGN; as inferred from narrow emission line diagnostics), with the aim of characterising the properties of its newly formed accretion flow.}
   {We classify AT~2020zso as a TDE based on the blackbody evolution inferred from UV/optical photometric observations, and spectral line content and evolution. We identify transient, double-peaked Bowen (N\,\textsc{iii}), He\,\textsc{i}, He\,\textsc{ii} and H$\alpha$ emission lines. We model medium resolution optical spectroscopy of the He\,\textsc{ii} (after careful deblending of the N\,\textsc{iii} contribution) and H$\alpha$ lines during the rise, peak and early decline of the light curve using relativistic, elliptical accretion disk models.}
   {We find that the spectral evolution before peak can be explained by optical depth effects consistent with an outflowing, optically thick Eddington envelope. Around peak the envelope reaches its maximum extent (approximately 10$^{15}$ cm, or $\sim$ 3000 -- 6000 gravitational radii for an inferred black hole mass of $5-10\times10^5 ~M_{\odot}$) and becomes optically thin. The H$\alpha$ and He\,\textsc{ii} emission lines at and after peak can be reproduced with a highly inclined ($i= 85\pm5$ degrees), highly elliptical ($e = 0.97 \pm 0.01$) and relatively compact (R$_{\rm in}$ = several 100 R$_g$ and R$_{\rm out}$ = several 1000 R$_g$) accretion disk. }
   {Overall, the line profiles suggest a highly elliptical geometry for the new accretion flow, consistent with theoretical expectations of newly formed TDE disks. We quantitatively confirm, for the first time, the high inclination nature of a Bowen (and X-ray dim) TDE, consistent with the unification picture of TDEs where the inclination largely determines the observational appearance.
   Rapid line profile variations rule out the binary SMBH hypothesis as the origin of the eccentricity; these results thus provide a direct link between a TDE in an AGN and the eccentric accretion disk. 
   We illustrate for the first time how optical spectroscopy can be used to constrain the black hole spin, through (the lack of) disk precession signatures (changes in inferred inclination). We constrain the disk alignment timescale to $>$ 15 days in AT2020zso, which rules out high black hole spin values ($a < 0.8$) for $M_{\rm BH} \sim 10^6$ $M_{\odot}$ and disk viscosity $\alpha \gtrsim 0.1.$ }
   {}


   \maketitle
%

\section{Introduction}
Double-peaked emission lines, usually seen in H Balmer and He\,\textsc{ii} optical transitions, are observed in a small fraction ($\sim$3\%) of active galactic nuclei \citep[AGN, ][]{Chen1989a, Chen1989b, Eracleous1994, Strateva2003}. Although a number of possible explanations for their origin exist in the literature, including binary supermassive black holes (SMBHs; \citealt{Begelman1980, Gaskell1983}), bipolar outflows \citep{Norman1984, Zheng1990}, or highly anisotropic continuum sources \citep{Goad1996}, the leading explanation is that they originate in the outer parts of an inclined accretion disk (several 1000 gravitational radii, where the gravitational radius $R_{\rm g} = \frac{G M_{\rm BH}}{c^2}$; see e.g. \citealt{Eracleous2003} for a detailed discussion). 

The original literature models (e.g. \citealt{Chen1989a, Chen1989b}) envisaged a circular accretion disk, and were successful in reproducing $\sim$40\% of the known samples \citep{Eracleous1994, Strateva2003}.
The observed sample morphology of double-peaked AGN is diverse, including both stronger blue than red peaks and vice versa (see e.g. \citealt{Eracleous2003} for an overview); the latter in particular cannot be explained with a circular accretion disk model alone. Some show line profile variability, observed on timescales from a few days (the {\it reverberation} timescale; \citealt{Schimoia2015}) up to months and years (the {\it dynamical} timescale; e.g. \citealt{Gezari2007, Schimoia2017}). Motivated by this diverse behaviour, more sophisticated, and more importantly non-axisymmetric, accretion disk models were developed \citep{Eracleous1995, Strateva2003}. Such elliptical models were able to reproduce the majority (but again, not all) of the sources where the circular models failed.

Two main hypotheses were put forward by 
\citet{Eracleous1995} for the formation of such eccentric accretion disks around SMBHs: binary SMBHs, where the eccentricity of the disk is pumped by the tidal torques of the secondary; and tidal disruption events, in which the formation of a highly eccentric structure is a natural expectation in the absence of a mechanism to efficiently and rapidly remove angular momentum from the stellar debris. The vastly different timescales involved provide a mechanism to discriminate between these hypotheses through spectroscopic monitoring. In particular, binary SMBH disks are expected to evolve on $\sim$1000s of years whereas in tidal disruption events (TDEs), evolution can be expected on weeks--months timescales.

There are some previous claims in the literature for the presence of disks with significant eccentricity following TDEs. \citet{Cao2018} model the proto-typical TDE ASASSN--14li, finding a large disk ($r_{\rm out} \sim 1700 ~R_{\rm g}$) and an eccentricity $e$ = 0.97. However, the profiles in this event are single peaked, and moreover can also be modelled as an optically thick, spherically symmetric outflow, where the line evolution is explained through electron scattering depth variations \citep{Roth2018}. Given the absence of significant asymmetries/double-peaked profiles, the evidence for an accretion disk origin of the emission lines is unclear.
Using the same model, \citet{Liu2017} model the TDE PTF--09djl (see also \citealt{Arcavi2014}); for this source the data is sparse and noisy, but the H$\alpha$ line does appear strongly asymmetric. It can be fit with a compact, highly elliptical ($e$ = 0.96) accretion disk. Unfortunately, no similar line profile is found in other lines (e.g. H Balmer lines, He\,\textsc{i}, He\,\textsc{ii}). The absence of He\,\textsc{ii} is explained through the inferred high inclination angle (an idea that is not compatible with the conclusions of this work), but the difference between the H$\alpha$ and H$\beta$ line profiles is more difficult to explain.

With increasing TDE detection rates and spectroscopic follow-up datasets, clearer evidence has emerged to associate optical emission line profiles directly to an accretion disk. \citet{Wevers2019b} and \citet{Cannizzaro2021} reported narrow Fe\,\textsc{ii} emission lines likely associated with a disk chromosphere, while \citet{Holoien201918kh} (PS18kh), as well as \citet{Short2020} and \citet{Hung2020} (AT~2018hyz) reported on flat-topped / double-peaked H$\alpha$ (and other H Balmer) emission line profiles that are very likely disk-related (although see \citealt{Hung2019} for an outflow scenario to explain the line profiles in PS18kh). In both PS18kh and AT~2018hyz, the inferred eccentricities are low ($\sim$ 0.1 -- 0.2) and uniform, while the disk inclinations are low to moderate (20 -- 60 degrees). 
This may appear somewhat surprising, given the current lack of understanding of the detailed dynamics of the post-disruption debris; in particular, from a theoretical point of view it is unclear how the stellar debris can shed its (expected) large amount of energy in such a short timescale to form a quasi-circular disk (e.g. \citealt{Krolik2020}). In the absence of such a mechanism, the naive expectation is for the debris to form a highly elliptical structure. Using hydrodynamical simulations, \citet{Shiokawa2015} found that the returning debris is unlikely to settle into a compact, circular disk, but instead forms an extended eccentric accretion flow. \citet{Piran2015} elaborated upon these results by showing that this is consistent with the relatively small amount of energy released in stream self-intersection shocks, and furthermore that such an elliptical disk scenario can provide a natural explanation of the observed properties (e.g. luminosity, temperature and line widths; see also \citealt{Krolik2016, Svirski2017, Ryu2020, Zanazzi2020}). \\

In this work we present the analysis of photometric and spectroscopic data of a new tidal disruption event, AT2020zso.
We describe the observations and their data reduction in Section \ref{sec:observations}. Our analysis methods and results are presented in Section \ref{sec:analysis}, and we discuss these results and their implications for accretion disk formation in tidal disruption events, as well as active galactic nuclei, in Section \ref{sec:discussion}. We summarise our conclusions in Section \ref{sec:conclusions}. Figures of the full posterior distributions for all model fitting results are provided in the Appendix, along with a table containing all the photometry used. We assume a flat $\Lambda$-Cold Dark Matter cosmology with $H_0=67.11~{\rm km\,s}^{-1}\,{\rm Mpc}^{-1}$, $\Omega_{\rm m}=0.32$ and $\Omega_\Lambda=0.68$ \citep{Planck2014a} throughout the article.

\section{Observations and data reduction}
AT~2020zso was first reported as a transient by the Zwicky Transient Facility (ZTF20acqoiyt, \citealt{alerce}), and also detected by ATLAS (ATLASbfok, \citealt{atlas}) and Gaia (Gaia20fqa, \citealt{Hodgkin21}). The host galaxy (SDSS J222217.13-071558.9) is an elliptical galaxy located at a redshift of $z$ = 0.0563. A classification spectrum \citep{atel} was obtained as part of the extended Public ESO Survey for Transient Objects (ePESSTO+; \citealt{smartt2015}), and further spectroscopic follow-up was triggered within ePESSTO+. 
A detailed observing log is presented in Table \ref{tab:obslog}. All phases are reported with respect to the phase of peak light (measured from the bolometric lightcurve) at MJD 59\,184. The optical spectroscopy will be made publicly available through WISErep.
\label{sec:observations}
\subsection{Spectroscopy}
\begin{table*}
 \caption{Observing log of spectroscopic observations. The phase is given with respect to the peak of the bolometric lightcurve, taken to be MJD 59\,184. The FWHM spectral resolution R is given at 4700$\AA$ for EFOSC2, FLOYDS and ALFOSC. }
 \label{tab:obslog}
 \begin{tabular}{ccccccccc}
  Instrument & Grism & Date & MJD & Phase & Slit width & Exposure time & Wavelength range & R \\
   & & & & (days) & (arcsec) & (Seconds) & ($\AA$) & \\\hline
EFOSC2 & Gr\#13 & 2020-11-17 & 59\,170 & --14 & 1.0 & 1500 &3685 -- 9315  & 850 \\
X-shooter & UVB & 2020-11-18 & 59\,171 & --13 & 1.0 & 1800 & 3000 -- 5600 & 5400 \\
 & VIS &  & & & 0.9 & 1920 & 5600 -- 10\,240 & 8900 \\
 & NIR & & & & 0.9JH & 1920 & 10\,240 -- 24\,800 & 5600\\
EFOSC2 & Gr\#11 & 2020-11-21 & 59\,174 & --10 & 1.0 & 1500 & 3380 -- 7520 & 1150 \\
EFOSC2 & Gr\#16 & 2020-11-21 & 59\,174 & --10 & 1.0 & 2700 & 6000 - 10\,000 & 1100\\
EFOSC2 & Gr\#11 & 2020-11-23 & 59\,176 & --8 & 1.0 & 2700 \\
FLOYDS & red/blue & 2020-11-26 & 59\,179 & --5& 2.0 & 3600 & 3200 -- 10\,000 & 250\\
X-shooter & UVB & 2020-11-28 & 59\,181 & --3 & 1.0 & 1200 \\
 & VIS & & & & 0.9 & 1320 \\
 & NIR &  & & & 0.9JH & 1320 \\
FLOYDS & red/blue & 2020-11-29 & 59\,182 & --2 & 2.0 & 3600 \\
EFOSC2 & Gr\#11 & 2020-12-09 & 59\,192 & +8 & 1.0 & 2700 \\
X-shooter & UVB & 2020-12-11 & 59\,194 & +10 & 1.0 & 1200 \\
 & VIS &  && & 0.9 & 1320 \\
 & NIR &  && & 0.9JH & 1320 \\

EFOSC2 & Gr\#11 & 2020-12-16 & 59\,199 & +15 & 1.0 & 2700 \\
ALFOSC & Grism 4 & 2020-12-17 & 59\,200 & +16 & 1.0 & 900 & 3200 -- 9600 & 360 \\
EFOSC2 & Gr\#11 & 2021-05-10 & 59\,344 & +160 & 1.5 & 2700 & & 750 \\

X-shooter & UVB & 2021-07-04& 59\,399 & +215 & 1.0 & 2600 \\
 & VIS & & & & 0.9 &  2720\\
 & NIR & && & 0.9JH & 2720 \\
  \hline
 \end{tabular}
\end{table*}

\subsubsection{New Technology Telescope / EFOSC2}
Low resolution optical spectra were taken with the EFOSC2 spectrograph mounted on the New Technology Telescope (NTT) at La Silla Observatory, Chile as part of the ePESSTO+ collaboration. We used the Gr$\#$11, Gr$\#$13 and Gr$\#$16 grisms and a 1 or 1.5 arcsecond slit width. The data reduction is performed using a dedicated pipeline \citep{smartt2015}, which includes standard tasks such as bias-subtraction, flat-fielding and a wavelength calibration based on arc frames and a comparison to sky emission lines. Cosmic rays are removed using the {\tt lacos} routine \citep{vandokkum12}. To minimise host galaxy contamination, the source extraction was performed using an extraction aperture of 1 arcsec. For some epochs (in particular, those with slid widths $> 1$ arcsec) the seeing was $> 1$ arcsec, which may lead to different galaxy light contamination in these spectra. The Gr$\#$16 observation is dominated above 7000$\AA$\ by second order contamination and is not used in our analysis. The flux calibration and extinction correction are performed using standard star observations.

\subsubsection{Very Large Telescope/ X-shooter}
Shortly after the first EFOSC2 spectrum we triggered target-of-opportunity (ToO) observations with X-shooter, mounted on the Very Large Telescope (VLT) Unit 3 (Melipal) at Paranal Observatory, Chile. A total of four spectra were obtained using slit widths of 1.0, 0.9 and 0.9 arcsec for the UVB, VIS and NIR arms, yielding a spectral resolution of $R = 5400$, $R = 8900$ and $R = 5600$, respectively.
The data were taken in on-slit nodding mode. To increase the signal to noise ratio (SNR) of the UVB and VIS arms, we reduce these data using the X-shooter pipeline with recipes designed for stare mode observations. The NIR arm is reduced with both the stare and the nodding mode X-shooter pipeline recipes. The latter method provides a slightly better sky subtraction. Regardless of the method used, no transient emission features are found in the NIR spectra; they are shown in Figure \ref{fig:xshooter} for completeness. For uniformity with the EFOSC2 spectra and to minimise host galaxy contamination, we adopt an extraction box with side 1 arcsec. Only for the last epoch (in which no TDE signal appears to be present) we use a 2 arcsec extraction box to boost the galaxy signal and determine the host galaxy properties. We use the {\it molecfit} software \citep{Smette2015} to calculate atmospheric profiles and subtract telluric absorption bands in the VIS arm, which contaminate a small region redward of the H$\alpha$ rest wavelength. The deep absorption band around 7300\AA\ is not well corrected, but does not contain any important emission lines. Figure \ref{fig:xshooter} in the appendix shows the flux calibrated X-shooter spectra.

\subsubsection{Las Cumbres Observatory / FLOYDS}
Two spectra were obtained with the low-resolution FLOYDS spectrograph mounted on the Las Cumbres Observatory 2m Faulkes Telescope North in Haleakala, Hawaii. The spectra were reduced using the \texttt{floydsspec} custom pipeline, which performs flux and wavelength calibration, cosmic-ray removal, and spectrum extraction\footnote{The pipeline is available at \url{https://github.com/svalenti/FLOYDS_pipeline/blob/master/ bin/floydsspec/}.}.

\subsubsection{Nordic Optical Telescope / ALFOSC}
One epoch of spectroscopy was obtained using a ToO program on the Nordic Optical Telescope (NOT) in La Palma, Spain. This spectrum was taken with the ALFOSC spectrograph in combination with Grism 4 and a 1 arcsec slit. This observation was reduced using custom scripts based on the pypeit Python package \citep{Prochaska20a, Prochaska20b}.\\

Following the standard data reduction recipes, we normalise all spectra to the continuum by fitting low order spline functions to the spectra, excluding known host galaxy and transient emission/absorption lines such as the He\,\textsc{ii} $\lambda 4686$, He\,\textsc{i} $\lambda 5876$ and H$\alpha$ regions. 

\subsection{Photometry}
\begin{figure*}
    \centering
    \includegraphics[width=\linewidth]{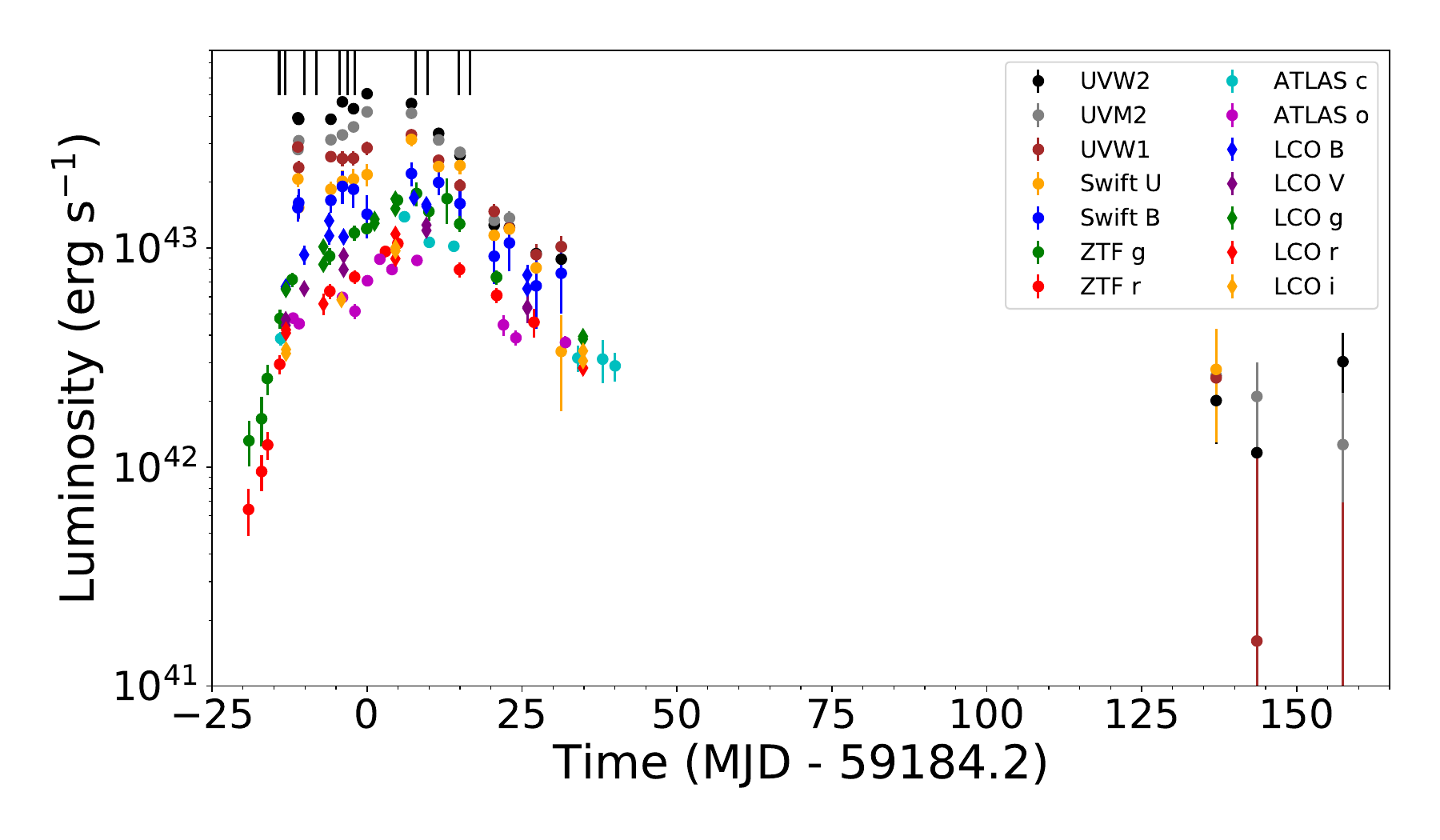}
    \caption{Host-subtracted lightcurves of AT~2020zso as observed by Swift, ZTF, ATLAS and LCO. Vertical lines indicate epochs of spectroscopic observations. As a result of colour evolution, the peak in the optical bands occurs significantly later than in the UV bands. This is consistent with the observed cooling of the blackbody temperature over time.}
    \label{fig:lc}
\end{figure*}

We retrieve the public ZTF photometry via the ZTF forced-photometry service \citep{Masci2019}. The multi-band lightcurves are shown in Figure \ref{fig:lc}.
Following the spectroscopic classification, Swift follow-up observations were triggered. The Swift/UVOT photometry is measured using the \textit{uvotsource} task in HEAsoft package v6.29 using a 5 arcsec aperture. Because no X-ray source was detected in the first observations, we derive an upper limit to the X-ray flux using the online XRT tool\footnote{https://www.swift.ac.uk/user$\_$objects/}. 
Combining all observations, we find an upper limit of 1.56$\times$10$^{-3}$ cts s$^{-1}$, which translates into a flux of 4.5$\times$10$^{-14}$ erg cm$^{-2}$ s$^{-1}$, assuming a thermal (blackbody) spectral model with a temperature of $kT = 75 ~{\rm eV}$\footnote{We used webPIMMS to simulate these values: https://heasarc.gsfc.nasa.gov/cgi-bin/Tools/w3pimms/w3pimms.pl}, typical for the soft X-ray emission in TDEs. This translates into a luminosity upper limit of 3.8$\times$10$^{41}$ erg s$^{-1}$ in the 0.3--10 keV band, uncorrected for foreground Milky Way extinction. Assuming instead an AGN like power-law spectral model (with power-law index $\Gamma$ = 1.7), this translates into an upper limit of 5.3$\times$10$^{41}$ erg s$^{-1}$ in the 0.3--10 keV band, and 4.3$\times$10$^{41}$ erg s$^{-1}$ in the 3--20 keV band.

\subsubsection{Las Cumbres Observatory}
Las Cumbres Observatory $BVgri$-band data were obtained using the Sinistro cameras on Las Cumbres 1m telescopes. PSF fitting was performed on host-subtracted images using the lcogtsnpipe pipeline \citep{Valenti16} which uses HOTPANTS \citep{Becker2015} for the subtraction, with template images obtained also at Las Cumbres after the event faded. BV-band photometry was calibrated to the Vega system using the AAVSO Photometric All-Sky Survey, then converted to the AB system using the corrections from \citet{Blanton07}, while $gri$-band photometry was calibrated to the AB system using the Sloan Digital Sky Survey \citep{Smith2002}.

\subsubsection{NIR photometry}
Two epochs of NIR photometry were taken on 2021 May 22 (MJD 59\,356) and 2021 July 15 (MJD 59\,416), at phases +172 and +232 days after peak light. The first epoch, comprising observations in the $H$ (three series of six dithered 20 second exposures, for a total of 1440 sec on source) and $K_{\rm s}$ (also 1440 sec  exposure time) bands, was taken using the Son of Isaac (SOFI) instrument mounted on the NTT in La Silla, Chile. The reduction and combination of dithered images were carried out with the PESSTO pipeline. 
The second epoch of observations, including $J$, $H$ and $K_{\rm s}$ band observations, was taken with the NOTCam instrument mounted on the NOT in La Palma via the NUTS2 programme. The NOTCam data were reduced using a version of the NOTCam Quicklook v2.5 reduction package\footnote{http://www.not.iac.es/instruments/notcam/guide/observe.html} with a few functional modifications (e.g. to increase the FOV of the reduced image).

In order to check for IR variability, we performed aperture photometry on the NIR images. We measured the brightness of the central regions of the host galaxy with a 2 arcsec aperture and calibrated the resulting magnitude against the magnitudes of field stars taken from the 2MASS catalogue. The measurements were consistent within the measurement uncertainties, which were typically $\sim$ 0.05 mag. We do not find any significant brightening in either epoch. Similarly, no brightening is observed in the NEOwise MIR lightcurves, that is there is no evidence for an on-going IR echo.

\subsection{Radio observation}
A weak radio detection by the Very Large Array was reported on 2020 December 21 (phase +20 days, \citealt{Alexanderzso}). 
A reported flux density of 22$\pm$7 micro-Jansky at 15 GHz corresponds to a monochromatic luminosity of 2.73$\pm$0.87 $\times$10$^{37}$ erg s$^{-1}$.
\subsection{Gaia astrometry}
AT~2020zso was detected by the Gaia Photometric Science Alerts (GSA; \citealt{Hodgkin21}) as Gaia20fqa at coordinates (RA, Dec) = (22:22:17.130, --07:15:59.08). This allows an accurate evaluation of the positional offset with respect to the host galaxy nucleus, which has positional coordinates listed in the Early Gaia Data Release 3 \citep{GaiaEDR3}. An offset of 42 milli-arcseconds (mas) is measured, which corresponds to 46 parsec (pc) at the host redshift. The per-transit accuracy of GSA is 55 mas \citep{Wevers2019b, Hodgkin21}, leading to an offset of 46$\pm$60 pc, consistent with a location in the nucleus of the galaxy.

\section{Analysis and results}
\label{sec:analysis}
\subsection{Host galaxy}
\subsubsection{Spectral energy distribution}
We compile the host galaxy (SDSS J222217.13-071558.9) spectral energy distribution (SED) using archival observations in the UV through IR bands (see Table \ref{tab:hostgalphot}). In the near-IR we use 2MASS \citep{Cutri_03} $J$, $H$ and $K_{\rm s}$ magnitudes, and we use the Pan-STARRS DR1 magnitudes in $g$, $r$, $i$, $y$ and $z$ optical bands. Finally, for the UV we perform aperture photometry in the GALEX \citep{Bianchi_11} $NUV$ and $FUV$ images with the {\tt gPhoton} package \citep{Million_16}. We perform forced aperture photometry in all available bands with two distinct apertures. For one set, we use 5 arcsec apertures; this is used to subtract the host contribution from the observed lightcurves (in particular the Swift photometry is performed with a 5" aperture). For the other set, we use an elliptical aperture with major and minor axis of 13 and 9 arcsec, optimised to include the entire host galaxy flux; this is used to model the host galaxy and derive its properties. 

We model the SED using the flexible stellar population synthesis \citep[FSPS, ][]{Conroy_09} module. We use the {\tt Prospector} \citep[][]{Johnson_21} software to run a Markov Chain Monte Carlo (MCMC) sampler \citep[][]{Foreman-Mackey_13}. We assume an exponentially decaying star formation history (SFH), and a flat prior on the five free model parameters: stellar mass ($M_{\star}$), stellar metallicity ($Z$), $V$-band extinction \citep[$A_V$, assuming the extinction law from][]{Calzetti_2000}, the stellar population age ($t$) and the e-folding time of the exponential decay of the SFH ($\tau_{\rm{sfh}}$).

Using the median and 1-$\sigma$ confidence intervals of the posteriors of the fit to the 9" by 13" aperture photometry, we derive a host stellar mass of $\log(M_{\star}/M_{\odot})= 10.11^{+0.02}_{-0.02}$, a metallicity of  $\log(Z/Z_{\odot})=-0.34^{+0.04}_{-0.07}$, $A_V=0.23^{+0.03}_{-0.04}$ mag, $t = 1.90^{+1.56}_{-0.94}$ Gyr, and $\log(\tau_{\rm{sfh}}) =2.42^{+0.35}_{-0.32}$ Gyr. The extinction is roughly consistent with the Galactic foreground extinction of $E(B-V) = 0.06$ \citep{Schlafly2011}. The estimated mass combined with the extinction-corrected rest-frame
colour $u-r = 1.89\pm0.02$ mag places the host galaxy near the “green valley” region \citep{Schawinski_2014} of the mass colour diagram, in which the TDE host galaxies seems to be over represented compared to the general galaxy population \citep[][]{Law-Smith_17,van_Velzen_21,Hammerstein_21}. 

To estimate the host galaxy fluxes in the UVOT bands, we similarly model the host galaxy SED but using the 5" aperture data. The host contribution is then subtracted from the measured photometry, which is also corrected for foreground Galactic extinction. The uncertainty on the host galaxy model is propagated into our measurement of the host-subtracted TDE flux (see Table \ref{tab:hostgalphot}).

\begin{table}
    \centering
    \begin{tabular}{c|cc}
        Band & Observed & Model\\
        & (AB mag) & (AB mag) \\\hline
       GALEX $FUV$ & 20.97 (0.23)& 21.04	(0.10) \\
       GALEX $NUV$ & 20.24 (0.10)& 20.50	(0.10) \\
       PS1 $g$ & 17.93 (0.02)&  17.87	(0.01) \\
       PS1 $r$ & 17.29 (0.01)& 17.29	(0.01) \\
       PS1 $i$ & 16.96 (0.01)&  16.97	(0.01) \\
       PS1 $y$ & 16.67 (0.04)& 16.66	(0.01) \\
       PS1 $z$ & 16.81 (0.01)& 16.77	(0.01) \\
       2MASS $J$ & 16.10 (0.08)& 16.47 (0.02) \\
       2MASS $H$ &16.48 (0.16) & 16.31 (0.02) \\
       2MASS $K_{\rm s}$ & 16.15 (0.11)& 16.52	(0.03) \\
       UVOT $U$ & ---& 19.42	(0.04) \\
       UVOT $B$ & ---& 18.26	(0.02)  \\
       UVOT $V$ & ---& 17.58	(0.01)  \\
       UVOT $UVW2$ & ---& 20.62	(0.09) \\
       UVOT $UVM2$ & ---& 20.52	(0.10) \\
       UVOT $UVW1$ & ---& 20.23	(0.08) \\
       
    \end{tabular}
    \caption{Results of the host SED model fitting of the 5" aperture data. Values between brackets indicate the uncertainties, which are propagated into the host subtracted photometry.}
    \label{tab:hostgalphot}
\end{table}

\begin{figure}
    \centering
    \includegraphics[width=\linewidth]{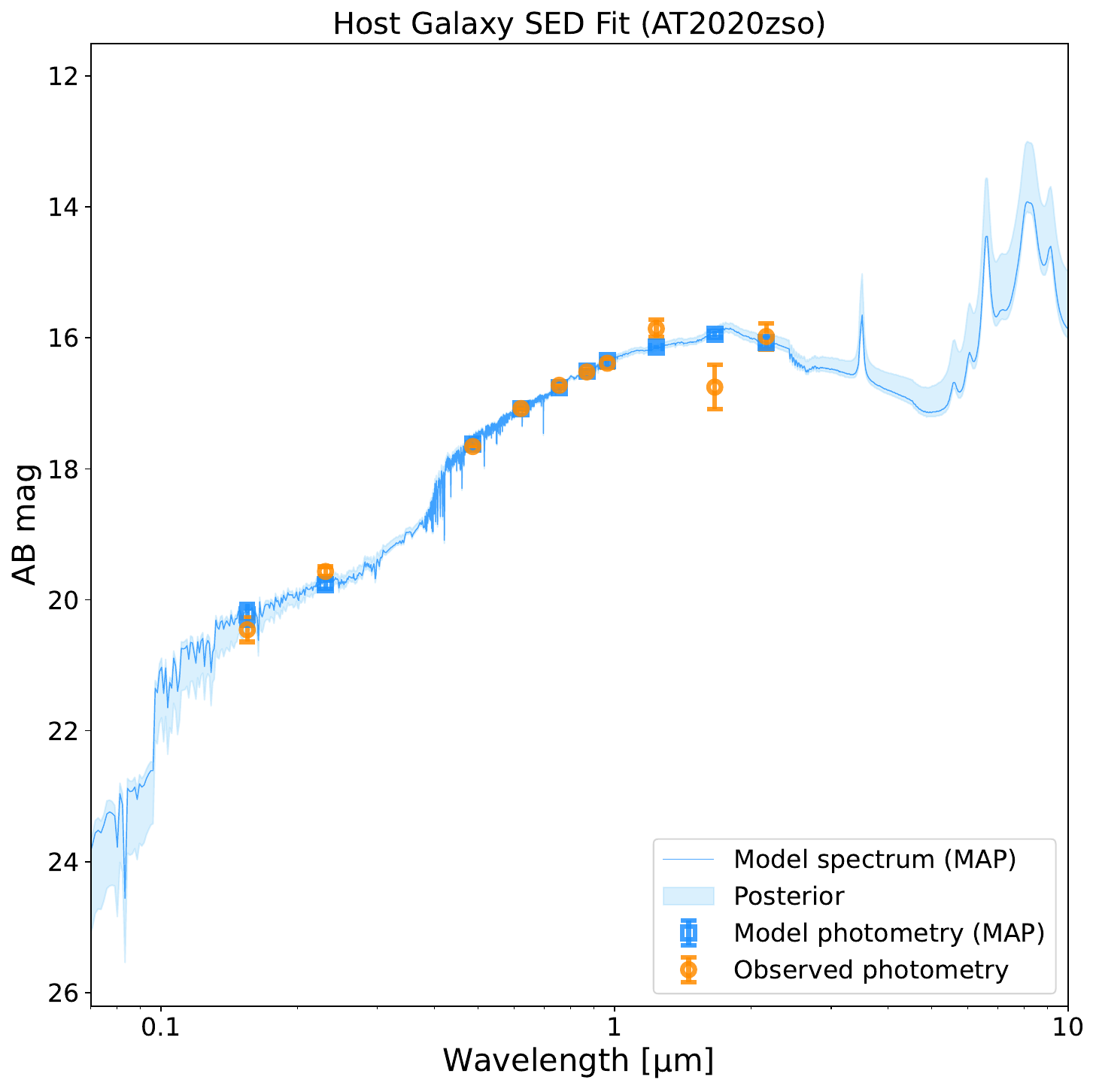}
    \caption{Host galaxy spectral energy distribution and best-fit template used to synthesise host galaxy magnitudes in the Swift bands. The data are tabulated in Table \ref{tab:hostgalphot}.}
    \label{fig:hostsed}
\end{figure}

\subsubsection{Black hole mass}
Using the late time X-shooter spectrum, in which no broad emission lines are apparent, we measure the velocity dispersion of the host galaxy following the method of \citet{Wevers2017}, using the penalized Pixel Fitting routine (PPXF, \citealt{Cappellari2017}). We find a velocity dispersion of $\sigma = 60\pm1$ km s$^{-1}$. Using the $M-\sigma$ relation from \cite{MM13} this translates to a black hole mass of $\log(M_{\rm BH}) = 5.3 \pm 0.4$, or alternatively $\log(M_{\rm BH}) = 6.2\pm0.3$ M$_{\odot}$ using the \citet{Kormendy13} relation, indicating a low mass black hole similar to many other UV/optical discovered TDEs \citep{Wevers2019}.

\subsubsection{Emission lines}
\label{sec:emlines}
From the narrow host galaxy emission lines, which are resolved in the X-shooter spectra, we measure a redshift of $z = 0.0563$, which corresponds to a luminosity distance of 263 Mpc.

\begin{figure*}
    \centering
    \includegraphics[width=\linewidth]{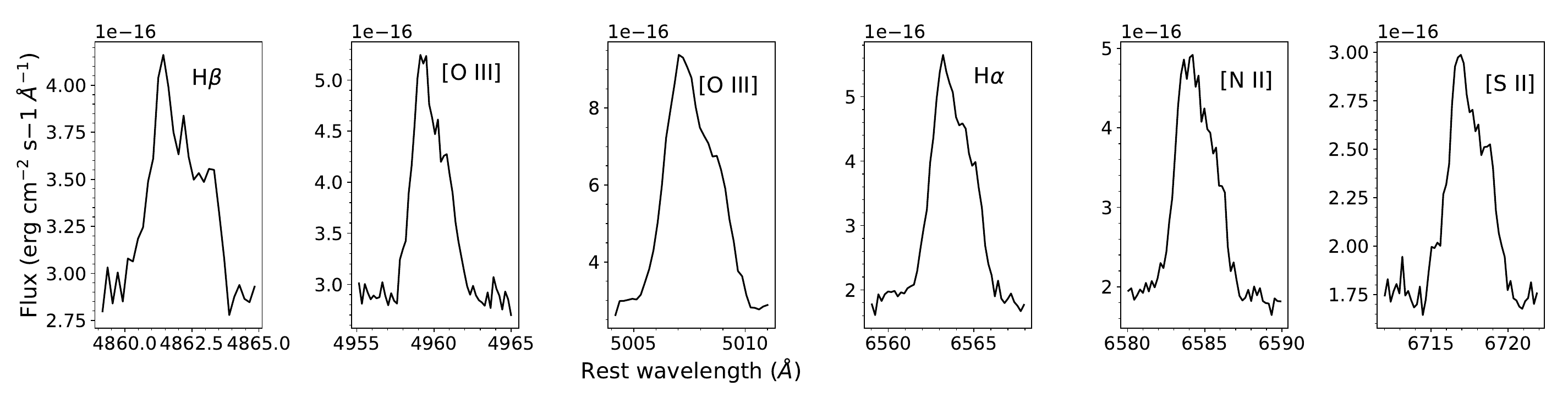}
    \caption{Insets of the narrow emission lines observed in AT2020zso.}
    \label{fig:narrowlines}
\end{figure*}

We identify a plethora of narrow emission lines originating in the host galaxy, including in order of increasing wavelength: the [O\,\textsc{ii}] $\lambda\lambda$3726, 3729 doublet, [Ne\,\textsc{iii}] $\lambda3869$, He\,\textsc{ii} $\lambda$4686, H$\beta$, the [O\,\textsc{iii}] $\lambda\lambda$4959, 5007 doublet, He\,\textsc{i} $\lambda5876$, [O\,\textsc{i}] $\lambda 6300$ line, the [N\,\textsc{ii}] $\lambda\lambda6548, 6584$ doublet, H$\alpha$, the [S\,\textsc{ii}] $\lambda\lambda6717, 6731$ doublet, a (very) weak [Ar\,\textsc{iii}] $\lambda7136$ line, the [S\,\textsc{iii}] $\lambda\lambda9069, 9532$ doublet lines, and Pa $\alpha$ $\lambda 1.875 \mu m$. High ionisation potential lines such as He\,\textsc{ii}, [Ar\,\textsc{iii}] and [S\,\textsc{iii}] indicate that a hard photo-ionising continuum source is present, while there is no sign of a broad component to any of these lines at late times. We measure a FWHM from the [O\,\textsc{iii}] line of 159$\pm$3 km s$^{-1}$, while for the narrow H$\alpha$, H$\beta$, N\,\textsc{ii} and S\,\textsc{ii} lines we measure an average of FWHM = 127$\pm$5 km s$^{-1}$. Closer inspection shows that some of these narrow lines are asymmetric/double-peaked, with a velocity separation of $\sim 70-80$ km s$^{-1}$. Figure \ref{fig:narrowlines} shows some of the prominent narrow emission line profiles. We measure the asymmetry of the [O\,\textsc{iii}] $\lambda$5007 line, the strongest narrow emission line, by using the nonparametric measurement of \citet{Liu2013b}, and find a very small asymmetry $A = 0.047$ (other lines yield similar values). This suggests that the narrow line region is rotation dominated, but probably not kinematically disturbed \citep{Blecha2013, Nevin2016}, which makes it very unlikely that the system hosts a dual AGN or a wide-separation SMBH binary. We measure an [O\,\textsc{iii}] line luminosity from the late time spectrum of L$_{\rm [O III]}$ = 1.17$\pm$0.05 $\times$10$^{40}$ erg s$^{-1}$. Assuming a correlation between L$_{\rm [O III]}$ and the 3--20 keV X-ray luminosity observed in AGN \citep{Heckman2005}, we expect an AGN X-ray luminosity of L$_X$ = 1.6$\times$10$^{42}$ erg s$^{-1}$. The upper limit for L$_X$ derived from Swift observations is 4.3$\times$10$^{41}$ erg s$^{-1}$ in the 3--20 keV band, i.e. marginally inconsistent, given the large scatter (a factor $\approx$ 3) in the correlation. Reconciling these two values (again assuming a power-law spectrum with index $\Gamma = 1.7$) would require an absorbing column of at least $\sim$ 1--1.5$\times$ 10$^{22}$ cm$^{-2}$. 

\begin{figure}
    \centering
    \includegraphics[width=\linewidth]{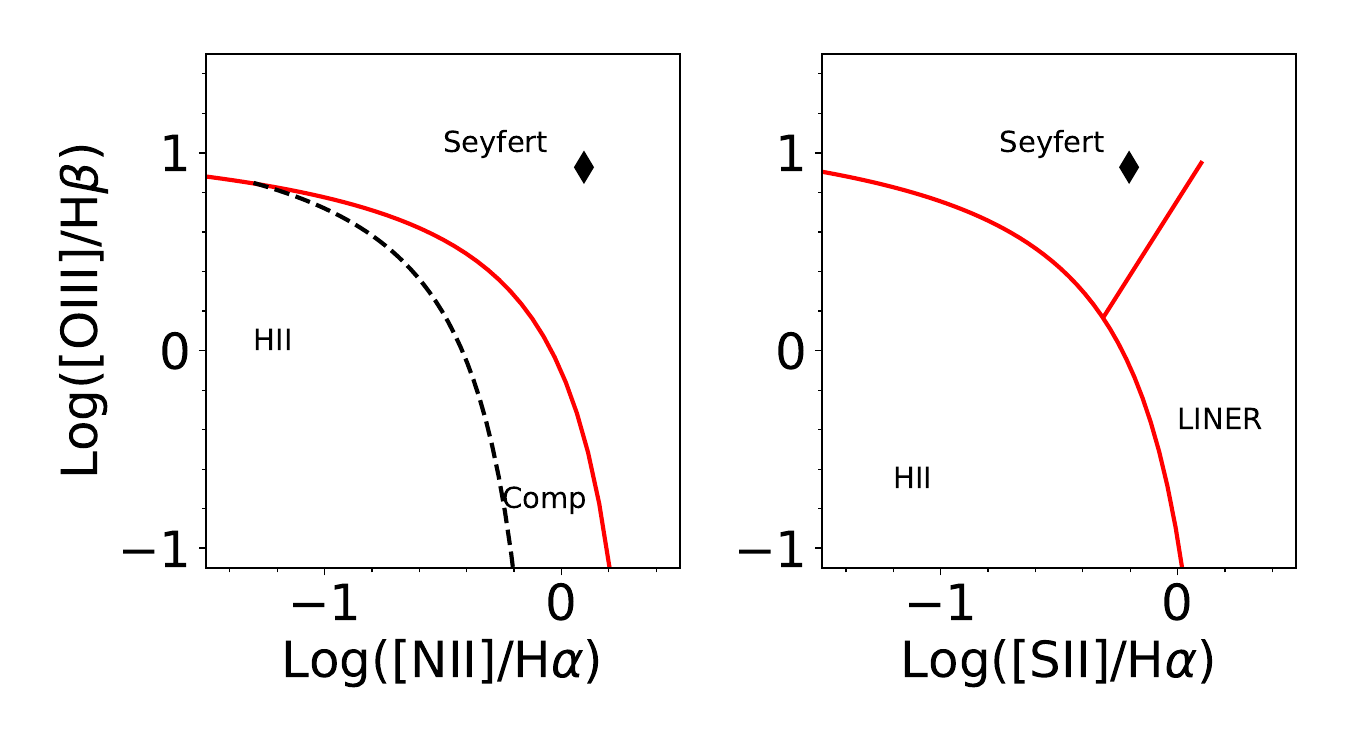}
    \caption{Narrow emission line ratios as measured from the late-time X-shooter spectrum. The results place the host galaxy firmly into the AGN (Seyfert) part of the diagram. Dividing lines are taken from \citet{Kewley2001} and (\citeyear{Kewley2006}). }
    \label{fig:bpt}
\end{figure}

We also use the narrow host galaxy line ratios to put the source on a Baldwin-Philips-Terlevic (BPT) diagram \citep{Baldwin}. We find that the source falls well into the AGN/Seyfert part of the diagram, in line with the presence of high ionisation narrow emission lines (Figure \ref{fig:bpt}). We conclude that this galaxy hosts a Seyfert AGN.

\subsection{Lightcurve evolution}
The host-subtracted lightcurves are shown in Figure \ref{fig:lc}. The best sampling is achieved in the ZTF bands, particularly at very early times. There appear to be three distinct phases in both the g- and r-band lightcurves: a very steep initial rise, followed by a break to a slower increase in brightness and finally a turnover to a decline in brightness. To characterise the lightcurve behaviour at early times, we fit a power-law model to the two parts of the rising ZTF lightcurve (before and after the break) independently, of the form 
\begin{equation}
    L = a + b \times (t-t_0)^{\alpha}
\end{equation}
We find that the early rising part of the lightcurve is consistent with L $\propto$ t$^2$ evolution ($\alpha$ = 1.9 $\pm$ 0.4). 
After the break (which happens around phase = --12 days), the slope flattens to $\alpha$ = 1.55 $\pm$ 0.25. We remark that the emission lines contribute $<$10$\%$ of the total light, and therefore do not significantly influence the lightcurve evolution nor the inferred parameters.

\subsubsection{Parameter inference with MOSFit and TDEMASS}
We fit the lightcurve using a TDE model in the MOSFit package \citep{Guillochon2017, Mockler2019}, employing the same free parameters and priors as in \citet{Mockler2019}. We used the \textsc{dynesty} dynamic nested sampling algorithm with default stopping criteria to explore the parameter space and sample the model posteriors \citep[see][for details]{Speagle2020}. The MOSFit TDE model only includes the fallback luminosity as an energy source, while at later times $\gtrsim$100 days (e.g. \citealt{Mummery2019}) there may be a significant contribution from an accretion disk, leading to a flattening of the light curve. We therefore exclude data points more than 100 days after peak luminosity from the fit, while noting that a fit including these data does not significantly alter the black hole mass and stellar mass estimates. Fits were run on the University of Birmingham BlueBEAR cluster.

\begin{figure}
    \centering
    \includegraphics[width=\linewidth]{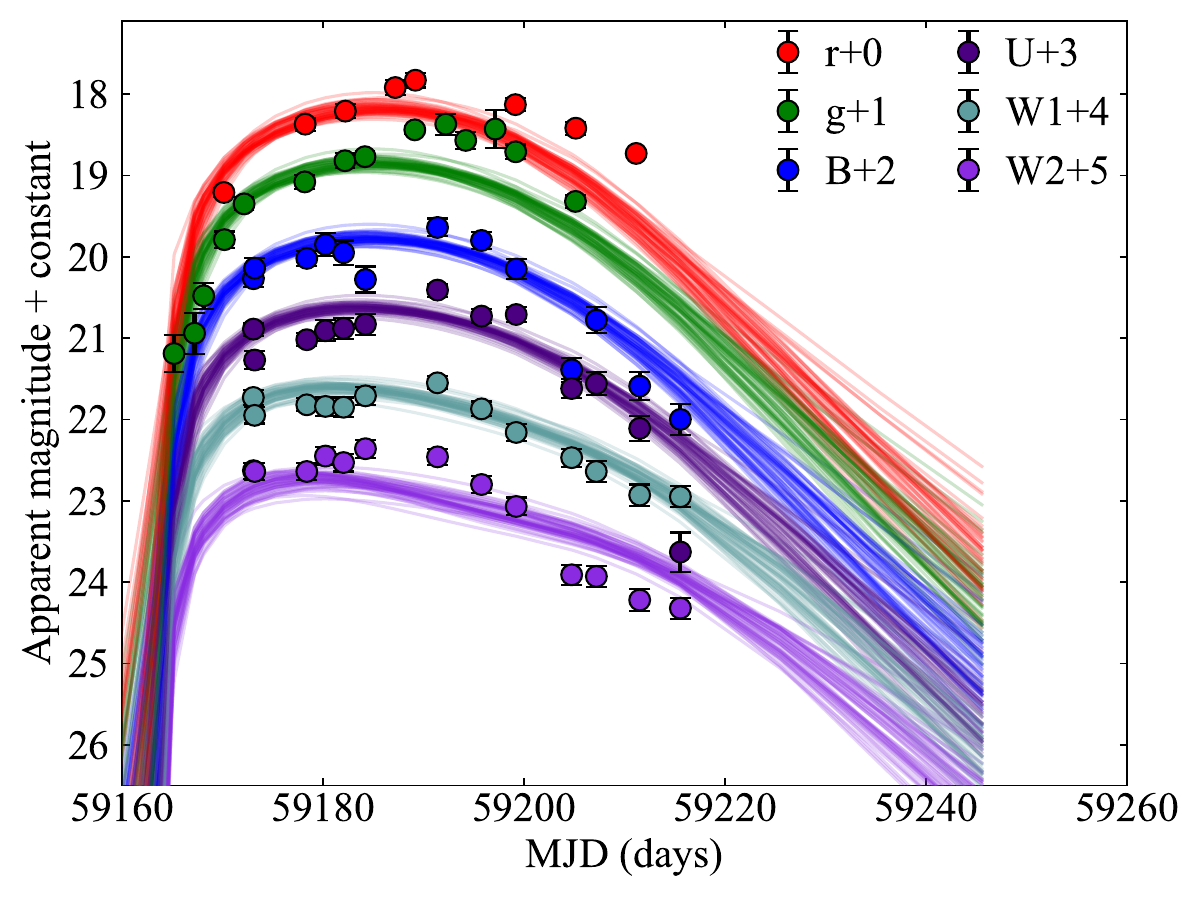}
    \caption{MOSFit model lightcurves overlaid on the data. The fits are not great, likely due to the relatively rapid temperature evolution and peculiar light curve behaviour (including a break in the rising part) which the model cannot accommodate. }
    \label{fig:mosfit}
\end{figure}

The results are a poor fit (Figure \ref{fig:mosfit}); there are short term variations that are not encapsulated by the model itself, so these are not expected to be reproduced, and the temperature variation is more rapid than the model can accommodate. These results should therefore be interpreted with some caution.
\begin{table}
    \centering
    \begin{tabular}{cc}
        Parameter & Value \\\hline \vspace{2mm}
        $\log_{10}(M_{\rm BH})$ [M$_{\odot}$] & 5.9 -- 6.1 [0.2] \\\vspace{2mm}
        Stellar mass [M$_{\odot}$] & 0.08 -- 0.13 [0.36]\\\vspace{2mm}
        Impact parameter $\beta$ & 0.58 -- 0.68 [0.35]\\\vspace{2mm} 
        $\epsilon$ & 0.05 -- 0.13\\\vspace{2mm}
        $\log_{10}(R_{\rm pho})$ & 2.65 -- 3.31 \\\vspace{2mm}
        Photospheric exponent l & 3.45 -- 3.91  \\\vspace{2mm}
        $\log_{10}(T_{\rm visc})$ [days] & --1 -- 0.82 \\\vspace{2mm}
        $t_{\rm expl}$ [days] & --10.2 -- --5  \\\vspace{2mm}
        $\log_{10}(\sigma)$ & --0.47 -- --0.39 \\
         &  \\\hline
        
    \end{tabular}
    \caption{Results of the lightcurve fitting with MOSFit (16--84 percentiles). The systematic uncertainties are taken from \citet{Mockler2019} and provided in square brackets, when available.}
    \label{tab:mosfit}
\end{table}
A black hole mass of $\log_{10}(M_{\rm BH})$ = 6.0 $\pm$ 0.3 is inferred from the lightcurve, which is consistent with the estimate from the stellar velocity dispersion. Furthermore, we obtain estimates of the disrupted stellar mass (at the lower allowed limit of $\sim0.1 ~M_{\odot}$) and the impact parameter $\beta \equiv \frac{R_p}{R_t}$ = 0.63 $\pm$ 0.05 (where R$_p$ is the orbital pericentre radius, and R$_t$ the tidal radius), although there are large systematic uncertainties of 0.66 dex or 0.36 in a linear scale for a value of 0.1 for the stellar mass, and 0.35 for the impact parameter (see \citealt{Mockler2019} for a detailed discussion of the systematic uncertainties produced by MOSFit). The results of these fits are reported in Table \ref{tab:mosfit}, and full posterior distributions for the fits can be found in the Appendix. These values are very similar to those found by \citet{Gomez2020} for the other double-peaked TDE AT~2018hyz, and indicate that AT~2020zso may likewise be the result of a partial, rather than a full, disruption (as inferred from the fact that $\beta <$ 0.9, e.g. \citealt{Guillochon2013}). Simulations by \citet{Ryu2020c} suggest that the surviving stellar remnant may have lost $\sim$ 40 per cent of its original mass (but keeping in mind the large systematic uncertainties, this could range from a few up to $>$ 60 per cent). 
Finally, MOSFit suggests an disruption date of $8\pm2$ days before the first datapoint, at MJD 59\,157 $\pm$ 2 days.

Alternatively, \citet{Ryu2020} presented a framework to infer the black hole and disrupted stellar masses on the basis of eccentric accretion disk dynamics. Using the peak bolometric luminosity of 7$\pm$1$\times$10$^{43}$ erg s$^{-1}$ and a peak colour temperature of 25000$\pm$5000 K, a black hole mass of M$_{\rm BH}$ = 1.7$^{+2.0}_{-0.9}$ M$_{\odot}$ and stellar mass of M$_{\star}$ = 0.92$^{+0.21}_{-0.13}$ M$_{\odot}$ are inferred; the former is consistent with alternative estimates from galaxy scaling relations, while the latter differs significantly from the MOSFit estimate. Given that the TDEMASS framework was explicitly developed on the basis of eccentric accretion disk dynamics, which appear to be particularly suited for application to AT2020zso, we give preference to these inferences.

\subsubsection{Blackbody modeling}
\begin{figure*}
    \centering
    \includegraphics[width=1.\linewidth]{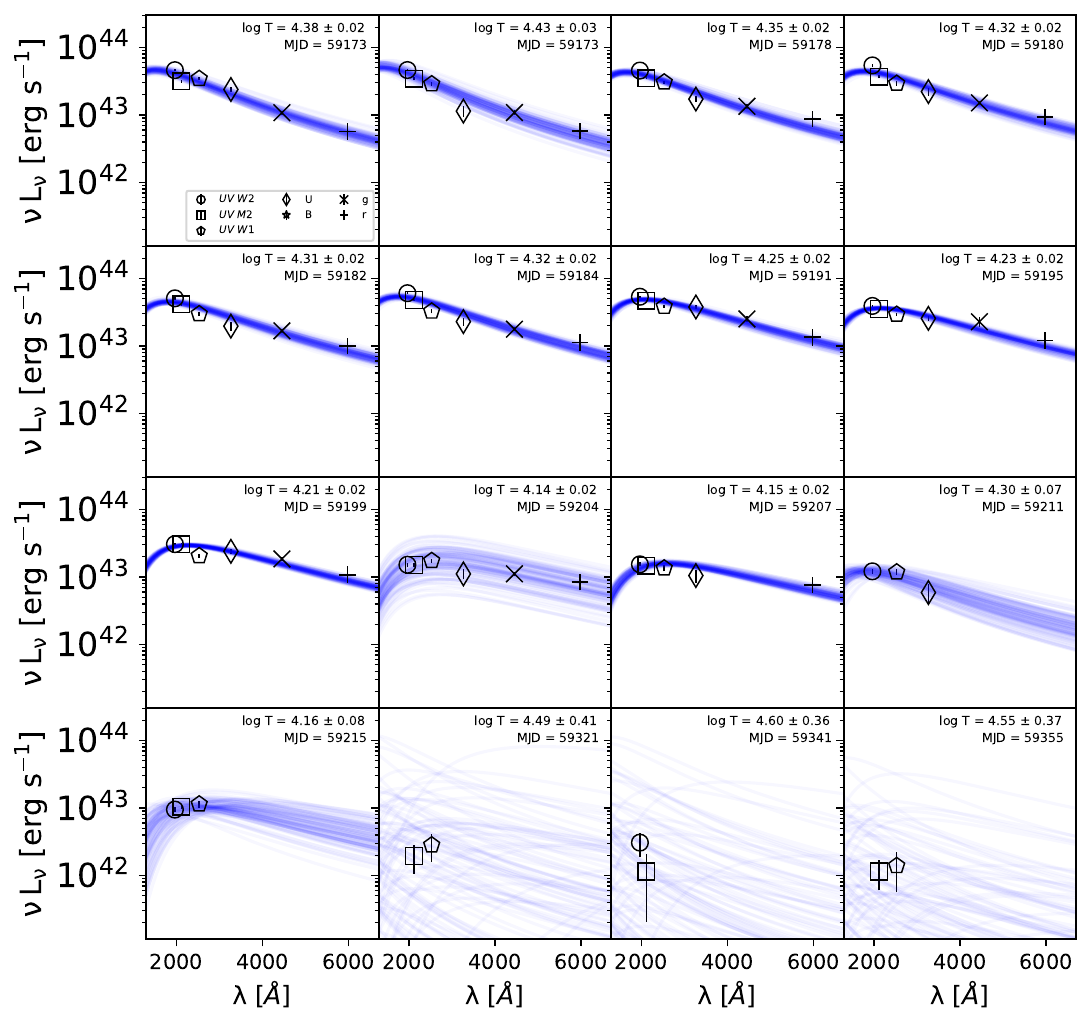}
    \caption{Blackbody fitting of the TDE spectral energy distribution over time. The uncertainties become very large for the last three epochs, as two datapoints do not provide much constraining power for the blackbody model. }
    \label{fig:seds}
\end{figure*}

The properties of UV/optical TDE flare emission can be empirically be well described using evolving blackbody models. This is somewhat surprising, given that there is growing evidence for a viewing angle dependence of the observational consequence of stellar destruction \citep{Dai2018, Leloudas2019}, implying that asymmetry is present in the ensuing structure. As a result, it is unclear to what degree the results of blackbody fitting the TDE SED can be physically interpreted. 

\begin{figure}
    \centering
    \includegraphics[width=\linewidth]{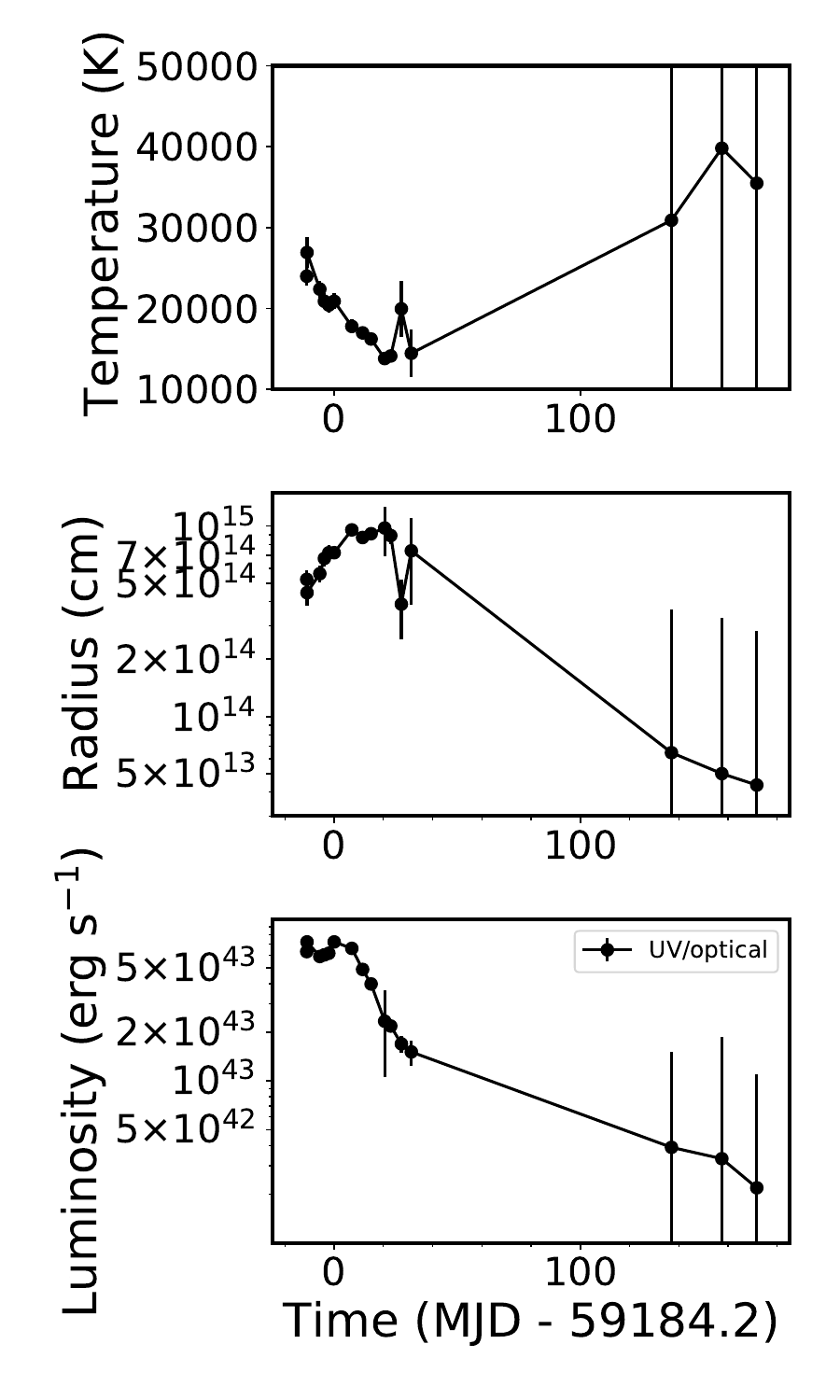}
    \caption{Blackbody parameter evolution of AT2020zso over time. Strong cooling is observed during the first part of the lightcurve.}
    \label{fig:templumrad}
\end{figure}

With this caveat in mind, for each epoch of Swift observations, we model the SED using a blackbody curve, although the physical interpretation of this model is unlikely to be straightforward, as the powering source of this emission remains unclear. We include all the host-subtracted Swift photometry, and linearly interpolate the ZTF $g$- and $r$-band measurements to these epochs to provide coverage from 2000 -- 7000 \AA. We do not extrapolate the ZTF measurements beyond their latest observing epochs. We use a maximum likelihood approach to fit a blackbody model to each epoch, assuming a flat prior for all parameters; the resulting fits are illustrated in Figure \ref{fig:seds}. Uncertainties are assessed by sampling from the posterior distributions of the parameters directly. Assuming isotropic emission also yields the characteristic blackbody radius. The temperature, radius and bolometric blackbody luminosity are shown in Figure \ref{fig:templumrad}; the values based only on ZTF (without temperature fit, but with a bolometric correction) data are shown as green triangles. 

The temperature decreases by $\sim$ 10\,000 K during the first part of the lightcurve. Several sources in the \citet{van_Velzen_21} sample show similar behaviour. Such cooling is typically seen in TDEs after peak, likely as a result of an expanding photosphere, although the effect is particularly strong for AT~2020zso (a similar effect was seen in AT2019qiz, \citealt{Nicholl2020}). The MOSFit results show a similar temperature evolution, although the temperature changes somewhat slower (this is intrinsic to the TDE model). There is some indication of an increasing trend at later times (again similar to typical TDE behaviour), but the uncertainties are large.

\begin{figure}
    \centering
    \includegraphics[width=\linewidth]{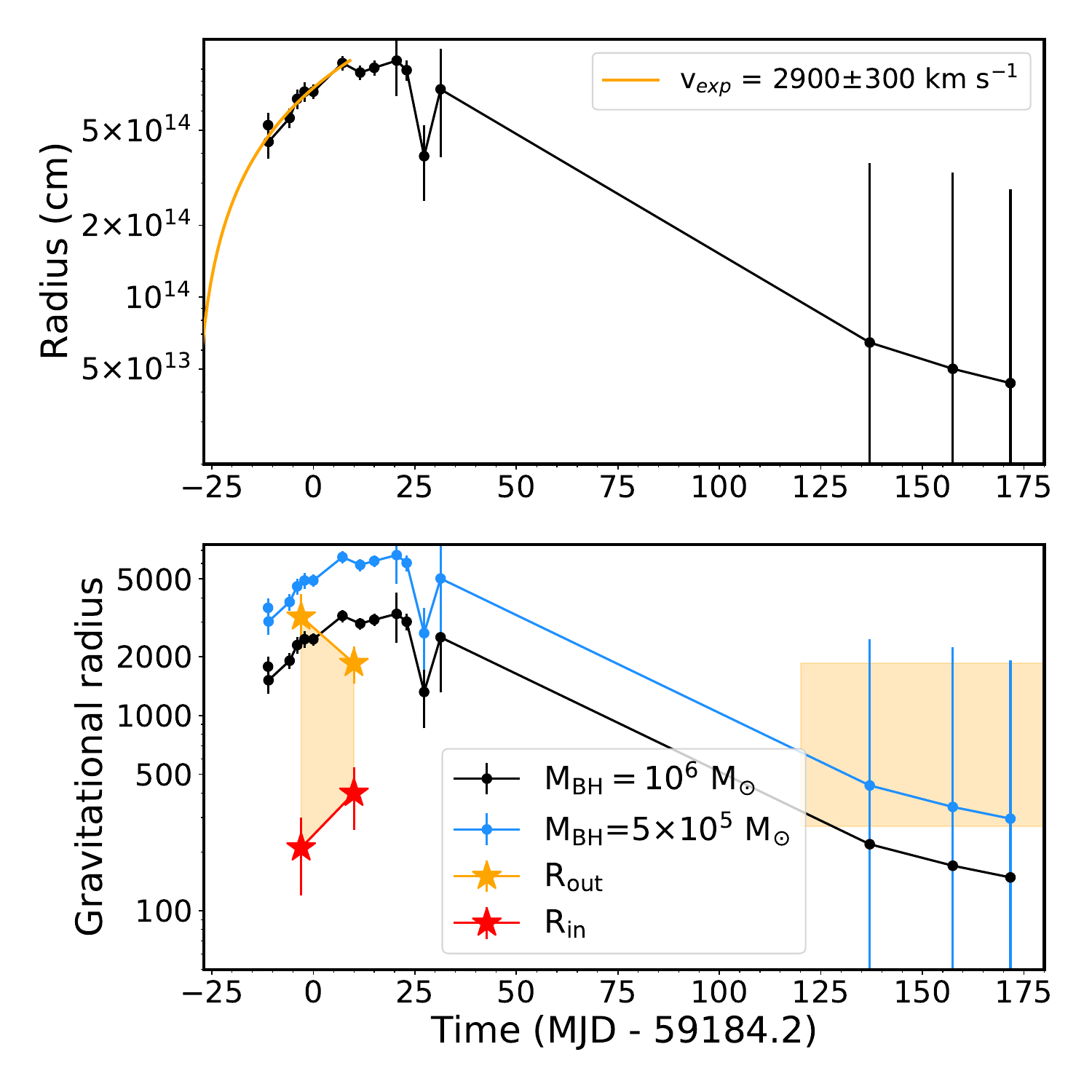}
    \caption{Top panel: evolution of the blackbody radius, overlaid with best-fit linear expansion model (orange line). At late times the radius reaches approximately $5\times10^{13}$ cm (but with large error bars), which corresponds to several 100 $R_{\rm g}$ for $M_{\rm BH} = 10^6$ $M_{\odot}$. Bottom panel: same evolution, but now converted to units of gravitational radius for two different black hole masses. The orange and red stars mark the inferred disk outer and inner radii, respectively (see Section \ref{sec:fitting}). At late times, the blackbody radii overlap with the estimated accretion disk size (highlighted with the shaded area, assuming that the disk size has not changed significantly from the last epoch that can be modeled).}
    \label{fig:radevol}
\end{figure}

The photoshere radius evolution follows a linear expansion profile initially. We measure an expansion velocity of $v_{\rm exp}$ = 2900 $\pm$ 300 km s$^{-1}$ before peak light (Figure \ref{fig:radevol}, top panel). Afterwards, the radius reaches a plateau before moving back inwards to scales $\approx$ $5\times10^{13}$ cm. This behaviour is very similar to that observed in AT2019qiz \citep{Nicholl2020} and AT2019ahk \citep{Holoien201919bt}. Based on the expansion velocity before maximum, we estimate that the first observations were taken approximately 15 days after expansion began, suggesting a disruption date around MJD 59\,149, and a rise time of approximately 35 days from disruption to peak (compared with an explosion date of MJD 59\,157 from MOSFit). This value is insensitive to the assumption about the temperature evolution before the first Swift observations.

We note that an expanding photosphere does not necessarily require a physical outflow (i.e. outward fluid motion) to be present. Alternatives to explain the photosphere expansion include the accumulation of matter around the peak of the mass fallback rate, which extends the photosphere to larger radii; it could be the result of time-dependent photon diffusion due to changing density and/or optical depth in the debris; or due to the orbital motion of heated matter, which at the inferred radius of $\sim$5$\times$10$^{14}$ cm is comparable ($\sim$5000 km s$^{-1}$) to the measured growth rate. 

\subsection{Transient emission features}
\begin{figure*}
    \centering
    \includegraphics[width=\linewidth]{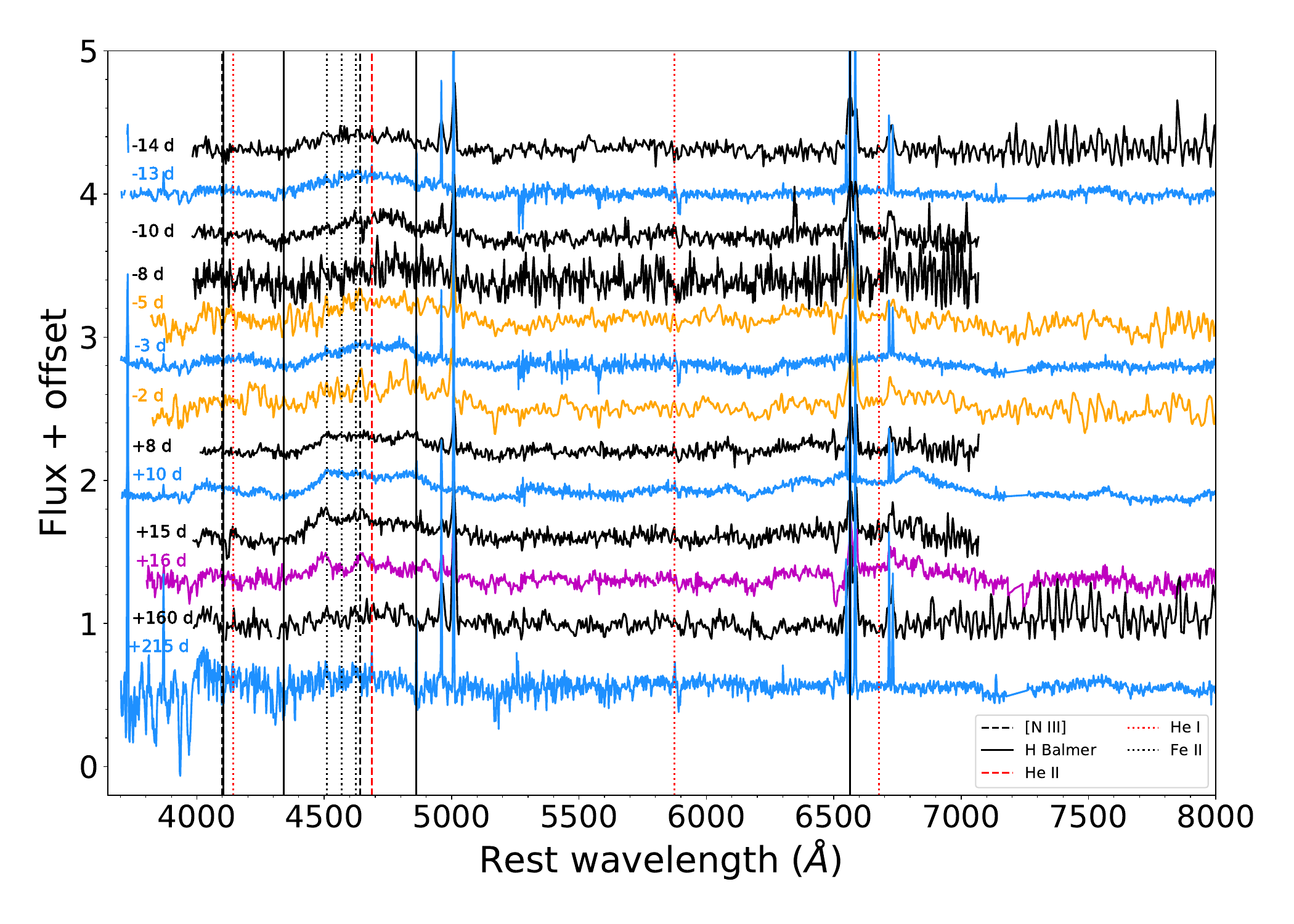}
    \caption{Spectral evolution of AT~2020zso from early to late times. All spectra are continuum normalised. The different emission features in the 4600~\AA\ blend are indicated with vertical lines: solid black lines indicate H Balmer lines, dashed black lines N\,\textsc{iii}, dotted black lines Fe\,\textsc{ii}, red dashed lines He\,\textsc{ii}, and red dotted lines He\,\textsc{i}. Black spectra were taken with NTT/EFOSC2; blue with VLTX-shooter, orange with LCO/FLOYDS; and magenta with NOT/ALFOSC. The X-shooter spectra are resampled at lower spectral resolution for clarity. The gap around 5200\AA\ is due to the low SNR of the data at the edges of the UVB and VIS arms, and the gap around 7300\AA is due to the poor correction of molecular bands.}
    \label{fig:normspec}
\end{figure*}

\begin{figure*}
    \centering
    \includegraphics[width=\linewidth]{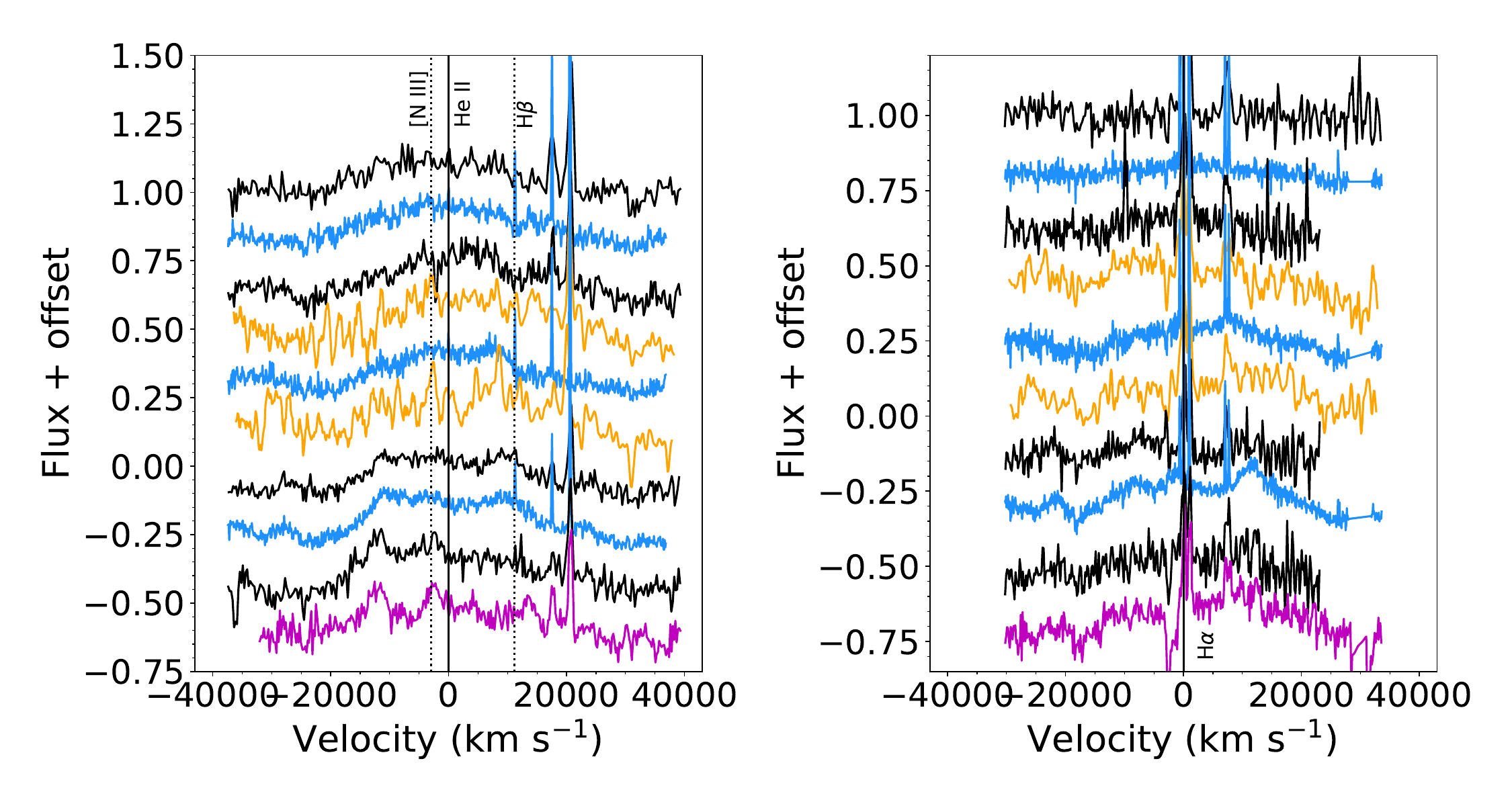}
    \caption{Same as Figure \ref{fig:normspec}, but now in velocity space. The left panel shows the He\,\textsc{ii} region, while the right panel shows the region around H$\alpha$.}
    \label{fig:normspec2}
\end{figure*}

In addition to the narrow host galaxy lines, broad evolving emission lines are present in the spectra, which are shown in Figures \ref{fig:normspec} and \ref{fig:normspec2}.
These lines are typically (quasi)-Gaussian in TDEs, with velocity shifts up to $\sim$15\,000 km s$^{-1}$. However, in AT~2020zso there are several emission features whose identification would be contrived, or completely unclear, when taking this approach. For example, a broad feature centred on 4500~\AA\ appears around --4 d. While this could in principle be broad Fe\,\textsc{ii} emission often seen in AGN, the profile appears smooth with a broad blue wing that would be atypical. Similarly, broad features centred on 4250 \AA, 5050 \AA, 6080 \AA\ and 6820 \AA\ are present in several of the spectra. No similar features have been readily identified in other TDEs to date, and no immediately obvious line identifications are available for these wavelengths.

\begin{figure*}
    \centering
    \includegraphics[width=1.1\linewidth]{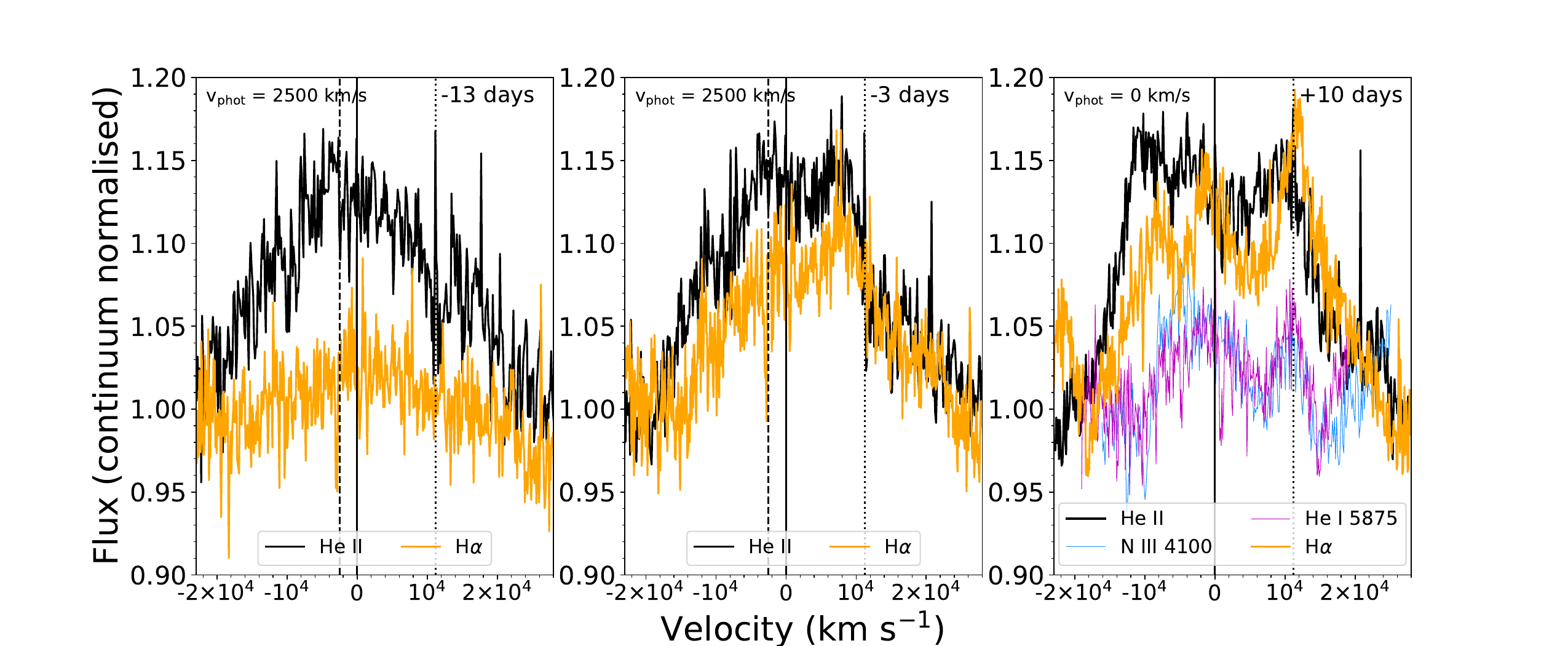}
    \caption{VLT/X-shooter spectra at various epochs, plotted in velocity space for He\,\textsc{ii} (black) and H$\alpha$ (orange). To guide the eye we show the rest velocity as a solid line; the photospheric velocity with a vertical dashed line; this happens to coincide with the expected position for N\,\textsc{iii} $\lambda$4640; and the position for H$\beta$ with respect to He\,\textsc{ii} as a black dotted line. For display purposes, narrow emission lines have been removed and the spectra are resampled to lower spectral resolution. }
    \label{fig:xsh_vel}
\end{figure*}

Noting that these broad emission features appear to be roughly symmetrical around rest wavelengths of $\sim$ 4700 \AA\ and 6560~\AA, we instead explore the idea that these are multi-peaked structures constituting a single feature, that is emission lines of He\,\textsc{ii}/H$\beta$ and H$\alpha$. This is motivated by previous studies that have identified double-peaked emission lines, attributed to accretion disk structures, in TDEs \citep{Arcavi2014, Short2020, Hung2020}. 

The velocity structure of these lines is shown in Figure \ref{fig:xsh_vel}; it is encouraging that the profiles appear very similar in this representation. We identify the emission features with two main contributors, consistent with rest wavelengths of He\,\textsc{ii} $\lambda$4686\ and H$\alpha$. Because H$\alpha$ appears centred near rest velocity, we disfavour an identification as H$\beta$ for the emission feature near 4700\AA; such an identification would require a large systematic blueshift (--11\,000 km s$^{-1}$), whereas for He\,\textsc{ii} the line would also be centred near rest velocity. Nevertheless, the broad feature around 5050~\AA\ may identify as the red wing of a similar double-peaked velocity profile consistent with H$\beta$, albeit much weaker than He\,\textsc{ii}. No other broad He\,\textsc{ii} emission lines (e.g. at $\lambda\lambda 3203, 10123\AA$) are evident in the spectra. We also note that there are other emission features seen in TDEs around this region, most notably the Bowen N\,\textsc{iii} 4640 line. We will explore possible contamination in more detail in Section \ref{sec:gauss}.

\begin{figure}
    \centering
    \includegraphics[width=\linewidth]{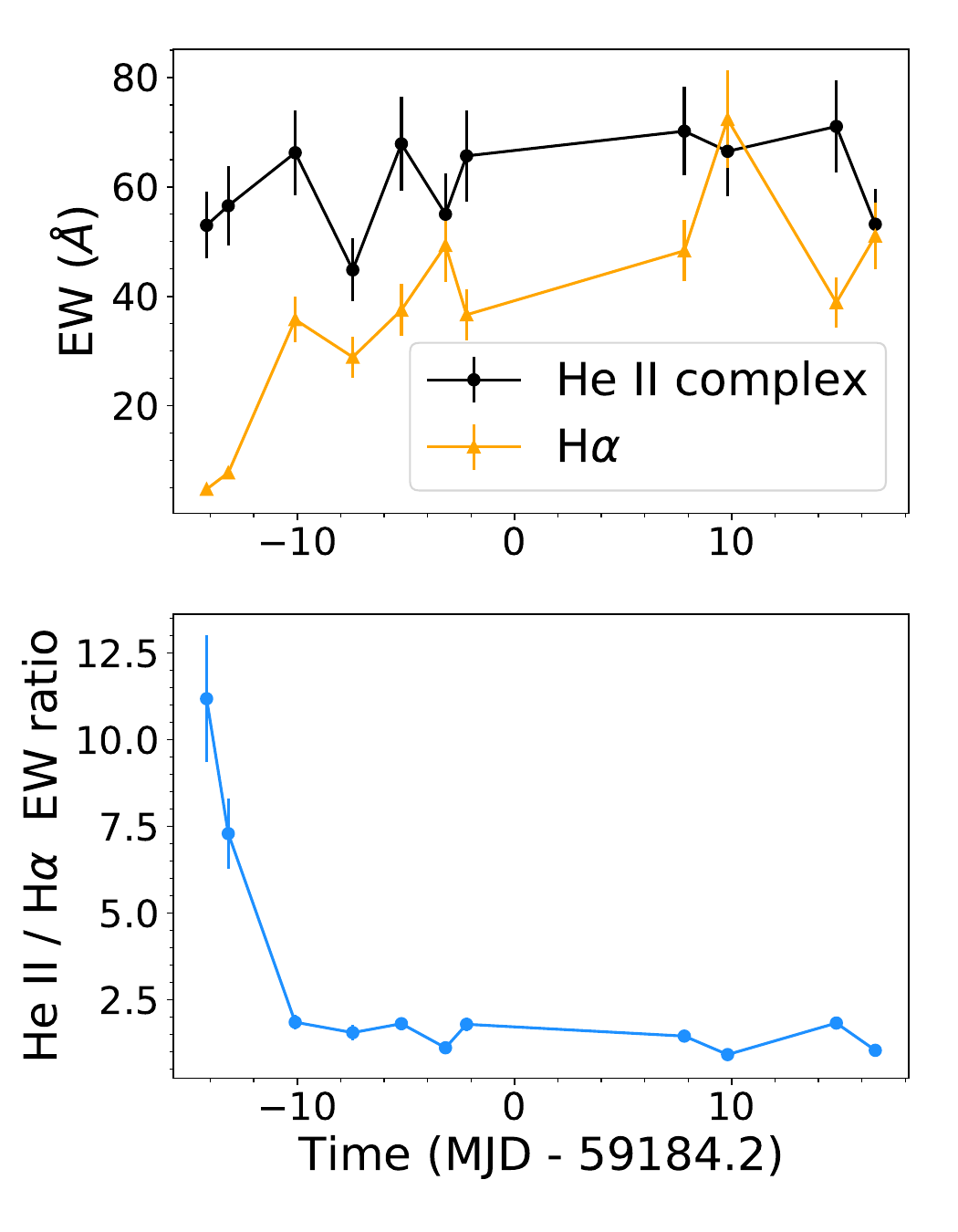}
    \caption{Top panel: equivalent width of the He\,\textsc{ii} complex (black circles ) and H$\alpha$ (orange triangles), measured through direct integration from the spectra. Bottom: their ratio (He\,\textsc{ii} / H$\alpha$) as a function of time.}
    \label{fig:ewevolution}
\end{figure}

Given the strongly non-Gaussian line profiles, we measure the emission line equivalent widths (EWs) of He\,\textsc{ii} complex and H$\alpha$ through direct integration. We mask out telluric absorption features in the spectra when present. The EW evolution and their ratio is shown in Figure \ref{fig:ewevolution}. The He\,\textsc{ii} / H$\alpha$ ratio decreases rapidly from $\approx$ 11 (phase --14 days), to 7 (phase --13 days), to stabilise around 1.5 (phases later than --10 days). For the earliest epochs, we also measure the EW ratio by using a Gaussian profile, which yields somewhat lower values ($\approx$ 8 and 4 at --14 and --13 d, respectively) but shows a consistent, rapidly decreasing trend.

The first four epochs are well described (reduced $\chi^2 < 1.3$) by a broad, single Gaussian, and for these spectra we also attempt to measure the line velocities and FWHM of He\,\textsc{ii} and H$\alpha$. The results indicate that both lines are likely at rest velocity. All measurements are consistent with 0 within 3$\sigma$, although we find large variations between different spectra that are unphysical given that they were taken only days apart, likely due to the relatively low SNR. The "best" early spectrum available was taken with X-shooter (phase --13 days), from which we measure line velocities of 800 $\pm$ 400 km s$^{-1}$ and --170 $\pm$ 135 km s$^{-1}$ for H$\alpha$ and He\,\textsc{ii}, respectively. From the same spectrum, we measure Gaussian FWHM values of 36\,000 $\pm$ 7000 (H$\alpha$) and 31\,000 $\pm$ 1000 (He\,\textsc{ii}) km s$^{-1}$ (for a reduced $\chi^2$ of 1.2).

\subsection{Emission line evolution}
\label{sec:evolution}
We plot the broad emission line profiles of He\,\textsc{ii} and H$\alpha$ in Figure \ref{fig:xsh_vel}. For clarity, we focus on the highest spectral resolution (X-shooter) spectra. We start by noting that there appears to be a delay between the emergence of He\,\textsc{ii} (which appears earlier) and H$\alpha$, which is very weak in the earliest epochs but strengthens $\sim$ 10 days after the first observation (as is also apparent from the EW evolution in Fig. \ref{fig:ewevolution}). The line profile of He\,\textsc{ii} is slightly asymmetric and peaks around --3000 km s$^{-1}$, which would be consistent with the photospheric outflow velocity measured through blackbody modeling. 

Near peak light (phase --3 days) the spectra become distinctly non-Gaussian. The He\,\textsc{ii} line profile is inconsistent with the presence of a very broad Gaussian (similar to the one observed at --13 days) -- this component would extend well beyond 4300 \AA, where no excess flux is observed. Both lines show a double-peaked structure, centred near rest wavelength; He\,\textsc{ii} appears roughly symmetric whereas H$\alpha$ shows a strong asymmetry, with a bright red peak and a broad blue shoulder (no clear blue peak is visible in the spectrum). The red peak of both lines occurs at similar velocities ($\sim$ 8000 km s$^{-1}$), and the red wing is nearly identical in velocity structure, extending out to $\approx$ 26\,000 km s$^{-1}$. On the blue side, the He\,\textsc{ii} profile extends roughly 5000 km s$^{-1}$ further blueward (out to 18\,000 km s$^{-1}$) compared to H$\alpha$, and is a factor of $\sim$ 2 brighter than H$\alpha$. This could indicate either that the He\,\textsc{ii} emission originates from a region with a different velocity profile, or potential contamination of other emission lines known to be present in this region (including N\,\textsc{iii} $\lambda$4640 and Fe\,\textsc{ii} lines), discussed in Section \ref{sec:gauss}.

After peak light (phase +10 days), the profiles have significantly evolved. H$\alpha$ shows a pronounced triple-peaked structure. In particular the red peak of the profile is remarkable, being significantly brighter and more narrowly peaked than its blue equivalent. Similar to the previous epoch, the red wings of H$\alpha$ and He\,\textsc{ii} have comparable velocity structures. However, the red peak has moved to higher velocities, particularly for H$\alpha$ ($\sim$ 11\,500 km s$^{-1}$). He\,\textsc{ii} remains broader and brighter in the blue wing, extending out to 21\,000 km s$^{-1}$. The blue H$\alpha$ peak is situated around --8000 km s$^{-1}$, whereas He\,\textsc{ii} peaks closer to --10\,000 km s$^{-1}$. The "central" peak of H$\alpha$ is more pronounced and near rest velocity. 

We note that the photospheric radius has reached a plateau at this phase (+10 d), that is the outflowing photosphere reaches its maximum radius before becoming optically thin and receding inwards. The origin of these central features could therefore be either from the accretion disk itself\footnote{See Figure \ref{fig:modelevolution} in the Appendix for an example of how large eccentricities lead to this third middle bump.}, or (less likely) in this outflowing component near maximum radius, consistent with the low observed velocities.

It is worth noting that there is no pronounced central peak in He$\textsc{ii}$, although some feature may be present. This is unlikely to be the blue wing of a double-peaked H$\beta$ line profile, because i) this would imply a velocity of --14\,000 km s$^{-1}$, significantly larger than both the H$\alpha$ and He\,\textsc{ii} blue peak velocities, and ii) for H$\alpha$ the red peak is brighter than the blue peak, whereas such an identification would imply a stronger blue peak for H$\beta$. Alternatively, this feature could be consistent with Bowen N\,\textsc{iii} $\lambda$4640. 

This remarkable triple-peaked structure is also seen in He\,\textsc{i} $\lambda$5876 emission lines (right panel of Figure \ref{fig:xsh_vel}).
An almost identical feature is also present near 4100$\AA$. This could be identified as H$\delta$, although this is inconsistent with the weakness of H$\gamma$ and H$\beta$. A more likely possibility is the N\,\textsc{iii} Bowen line at 4100\ \AA\ (see Section \ref{sec:gauss}). These profiles may also be present in the earlier X-shooter epochs, albeit very weak. The wings of these profiles extend from --10\,000 km s$^{-1}$ to 12\,000 km s$^{-1}$, so are significantly more narrow than H$\alpha$ and He\,\textsc{ii}. This could imply that they originate in lower velocity regions of the disk, that is further out than H$\alpha$ and He\,\textsc{ii}.  Their central peak also appears near rest velocity. We note that He\,\textsc{ii} shows broader peaks compared to the profiles of the lower ionisation lines. At the same time, the He\,\textsc{ii} blue peak appears brighter than the red peak, which is opposite to the H$\alpha$ profile. This is likely due to contamination of the Bowen N\,\textsc{iii} line at 4640\ \AA, as quantified below.

\subsection{Deblending the He\,\textsc{ii} complex}
\label{sec:gauss}
As noted above, the line profiles of He\,$\textsc{ii}$ differ somewhat from those of H$\alpha$: in particular, the blue peak is either equal or brighter than the red peak, whereas in H$\alpha$ the red peak is always stronger than the blue peak. 

The region around He\,\textsc{ii} contains a number of emission lines that are observed in TDEs, including N\,\textsc{iii} $\lambda 4640$, H$\beta$ and Fe\,\textsc{ii} lines. To investigate this in more detail, we have tried to fit the entire emission feature with a superposition of Gaussians for the aforementioned elements. We are not able to find consistencies in the derived parameters for the respective lines (e.g. line identification, velocity, FWHM) if we include single Gaussian components in addition to He\,\textsc{ii}. 

\begin{figure}
    \includegraphics[width=0.45\paperwidth]{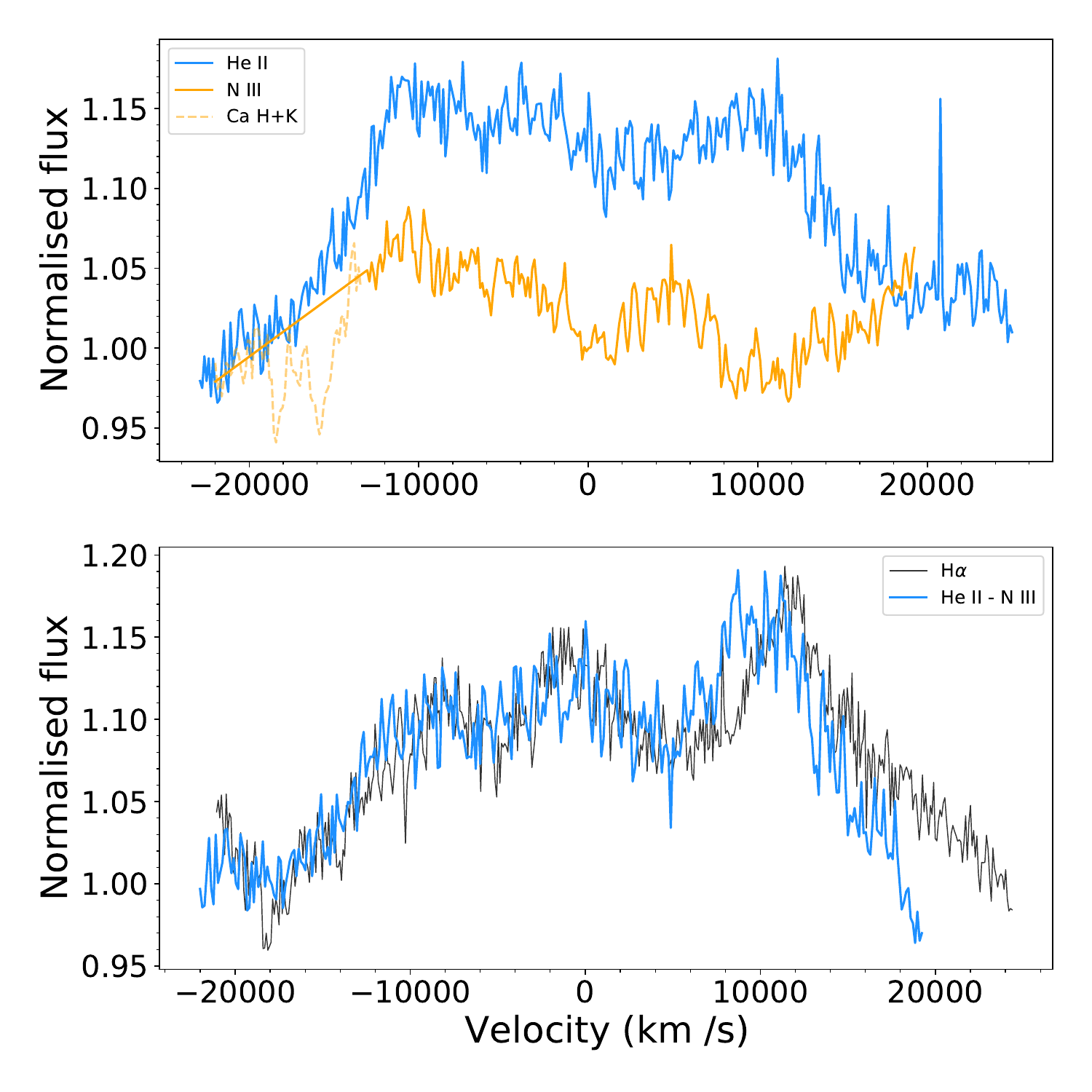}
    \caption{Top panel: He\,\textsc{ii} (blue) and N\,\textsc{iii} 4100\ \AA\ lines from the X-shooter spectrum at +10 days. Bottom panel: Subtracted spectrum (He\,\textsc{ii} -- N\,\textsc{iii}, blue) and H$\alpha$ for comparison (in black). The line profiles are nearly identical, suggesting the presence of double-peaked Bowen N\,\textsc{iii} lines, the first such line profiles seen in a TDE to date. The discrepancy in the red wing (compared to H$\alpha$ is likely related to the imperfect normalisation and subtraction of the N\,\textsc{iii} profile.}
   \label{fig:he2subtraction}
\end{figure}

Instead, we turn to the emission lines observed during the epoch at +10 days. In particular, the double-peaked feature centred on 4100\AA\ is unlikely to be H$\delta$, given the absence of both H$\beta$ and H$\gamma$. Instead, this line can be identified as a N\,\textsc{iii} Bowen line. From a spectroscopic study of a sample of TDEs, it was found that the N\,\textsc{iii} 4100 and 4640\AA\ lines have a roughly 1:1 flux ratio \citep{Panos2021} when they are present. Given that the 4100\AA\ line is readily identified in the X-shooter spectrum at +10 days, we attempt to subtract this line profile from the He\,\textsc{ii} line, assuming a 1:1 flux ratio and identical velocity structure (so we assume that N\,\textsc{iii} 4640 is identical to N\,\textsc{iii} 4100). The red wing of the N\,\textsc{iii} 4100 appears to be contaminated by another potential emission line. This feature is centred around 4250\AA, and it is not immediately clear to which element this line can be attributed. In order not to overfit the spectra, we have normalised the entire UVB spectrum (in the range 3700 -- 5250 \AA) using a low order spline function, and this feature therefore remains in the normalised spectrum. The N\,\textsc{iii} 4100 and He\,\textsc{ii} features are shown in the top panel of \ref{fig:he2subtraction}. The contamination of N\,\textsc{iii} manifests as the increasing flux trend at velocities above 10\,000 km s$^{-1}$, and may lead to some systematic uncertainties in the fitting described later. To avoid biasing the results in the blue wing, we linearly interpolate over the Ca H+K region (shown by the dashed line in the top panel). The bottom panel of Figure \ref{fig:he2subtraction} shows the result of the subtraction (in blue) overlaid on the (unmodified) H$\alpha$ line profile. The result of the red wing contamination leads to a somewhat narrower red wing profile in velocity space. Notwithstanding the simplifying assumptions for the various line profiles, the subtracted He\,\textsc{ii} profile is remarkably similar to the H$\alpha$ line profile. We conclude that AT2020zso belongs to the spectroscopic class of He+Bowen line TDEs, and that the He\,\textsc{ii} line likely originates from the same physical region as the H$\alpha$ and Bowen N\,\textsc{iii} lines. 

We perform the same subtraction procedure also for for the X-shooter spectrum at phase --3 days, and present the results in Figure \ref{fig:he2subtraction_app} in the Appendix. We will use these subtracted He\,\textsc{ii} line profiles for the accretion disk modeling. It is likely that the subtraction procedures introduce some systematic errors (e.g. the contamination in the red wing of N\,\textsc{iii}), so we treat the fits to the H$\alpha$ lines as our primary results. We will also present the He\,\textsc{ii} fitting results, but keeping in mind the caveats mentioned here.

\subsection{Elliptical accretion disk fitting}
\label{sec:fitting}
Before we describe our model fitting results, we remind the reader that a general prediction of relativistic circular accretion disk models is that the blue peak is equal to, or brighter than, the red peak, due to Doppler boosting of emission from the region moving in our direction (e.g. \citealt{Chen1989a, Chen1989b}). This is no longer true for eccentric accretion disks: for non-negligible eccentricities, there exist orientation and inclination angle combinations such that the red peak can appear brighter than the blue peak. In other words, when interpreting a line profile as originating from an accretion disk, a double-peaked line profile with a dominant red peak is a strong indication of a non-axisymmetric (eccentric) configuration. There exist other mechanisms that can produce similar asymmetric line profiles, which will be discussed in Section \ref{sec:discussion}.

Double-peaked emission lines have been observed in other TDEs, where they were interpreted as signatures of a spiral wave or an accretion disk. For the TDE with the most convincing double-peaked emission profiles so far (AT2018hyz; \citealt{Short2020, Hung2020}), an almost circular geometry was derived through model fitting. In our analysis we take a similar approach to \citet{Hung2020}, and attempt to fit a general relativistic accretion disk model of \citet{Eracleous1995} to the H$\alpha$ and He\,\textsc{ii} lines. We do not model the N\,\textsc{iii} and He\,\textsc{i} lines because of their limited SNR, and furthermore we do not present modeling results for the lines in low resolution spectra because this leads to degeneracies and inconsistent results.

This model has seven free parameters, including the emissivity power-law index $q$, the broadening parameter $\sigma$, the major axis orientation of the elliptical rings $\phi_0$ (for 0 degrees, the nodal line is along our line of sight, and the apocentre is in our direction), the inclination angle $i$ (where 0 degrees is face-on), the eccentricity $e$, and the inner and outer pericentre distances (that is the line emitting region of the disk is bounded by elliptical annuli of radii $r_1$ and $r_2$), and is described by the following specific intensity profile $I_{\nu}$:
\begin{equation}
\centering
    I_{\nu} = \frac{1}{4 \pi} \frac{\epsilon_0\ \xi^{-q}}{\sqrt{2\pi} \sigma} {\rm exp} \left[ - \frac{(\nu^2 - \nu_0^2)}{2\sigma^2} \right]
\end{equation}
The line profile flux $F \equiv \int d\nu \int\int d\Omega I_{\nu}$, that is the integral of $I_{\nu}$ over frequency, specific intensity and solid angle, is calculated by numerical integration and rescaled to fit the observations. The fit is performed by varying two normalisation constants $A$ and $B$, $A + B \times F$ (where $A\approx 1$ is the adjacent continuum level for each spectrum and $B$ is the amplitude of the profile) to the data to account for small (few per cent) differences in the normalisation level.

We focus our analysis on the X-shooter spectra as they have superior spectral resolution and signal-to-noise ratio. The first epoch (at --13 days) is omitted, because at these early times the line emitting regions likely originate in an outflowing photosphere rather than in an accretion disk like structure (as is implied by the EW evolution and will be discussed later). We nevertheless attempted a fit for this epoch, but the results are highly degenerate and do not allow robust parameter inference.
We fit each emission line profile (He\,\textsc{ii} and H$\alpha$) separately, for each epoch, for a total of four fits. 
We use a nested MCMC sampling approach implemented in {\tt dynesty}, with uniform priors for all parameters as summarised in Table \ref{tab:priors}. Because the computational time of this approach is proportional to the size of the parameter space, we first create a grid of $\sim$ 231\,000 models and perform a least-squares minimisation to assess the best-fit models. These results are used to inform the prior ranges, which are nevertheless taken very conservatively to encompass most of the plausible parameter space. Following the results presented in \citet{Hung2020}, we report on the results of a composite accretion disk + outflowing component model fit to the data, where the outflow is represented by a Gaussian component. 

\begin{table}
    \centering
    \begin{tabular}{c|cc}
        Parameter & Prior range\\\hline
        Emissivity index ($q$) & 2 -- 3\\
        Line broadening ($\sigma$) & 500 -- 4500 km s$^{-1}$ \\
        Inclination ($i$) & 0 -- 90 degrees\\
        Eccentricity ($e$) & 0 -- 1\\
        Orientation angle ($\phi_0$) & 0 -- 360 degrees \\ 
        Inner radius ($r_1$) & 100 -- 550 $R_{\rm g}$\\
        Outer radius ($r_2$) & 750 -- 4750 $R_{\rm g}$\\
        A & 0.95 -- 1.05\\
        B & 0 -- 0.2\\\hline
        Amplitude & 0 -- 0.125 \\
        FWHM (narrow) & 0 -- 3000 km s$^{-1}$ \\
        Velocity shift (narrow) & -3000 -- 1000 km s$^{-1}$\\
        FWHM (broad) & 22\,500 -- 40\,000 km s$^{-1}$\\
        Velocity shift (broad) & --4000 -- 4000 km s$^{-1}$\\\hline
        
    \end{tabular}
    \caption{Priors used for MCMC accretion disk + Gaussian model fitting. Values below the horizontal line pertain to the Gaussian component. An exploratory study using a grid of models covering a wide parameter range was used to inform upon the range of some priors. }
    \label{tab:priors}
\end{table}

\begin{table*}
    \caption{Results of the MCMC accretion disk + Gaussian model fitting. For completeness we also provide the results of the Gaussian-only model for the first epoch, where the accretion disk model does not provide a good fit.}
    \centering
    \begin{tabular}{c|ccccccccccc}
        Line & $q$ & $\sigma$ & $i$ & $e$ & $\phi_0$ & r$_1$ & r$_2$ & Vel. & FWHM & Ampl. \\
        & & (km s$^{-1}$) & ($^{\circ}$) & & ($^{\circ}$) & (R$_g$) & (R$_g$)  & (km s$^{-1}$) & (km s$^{-1}$) & \\\hline
        H$\alpha$, --13d & --- & --- & --- & --- & --- & --- & ---  & 800$\pm$400& 37\,500$\pm$6500 & 0.02$\pm$0.01 \\
        He\,\textsc{ii}, --13d & --- & --- & --- & --- & --- & --- & ---  & --170$\pm$135 & 31\,000$\pm$850 & 0.13$\pm$0.01\\
        H$\alpha$, --3d & 2.41$\pm$ 0.04 & 1018$\pm$273 & 85$\pm$6 & 0.96$\pm$0.02 & 256$\pm$2 & 302$\pm$13 & 4095$\pm$315  & --944$\pm$1019 & 1014$\pm$957 & 0.02$\pm$0.01\\
        He\,\textsc{ii}, --3d & 2.12$\pm$0.05 & 1570$\pm$205 & 89$\pm$1 & 0.97$\pm$0.01 & 219$\pm$2 & 121$\pm$12 & 2250$\pm$375  & 675$\pm$380 & 1750$\pm$650 & 0.02$\pm$0.01\\
        H$\alpha$, +10d & 2.28$\pm$0.04 & 1270$\pm$145 & 78$\pm$5 & 0.96$\pm$0.02 & 242$\pm$2 & 270$\pm$11 & 1453$\pm$74 & --494$\pm$935 & 2169$\pm$818 & 0.02$\pm$0.01  \\
        He\,\textsc{ii}, +10d & 2.92$\pm$0.06 & 1330$\pm$100 & 86$\pm$3 & 0.97$\pm$0.01 & 208$\pm$1 & 530$\pm$10 & 2100$\pm$90 & 400$\pm$90 & 2400$\pm$110 & 0.07$\pm$0.01 \\\hline

    \end{tabular}

    \label{tab:elldisk_result}
\end{table*}

\begin{figure*}
    \centering
    \includegraphics[width=0.49\linewidth]{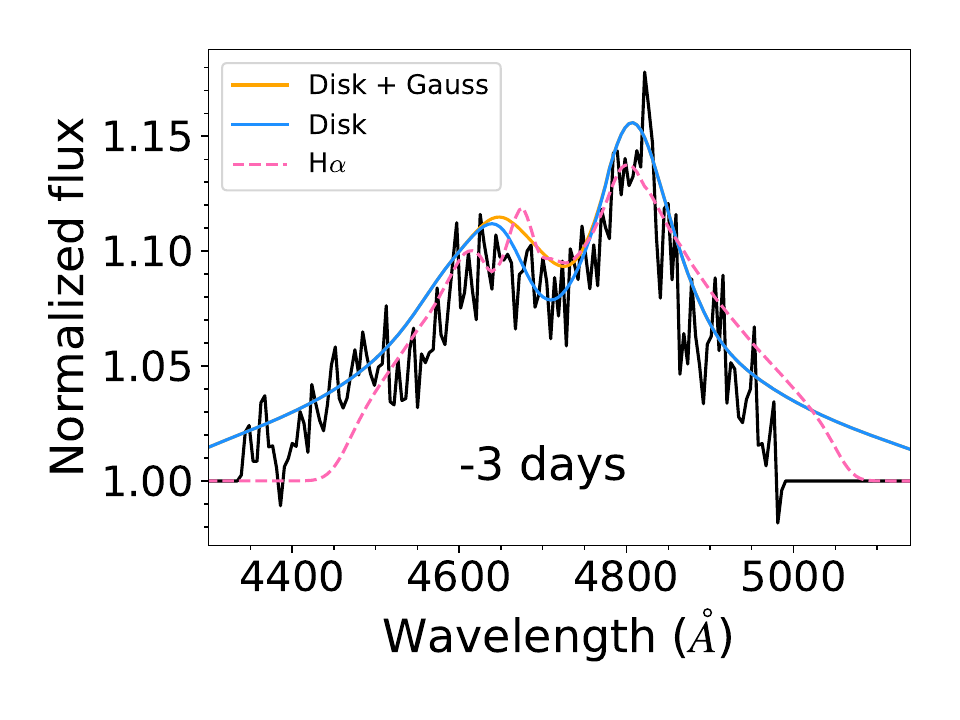}
    \includegraphics[width=0.49\linewidth]{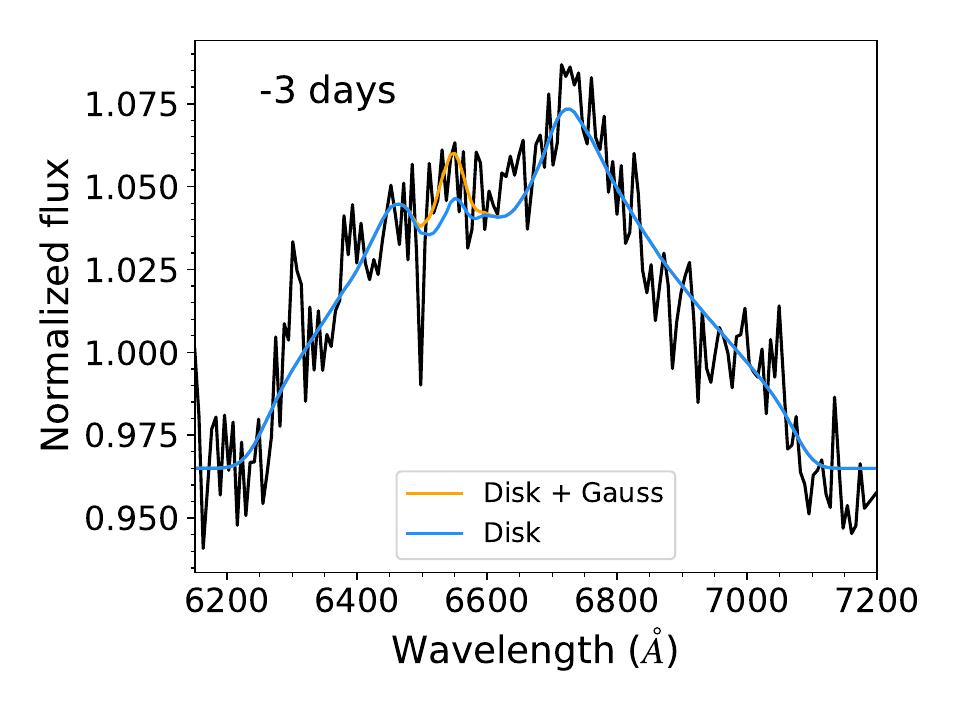}
    \includegraphics[width=0.49\linewidth]{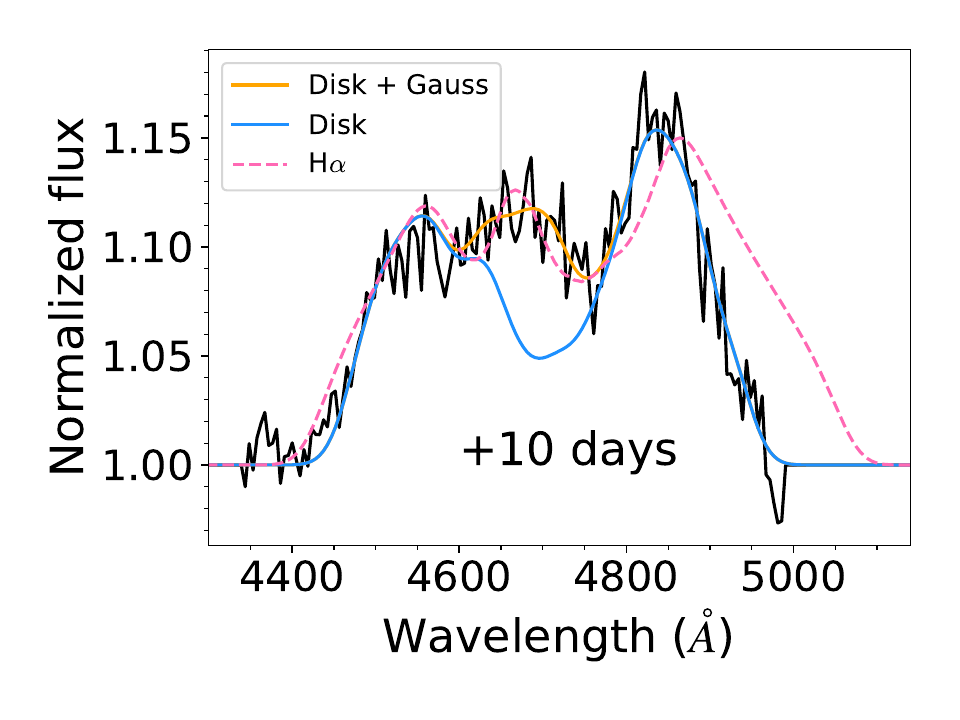}
    \includegraphics[width=0.49\linewidth]{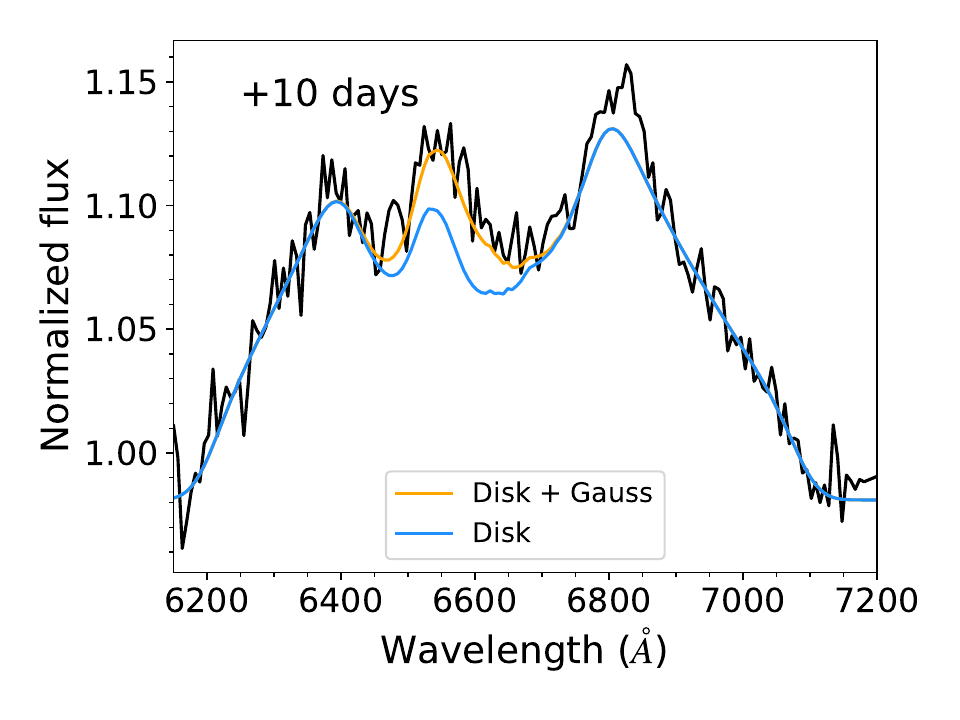}

    \caption{Best-fit accretion disk + Gaussian models, overlaid on the emission line profiles of He\,\textsc{ii} (left panels) and H$\alpha$ (right panels). Black represents the data, blue the accretion disk model and orange the accretion disk + outflow model. The best fit H$\alpha$ models are overplotted in the He\,\textsc{ii} profiles as dashed lines for comparison. The red wing of He\,\textsc{ii} is poorly subtracted, leading to some differences with the H$\alpha$ profiles, but overall the results are consistent. The model parameters can be found in Table \ref{tab:elldisk_result}.}
    \label{fig:elldisk_fit}
\end{figure*}

The best-fit model parameters are presented in Table \ref{tab:elldisk_result}, and they are overlaid on the data in Figure \ref{fig:elldisk_fit}. The posterior distributions of the parameters for each fit can be found in Appendix \ref{sec:appendix}.

For each individual epoch, we infer global accretion disk model parameters that are very similar for the H$\alpha$ and the He\,\textsc{ii} profiles. This may appear somewhat surprising, given that the He\,\textsc{ii} profile is significantly contaminated by N\,\textsc{iii}. This indicates that the subtraction (and the 1:1 flux ratio assumption for the N\,\textsc{iii} lines) performs well. 

We find that the disk must be highly inclined with respect to our line of sight, that is we are seeing it nearly edge-on ($i = 85\pm5$ degrees for H$\alpha$, and $i = 88\pm2$ for He\,\textsc{ii}). The inferred eccentricity is high and consistent between H$\alpha$ and He\,\textsc{ii}, $e = 0.97\pm0.01$. This is very similar to the characteristic eccentricity of ballistic orbits travelled by the tidal debris, $\sim$0.98 [$\frac{(M_{\rm BH} / M_{\star})}{10^6}$]$^{1/3}$, expected for returning debris whose orbit has not been significantly altered by hydrodynamics. We furthermore find largely consistent orientation angles ($\phi$ between 210 and 260 degrees) and line broadening parameters ($\sigma \sim$ 1000 -- 1500 km s$^{-1}$) for all epochs/lines. There is some scatter in the inner and outer disk radii, but this may not be surprising given the limited SNR of the data. The inner disk radius is several 100 gravitational radii, while for the outer radius we find values between 1400--2200 R$_g$, with one outlier at 4100 R$_g$. Finally, we note that while we have added a Gaussian component to the line profile fitting, in most cases the amplitude/contribution of this component is small. Only in the last epoch is there a clear triple-peaked structure. This additional component is found to be consistent with being at rest velocity, with a FWHM of 1000 -- 2500 km s$^{-1}$. 

In summary, we find that a highly inclined, highly elliptical accretion disk model can reproduce the H$\alpha$ and He\,\textsc{ii} line profiles of both epochs, with general disk parameters that are largely consistent within their uncertainties. Given the contamination He\,\textsc{ii}, the fitting results are reasonably similar to those inferred from H$\alpha$. 

\section{Discussion}
\label{sec:discussion}
\subsection{Alternative origins of the broad double-peaked emission lines}
As previously indicated, there exist multiple mechanisms / structures that can explain the presence of double-peaked broad emission lines. We now discuss each of these in more detail, and why we prefer the accretion disk model as an explanation.
\subsubsection{Supermassive black hole binary}
An SMBH binary would spend most of its time in the hard binary phase, where the separation is typically $0.1-1$ pc ($10^{17-18}$ cm, \citealt{Eracleous1995}). In this scenario, the tidal field of the secondary would drive a disk around the primary to become eccentric if the mass ratio is $>$~4 \citep{Eracleous1995}. However, the timescale for this eccentricity to evolve is $\sim$ 1000s of years, clearly inconsistent with the observed line profile variations on timescales of $\sim$ weeks. 

We conclude that an SMBH binary alone can not provide an explanation for the observed properties of AT~2020zso, in particular the rapid evolution of the line profiles. 

 \subsubsection{Turn-on / changing-look AGN}
One scenario that could help explain the observed properties is that rather than a newly formed accretion disk, we are seeing either a turn-on AGN (without the need for a tidal disruption) or a dormant, eccentric accretion disk being {\it reinvigorated} with fresh material, so an AGN {\it turning on} where a fossil accretion disk is being resupplied by the debris of a star. In either case, the binary SMBH hypothesis is then required to explain the observed eccentricity. This scenario could be consistent with the hypothesis that the primary black hole was a {\it dormant} AGN that shut off in the recent past, as inferred from the narrow line region diagnostics. In other words, the AGN must have shut off within a light travel time (typically a few 100 years) to the narrow line region. Such timing would be coincidental; the high inferred eccentricities also imply that the origin of the emission lines is very unlikely to be a pre-existing BLR that is re-activated by the flare, as the values inferred from double-peaked AGN sources are typically much more modest (e.g. \citealt{Eracleous1995, Strateva2003}).

\subsubsection{Outflows and spiral arms}
Bipolar outflows can also result in double-peaked line profiles, but the brighter red-than-blue peaked profile cannot be reproduced through Doppler boosting of emission. Nevertheless, profiles similar to AT2020zso (in that they have a brighter red than blue wing) have been observed in some supernovae \citep{Smith15, Bose19}, although they are generally seen in H$\alpha$, not He\,\textsc{ii}. It is unclear what would power the Bowen fluorescence lines in this scenario. This scenario would require an ad-hoc adjustment of the relative brightness of the blue and red peaks to produce the observed variability. Furthermore, it would require that the UV/optical blackbody photosphere expands and recedes independently from the outflow (as we see the former moving inwards after peak, which is incompatible with an outflow scenario powering the lines at those times).
The quasi-Gaussian line profile, combined with the lightcurve evolution at early times suggests that any outflow present in AT~2020zso would likely have a near spherical geometry.
While a wide-angle bipolar outflow can therefore not be excluded based on current data, an aspherical structure could be detectable in polarimetric observations.

Spiral structures have also been invoked to help explain the variability in double-peaked AGN sources (e.g. \citealt{Storchi2003}). This variability is typically associated to rotation of the gas (or precession of the spiral structure) on timescales larger than several dynamical timescales. This dynamical timescale is roughly 
\begin{equation}
    t_{\rm dyn} = 2 M_6 \xi^{3/2} \ \rm days
\end{equation}
where $M_6$ is the black hole mass in units of $10^6 M_{\odot}$ and $\xi$ is the disk outer radius in units of 1000 $R_{\rm g}$. This yields a value of $\sim$ 5 -- 15 days for AT~2020zso.
While this appears compatible with the observed timescales for variability, it remains unclear how such a spiral structure would form in the very brief period of time between disruption and peak light. Furthermore, this scenario generally invokes an axisymmetric disk configuration, and hence circularisation would have to be extremely rapid -- not accounting for the formation timescale of the spiral structure itself. We therefore deem it unlikely that spiral arm patterns can provide a plausible explanation of the observed behaviour.

\subsection{Lightcurve and blackbody evolution}
The lightcurve and radius evolution are remarkably similar to other TDEs with pre-peak observations, including AT~2019qiz \citep{Nicholl2020} and AT2019ahk \citep{Holoien201919bt}: consistent with $L\propto t^2$ and constant outflow velocities of a few 1000 km s$^{-1}$. A quasi-spherical outflow with constant velocity and temperature will lead to the observed $L\propto t^2$ behaviour. The fact that the early evolution (before the first Swift observations) is consistent with $L\propto t^2$ suggests that the temperature was roughly constant during this phase.

The temperature cools significantly over the first 40 days, behaviour that is similar to AT~2018hyz and ASASSN--14ae \citep{Gomez2020} as well as ASASSN--15lh \citep{Leloudas2016}. This may be related to a comparatively low amount of debris due to a partial tidal disruption, leading to shorter diffusion times \citep{Short2020, Gomez2020} and therefore faster temperature evolution. This cooling phase may also help explain the transition from initially broad Gaussian line profiles dominated by He\,\textsc{ii} to the appearance of H$\alpha$ slightly later, and finally to the emergence of the double-peaked disk profiles.

The peak bolometric UV/optical luminosity reaches $7\times 10^{43}$ erg s$^{-1}$ -- this corresponds to roughly the Eddington limit of a $\sim$5$\times$10$^5$ $M_{\odot}$ black hole, or an Eddington ratio of 0.4 for a $\sim$10$^6$ M$_{\odot}$ black hole.
The blackbody radius (Figure \ref{fig:radevol}) reaches a maximum around $10^{15}$ cm, then rapidly decreases after peak light, and asymptotes to $5\times10^{13}$ cm at late times. Assuming a $10^6$ ($5\times 10^5$) $M_{\odot}$ black hole, the peak and late-time values correspond to approximately 3000 (6000) and 200 (400) gravitational radii, respectively. This latter value is similar to that inferred for the inner edge of the accretion disk at +10 days, indicating that the UV/optical emission at these epochs (+150 days) is consistent with being produced directly by the accretion disk. Here it is assumed that the inner disk radius does not significantly increase in size between the +10 days and +160 days epochs. We justify this assumption by noting that the accretion disk emission is dominated by the hottest, inner regions of the disk, and there are no obvious accretion related processes that would lead to a $\sim$ order of magnitude increase in the inner disk radius. Similar behaviour has been observed in other TDEs, where the late-time UV emission (in this case meaning several years after peak light) is also found to be consistent with an accretion disk origin \citep{vanvelzen2019}. In AT~2020zso, however, the accretion disk emission appears to dominate the UV bands already much earlier, similar to the TDE AT~2018fyk \citep{Wevers2021}, where rapid disk formation was inferred from other spectroscopic emission features \citep{Wevers2019b}. 

Emission lines originating in an accretion disk may be collisionally excited rather than through photo-ionisation. This can have a profound effect on the Balmer decrement (H$\alpha$ / H$\beta$ line ratio), as this depends sensitively on the temperature (see e.g. Fig. 11 in \citealt{Short2020}). If the temperature in AT2020zso was lower than in AT2018hyz, this may help to explain the weakness / absence of both H$\beta$ and H$\gamma$ lines. As noted in \citet{Short2020}, the typical blackbody temperatures in TDEs (and also in AT2020zso) are much higher than those inferred from the Balmer decrement, so the accretion disk must be significantly cooler than the blackbody emission. An alternative explanation may be that the blackbody modeling, while empirically a good fit to the data, is not intrinsically related to the observed SED shape. In this case, the inferred blackbody temperatures do not represent physical temperatures of the emitting regions.

Finally, we highlight the peculiar early time lightcurve evolution, with evidence for a clear {\it break} at very early times in the ZTF g- and r-band lightcurves from a L$\propto t^2$ behaviour to a slower evolution afterwards. Unfortunately, we do not have temperature information at the earliest times before the change in behaviour. If the outflow did not cool significantly initially, a homologously expanding outflow would be consistent with the L$\propto t^2$ evolution. This could imply the presence of an additional source of energy injection to keep the material from cooling at the earliest times. We speculate that this may be provided by the initial debris self-intersection and/or disk formation processes. Once the bulk of this energy is radiated, the further evolution is dominated by cooling as the envelope expands. Verifying such a scenario will necessitate observational constraints on the temperature evolution shortly after disruption in future TDEs. Alternatively, non-spherical expansion may also result in differences from the canonical L$\propto t^2$ evolution before peak in TDEs.

\subsection{Evolution in the context of the accretion disk model}
The {\it blackbody} emission likely has 2 components: a reprocessing envelope and an accretion disk. 

At phase -14 days, the expanding outflow is very likely reprocessing the X-rays produced at very small scales (whose presence is inferred from the presence of Bowen lines). This outflow provides the dominant contribution to the emission lines before peak, gradually decreasing as the material becomes optically thin. The spectrum is hence dominated by very broad Gaussian-like signatures, the hall-mark sign of TDEs. The accretion disk at this time is weak, either contained within the expanding photosphere, or alternatively because it is still assembling. For this reason, there are not yet any double-peaked signatures in the spectra. 

This evolution is consistent with the evolution of the EW ratio of He\,\textsc{ii} / H$\alpha$. Shortly after our observations begin, the H$\alpha$ line emerges, that is the EW ratio decreases rapidly. This apparent evolution from H-poor to H-rich is a natural consequence of an expanding reprocessing envelope \citep{Roth2018}, where H$\alpha$ suffers from more self-absorption when the envelope is more compact (while He\,\textsc{ii} photons can escape unimpeded). As the envelope expands and cools, H$\alpha$ becomes less self-absorbed and its equivalent width increases, while the He\,\textsc{ii} emitting region, located closer to the central ionising source, does not change significantly. This process has been observed in several TDEs to explain the evolution of H-rich to H-poor as the outflowing photosphere contracts after peak light \citep{Nicholl2019, Panos2021}, but here we show that this process is very likely also at work at very early times when the envelope first starts expanding.

Around peak light, when the envelope has expanded sufficiently such that it becomes optically thin, the reprocessing becomes much more inefficient. The blackbody (continuum) emission is now a superposition of both the reprocessing outflowing layer (weakening contribution) and the accretion disk (increasing contribution). The envelope is no longer optically thick, so the spectra are dominated by the accretion disk, showing broad double-peaked emission lines which are now visible due to the large contrast with the host galaxy at peak brightness. If the outflowing photosphere was still partially optically thick at --3 days, it may contribute to the spectrum as a low amplitude, broad Gaussian. We speculate that this could help explain the peculiar outer disk radius evolution. As shown in Figure \ref{fig:radevol}, this outer radius is very similar to the blackbody radius at that epoch. Given the limited SNR of the spectrum, the contribution of the outflow may be below the level that can be detected during the fitting. When the outflow reaches its maximum extent and becomes completely optically thin at $\approx+5$ days, there is no more contamination in the last X-shooter spectrum at +10 days, and the outer radius as inferred from the modeling reflects the true accretion disk outer edge. 

Because the mass fall-back rate scales as a negative power-law with time (and this is what mainly powers the accretion disk emission), after peak the contrast with the host starts to decrease. The continuum emission of the accretion disk remains visible in the UV (even at phases +150 days, see Figure \ref{fig:lc}) because of the higher contrast with the host (see the host SED in Figure \ref{fig:hostsed}), whereas the optical (continuum as well as line) emission falls below the host level. As a result, the blackbody UV (continuum) emission remains visible, but no the optical spectra no longer shows emission line signatures of the disk.

\subsection{A rapidly formed, elliptical accretion disk}
The H$\alpha$ and He\,\textsc{ii} emission line profiles display a prominent asymmetry, contrary to predictions from relativistic, circular accretion disks. Warped disks are also able to produce brighter red-than-blue profiles if the warp preferentially obscures the blue-shifted side of the disk. Our fitting results show that the line profiles can be well reproduced by eccentric, inclined relativistic disk models. We fit all epochs and lines independently and find consistent values for the main disk parameters in spite of the significant line profile variability that is observed on approximately two week timescales. We focus below on the results from H$\alpha$ modeling, given the potential systematic uncertainties introduced by deblending the He\,\textsc{ii} region.

The line broadening parameter $\sigma$ is not expected to strongly influence the line profiles \citep{Eracleous1995}; we find values $\sim1000-2000$ km s$^{-1}$ for both H$\alpha$ and He\,\textsc{ii}. Similarly, the inferred inclinations and orientation angles agree well for all epochs and lines (average $i = 85$ degrees and average $\phi_0 = 240$ degrees). The inner and outer radii are largely consistent with theoretical expectations for TDE disks, predicted to be more compact in nature than e.g. AGN disks. Figure \ref{fig:radevol} shows that the peak of the inferred (expanding) blackbody radius coincides roughly with the outer extent of the disk at a similar epoch. Similarly, at late times the blackbody radius is of the same order as the disk inner and outer radii.
This comparison is somewhat ambiguous, as we compare the radii of an elliptical structure (the accretion disk) with a spherical structure (implicitly assumed when calculating the blackbody radius). We stress that this is an order of magnitude comparison only.

It has been shown, on theoretical grounds, that the initial debris following the tidal disruption of a star is distributed on highly eccentric rings, and that this eccentricity may be long-lived (e.g. \citealt{Syer1992, Syer1993, Zanazzi2020}). Hydrodynamical models have further corroborated the picture where an eccentric, extended disk forms around the time when the mass return rate peaks \citep{Shiokawa2015, Piran2015, Krolik2016}.
A complicating factor in the identification of such a structure is the presence of an optically thick wide-angle outflow at early times (e.g. \citealt{Sadowski2016}). While it is theoretically unclear if, and if so how quickly, the debris can shed its orbital energy and form a disk \citep{Guillochon2014, Bonnerot2020}, observationally it is now well-established that an accretion disk can form on $\sim$month timescales \citep{Short2020, Hung2020, Cannizzaro2021, Wevers2019b, Wevers2021} in line with hydrodynamical simulations. In the case of AT~2018hyz, accretion disk modeling similar to that performed here was used to infer a quasi-circular structure (e $\approx$ 0.1) around 50 days after the peak of the lightcurve \citep{Hung2020}. Here, we establish the presence of an elliptical accretion disk around peak light, around one month after disruption, providing further evidence that in spite of the theoretical uncertainties and our lack of understanding of the post-disruption dynamics, an accretion disk can form very quickly. 

The compact, elliptical nature of the accretion disk in AT2020zso is in contrast with the majority of the literature, in which it is often assumed that apsidal precession (or some other mechanism) will quickly remove orbital energy from the debris, leading to a nearly circular orbit on the scale of the tidal radius. Instead, our findings suggest a highly eccentric structure with a semi-major axis of $\sim$100 times the tidal radius, similar to that found by hydrodynamical simulations (e.g. \citealt{Shiokawa2015}). 

AT~2020zso is the first TDE where it can be quantitatively confirmed that the initial debris maintains highly eccentric orbits for a significant amount of time. Due to the lack of observational data before peak in other TDEs, it remains unclear how often this occurs, i.e. if inefficient circularisation is common among TDEs. \citet{Krolik2020} argue that whether or not circularisation is efficient depends sensitively on the pericentre radius of the fatal orbit, with circular accretion disks being a rare occurrence. 

Future observations of double-peaked emission lines covering pre- and post-peak phases, where disk signatures can be identified unambiguously throughout the evolution, are necessary to establish the eccentricity evolution of TDE disks in more detail. 

\subsection{Bowen lines and the TDE unification model}
We have confirmed AT2020zso as a TDE with Bowen emission features -- and the first with double-peaked Bowen lines. The excitation of these lines requires a strong soft X-ray / EUV source, as they are powered through a recombination cascade including He\,\textsc{ii}. However, \citet{Leloudas2019} found that the majority of Bowen-strong TDEs were not detected at X-ray wavelengths. In the TDE unification model of \citet{Dai2018}, the properties of Bowen-strong TDEs can be explained if the inclination of the newly formed accretion disk is closer to edge-on than face-on. For Eddington ratios of L$_{Edd} > 0.1$, the accretion disk is likely slim rather than thin, leading to an optically thick barrier (potentially aided by an optically thick outflow) that results in strong suppression of X-ray photons for an outside observer.

The presence of double-peaked Bowen lines implies that they are formed very close to the accretion disk surface, most likely in the same region as the He\,\textsc{ii} and H$\alpha$ emitting regions. With a peak UV/optical Eddington ratio of $\approx$ 0.5 for AT2020zso, the disk likely has a slim geometry. Combined with the very high inclination angle ($\sim$85 degrees) this may provide the dense gas that produces the Bowen lines through X-ray irradiation, while at the same time explaining the lack of observed X-ray emission by Swift.
This provides the first direct confirmation (albeit for a single source) that the orientation of Bowen-strong TDEs is indeed near edge-on, and the unification model laid out by \citet{Dai2018} and \citet{Leloudas2019} is consistent with these results. 
The high inclination of the newly formed disk may also help to explain the lack of intrinsic (as well as TDE) X-ray emission at early and late times -- we derived a lower limit for the column density of 10$^{22}$ cm$^{-2}$ to reconcile the observed X-ray upper limit with the expected AGN X-ray luminosity at late times. Such a column could be provided by a high inclination compact accretion disk. 

\subsection{Comparison to double-peaked TDEs and AGNs}
\begin{figure}
    \centering
    \includegraphics[width=\linewidth]{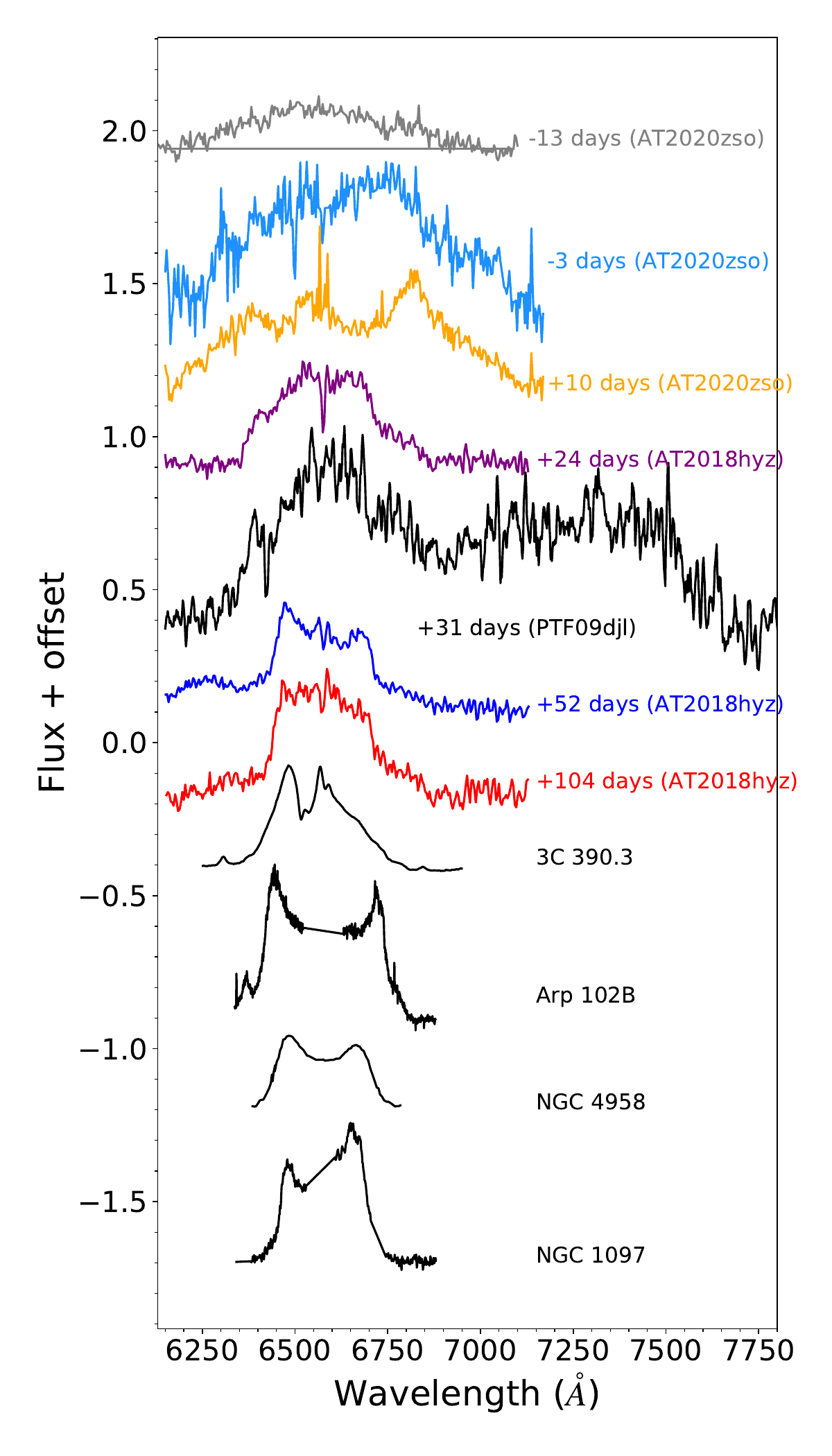}
    \caption{Comparison between the known double-peaked TDE spectra at various epochs and double-peaked AGN sources. The grey horizontal line denotes the continuum level for clarity in the top spectrum. Data for NGC 1097 are from \citet{Schimoia2015}, for NGC 4958 from \citet{Ricci}, for Arp 102B from \citet{Couto}, for AT2018hyz from \citet{Short2020} and the data from PTF--09djl are from \citet{Arcavi2014}.}
    \label{fig:spectralcomparison}
\end{figure}

The elliptical accretion disk model that we have employed has been extensively used in the literature for fitting AGN optical emission lines. Typically, it is applied to the (low-ionisation) double-peaked Balmer emission lines, from which parameters are extracted and analysed. We can therefore compare our own results, obtained from fitting H$\alpha$ in particular, to the typical values inferred for AGN accretion disks. 
Figure \ref{fig:spectralcomparison} compares double-peaked TDE spectra at different phases with some double-peaked AGN spectra. Considering only the TDE spectra, it becomes apparent that different disk parameters have very different observational signatures. In particular the inclination of the system with respect to the line of sight can dramatically alter the line widths (AT~2018hyz has an inferred inclination angle of $\sim50-60$ degrees, whereas both AT~2020zso and PTF--09djl have inclinations $>80$ degrees).

While the emissivity profile indices (between 2 -- 3), line broadening parameters (1000 -- 2000 km s$^{-1}$) and the outer radii (1500 -- 10\,000 $R_{\rm g}$) we find are typical of AGN samples \citep{Eracleous1995, Strateva2003, Storchi2017}, the inner radii we find are significantly smaller (typically $> 500 ~R_{\rm g}$ for AGN). This is in line with theoretical expectations, which predict that the stellar debris will form an accretion disk with a size of about twice the fatal orbit pericentre (on the order of several tens of gravitational radii). It is also consistent with the much broader emission lines observed in AT~2020zso.

One notable feature of the transient emission is the presence of double-peaked high-ionisation lines, He\,\textsc{ii} and N III $\lambda 4640$. To our knowledge, these are the first high ionisation line with an observed geometry similar to H$\alpha$ in both TDEs and AGNs. 
In AGNs, double-peaked profiles are observed in low-ionisation lines such as the Balmer lines and sometimes Mg\,\textsc{ii}, but no clear double-peaked profiles have been found yet in the (high ionisation) UV resonance lines \citep{Eracleous2009}. This may be related to the (typically) much higher optical depth in the high ionisation lines, which are thought to form in the densest parts of the accretion disk (wind). 
The likeness of He\,\textsc{ii} to H$\alpha$ suggests that the optical depth in both lines is similar, and hence they originate from largely the same physical region, as we infer from our accretion disk modeling.

Finally, we find marginal evidence for an additional Gaussian component near systemic velocity, with a width of $\sim 1000-2000$ km s$^{-1}$, for the epoch at +10 days (Fig. \ref{fig:elldisk_fit}). This component is very similar to the geometries found for Seyfert 1 galaxies with double-peaked emission lines \citep{Ho1997, Schimoia2017, Storchi2017}; it can be interpreted as originating in clouds outside of the accretion disk or outside of the disk plane (e.g. produced from a slow accretion disk wind), and hence they have lower velocity widths and are found near rest velocity. This component appears variable, as it does not appear prominently in the spectra around peak light whereas the triple peaked structure is very clear at +16 days. The difference of this component between the H$\alpha$ and He\,\textsc{ii} spectra is likely a result of a degeneracy with the additional Gaussian model, leading to a slightly different inferred orientation angle (this component prominently appears for $\phi >$230 degrees). In the interpretation as a disk wind, this variability could be intrinsic: if the wind is initially optically thin (and hence appears as a weak contribution to the overall flux) but over time intensifies (e.g. if the disk reaches a steady state) and becomes optically thick, it will become stronger over time. Alternatively, the wind component may remain steady, but because it is superposed onto the accretion disk contribution (which is observed to diminish over time) it appears variable, becoming more prominent as the accretion disk flux decreases. Our spectra are not of sufficient quality (in terms of SNR) to distinguish between these scenarios. 

Higher signal-to-noise ratio, medium resolution spectroscopy (R$\gtrsim$ 5000) of future double-peaked TDEs can help to shed more light on the presence and evolution of this component.

\subsection{Constraints on the precession / alignment timescales and black hole spin}
\label{ref:discussion:misalignment}
On the one hand, the late-time optical spectrum taken with X-shooter (215 days after peak) shows that there is no evidence for persistent broad emission lines, neither in the optical nor in the NIR. 
On the other hand, the blackbody radii that we derive for the UV/optical emission component is $10^{14-15}$ cm (Figure \ref{fig:templumrad}), well inside typical torus size scales ($\gtrsim 10^{16-18}$ cm, \citealt{Suganuma2006, Hickox}). This suggest that any obscuration must be comparatively low; certainly lower than values observed in Seyfert 2 AGN (typically in excess of $A_V > 5$ mag, e.g. \citealt{Burtscher2016, Schnorr2016}). Hence a type 2 (heavily obscured) AGN configuration is hard to reconcile with the observed photometric and spectral evolution. 
The implication is that any pre-existing structure, assuming it is aligned perpendicular to the BH spin vector, has at most a moderate inclination ($< 40-60$ degrees) with respect to our line of sight. 

From our modeling results, we have inferred a very high ($\sim 80$ degrees) inclination for the newly formed TDE disk. 
If it is indeed the case that the torus and any pre-existing accretion disk structures are located in a plane perpendicular to the black hole spin vector, this implies a significant misalignment between the BH spin axis and the new disk, even after peak light. 
Such misaligned configurations lead to relativistic torques and so-called Lense-Thirring precession, which will tend to align the newly formed disk with the black hole spin vector as the black hole mass is orders of magnitude larger than the disrupted star.

Our spectroscopic observations can therefore be used to constrain a minimum decay timescale for the misalignment, which depends sensitively on the black hole spin and disk viscosity, as well as weakly on the black hole mass \citep{Franchini2016, Zanazzi2019}. Rigid body precession occurs if the local precession period is longer than the sound crossing time, and simulations suggest that this is the case for a large part of parameter space typical for TDEs \citep{Franchini2016, Zanazzi2019}.
The time between the two observing epochs ($\sim 15$ days), for which we infer no significant changes in inclination, eccentricity or disk orientation angle, can therefore be interpreted as an absolute lower limit to the alignment timescale. Ignoring for the moment the (small) effect of black hole mass, with these (very) conservative estimates we can already rule out spin values in excess of $a > 0.8$ for the central black hole, regardless of disk viscosity (see e.g. Figure 12 in \citealt{Franchini2016}). 

Longer spectral series may in the future be used to constrain changes in disk inclination. This can, in turn, provide an alternative way to constrain the disk alignment timescales (typically constrained through the detection of X-ray variability such as quasi-periodic oscillations, e.g. \citealt{Pashamqpo}). These alignment timescales then provide a new method (see e.g. \citealt{Leloudas2016, Pashamqpo, Mummery2020}) to constrain the black hole spin and/or disk viscosity.

\section{Summary}
\label{sec:conclusions}
AT~2020zso is a nuclear transient classified as a TDE based on its UV/optical photometric and spectroscopic properties and evolution. We summarise the main results reported in this study as follows:

\begin{itemize}
\item The host galaxy has an active galactic nucleus, based on the emission line content and narrow line region diagnostics (BPT diagram). From the host galaxy velocity dispersion, as well as lightcurve modeling, we infer a black hole mass of $5-10\times10^5$ $M_{\odot}$.

\item NIR photometric observations show no evidence for a light echo $\sim 180$ days after UV/optical peak. 

\item The spectra show transient, asymmetric double-peaked line profiles in He\,\textsc{ii} and H$\alpha$. These lines significantly evolve from the --14 days before peak light to +14 days post-peak for which there is spectral coverage, and similar line profiles also become apparent in the He\,\textsc{i} $\lambda$5876 and N III Bowen lines ($\lambda \lambda 4100, 4640$). The He\,\textsc{ii} and Bowen profiles are, to our knowledge, the first high ionisation double-peaked lines observed in an accreting SMBH.

\item Lightcurve modeling indicates that AT~2020zso may have been the result of a partial stellar disruption. This was also the case for AT~2018hyz, another TDE that showed prominent double-peaked Balmer emission lines. This commonality may suggest that the accretion disk was visible directly due to a low amount of obscuring material, compared to more typical full stellar disruptions.

\item The low amount of debris resulting from a partial disruption can also help explain the relatively rapid observed blackbody cooling as a consequence of shorter diffusion times in the optically thick envelope. At the same time, this can also explain the rapid spectroscopic evolution: the spectra are initially He-dominated while the envelope is hottest and densest; as it expands and cools H$\alpha$ strengthens; and finally, when the debris becomes optically thin near peak light the double-peaked disk profiles appear.

\item Modeling the emission line profiles with an elliptical accretion disk model, we find that the system is highly elliptical and highly inclined (nearly edge-on). From independent fits of He\,\textsc{ii} and H$\alpha$ at two epochs ($-3$ and $+10$ days with respect to peak light), we infer consistent disk parameters such as the inclination and orientation angles, emissivity profiles, line broadening parameters and inner and outer radii. The consistency between independent fits strengthens the conclusion that the eccentricity in the accretion disk is long-lived.

\item The high inferred inclination (a nearly edge-on orientation), combined with the presence of Bowen fluorescence lines and the lack of observed X-ray emission are consistent with the unification picture of TDEs, where the inclination angle largely determines the observational appearance across wavelength. To our knowledge this is the first direct confirmation of this theoretical picture.

\item The presence of double-peaked emission lines originating in an accretion disk already before peak light confirms that an accretion disk can form very quickly and efficiently ($\sim 1$ month after disruption), in contrast to theoretical predictions and simulations. This indicates that our current knowledge of the post-disruption debris is far from complete. 

\item Around 150 days after peak, the size of the blackbody radius, as inferred from the lightcurve, is consistent with the size of the accretion disk inferred from spectroscopic modeling (assuming it is similar to the values inferred at +14 days). This suggests that the UV emission may be dominated by the accretion disk already early-on in the evolution. 

\item We use, for the first time, the lack of change in inclination as inferred from the spectroscopic signatures to constrain the alignment timescale of the newly formed disk with the black hole spin vector. This provides a novel way to probe disk precession in tidal disruption events through spectroscopic monitoring. We find that high black hole spin values (a $>$ 0.8) can be ruled out for the inferred black hole mass.
\end{itemize}

This work provides a strong link between tidal disruption events and the elliptical accretion disks that are often inferred to explain the asymmetric double-peaked profiles in AGN. However, the timescales for which these emission lines are visible for both types of sources are vastly different (100s of days for TDEs versus decades for AGNs), likely as a result of the rapidly decreasing mass accretion rate following a TDE. Hence it remains unclear at present if TDEs can also be held responsible for the long-lived asymmetric disk structures observed in some AGNs -- a steady influx of material into the elliptical disk is required for the emission lines to remain visible for prolonged periods of time. 

Future observations of TDEs with double-peaked line profiles will help to shed more light on the post-debris dynamics, including the efficiency of disk formation and subsequent circularisation, as well as the connection between tidal disruption events and double-peaked AGN sources. 

\begin{acknowledgements}
We are grateful to T. Hung for sharing a Python implementation of the relativistic accretion disk model, and we thank T. Ricci, J. Schimoia, and G. Couto for providing spectra of NGC4958, NGC1097 and Arp102B. We also thank J. Krolik, T. Piran and T. Ryu for insightful comments, and the anonymous referee for comments and suggestions that improved the paper.

This work is based on observations collected at the European Southern Observatory under ESO programmes 106.216C, 106.2169.001 and 106.2169.002 (PI: Inserra). The spectra will be made publicly available through WISErep.
We thank the Swift team for scheduling the requested ToO observations. The Swift data are publicly available from the Swift science archive.
This work is partly based on the NUTS2 programme carried out at the NOT. NUTS2 is funded in part by the Instrument Centre for Danish Astrophysics (IDA).
Based on observations made with the Nordic Optical Telescope, owned in collaboration by the University of Turku and Aarhus University, and operated jointly by Aarhus University, the University of Turku and the University of Oslo, representing Denmark, Finland and Norway, the University of Iceland and Stockholm University at the Observatorio del Roque de los Muchachos, La Palma, Spain, of the Instituto de Astrofisica de Canarias.
This work makes use of observations from the Las Cumbres Observatory global telescope network.
Based  on  data  products  created  from  observations  collected  at  the  European  Organisation  for Astronomical Research in the Southern Hemisphere under ESO programme 179.A-2010 and made use of data  from  the  VISTA  Hemisphere  survey \citep{Mcmahon2013}. 

T.M.B. was funded by the CONICYT PFCHA / DOCTORADOBECAS CHILE/2017-72180113. S.Y. is funded through the GREAT research environment grant 2016-06012. P.C is supported by a research grant (19054) from VILLUM FONDEN. T.-W.C. acknowledges the EU Funding under Marie Sk\l{}odowska-Curie grant H2020-MSCA-IF-2018-842471. The LCO team was supported by National Science Foundation (NSF) grants AST-1313484, AST-1911225, and AST-1911151, as well as by National Aeronautics and Space Administration (NASA) grant 80NSSC19kf1639. M.N. acknowledges support from the European Research Council (ERC) under the European Union’s Horizon 2020 research and innovation programme (grant agreement No. 948381). N.I. is partially supported by Polish NCN DAINA grant No. 2017/27/L/ST9/03221. S.S. acknowledges support from the G.R.E.A.T research environment, funded by Vetenskapsr\aa det, the Swedish Research Council, project number 2016-06012. IA is a CIFAR Azrieli Global Scholar in the Gravity and the Extreme Universe Program and acknowledges support from that program, from the ERC under the European Union’s Horizon 2020 research and innovation program (grant agreement number 852097), from the Israel Science Foundation (grant number 2752/19), from the United States - Israel Binational Science Foundation (BSF), and from the Israeli Council for Higher Education Alon Fellowship.
\end{acknowledgements}

%
%
\bibliographystyle{aa} 
\bibliography{bib} 

\appendix{}
\section{Figures}
\label{sec:appendix}
\begin{figure*}
    \centering
    \includegraphics[width=1.\linewidth]{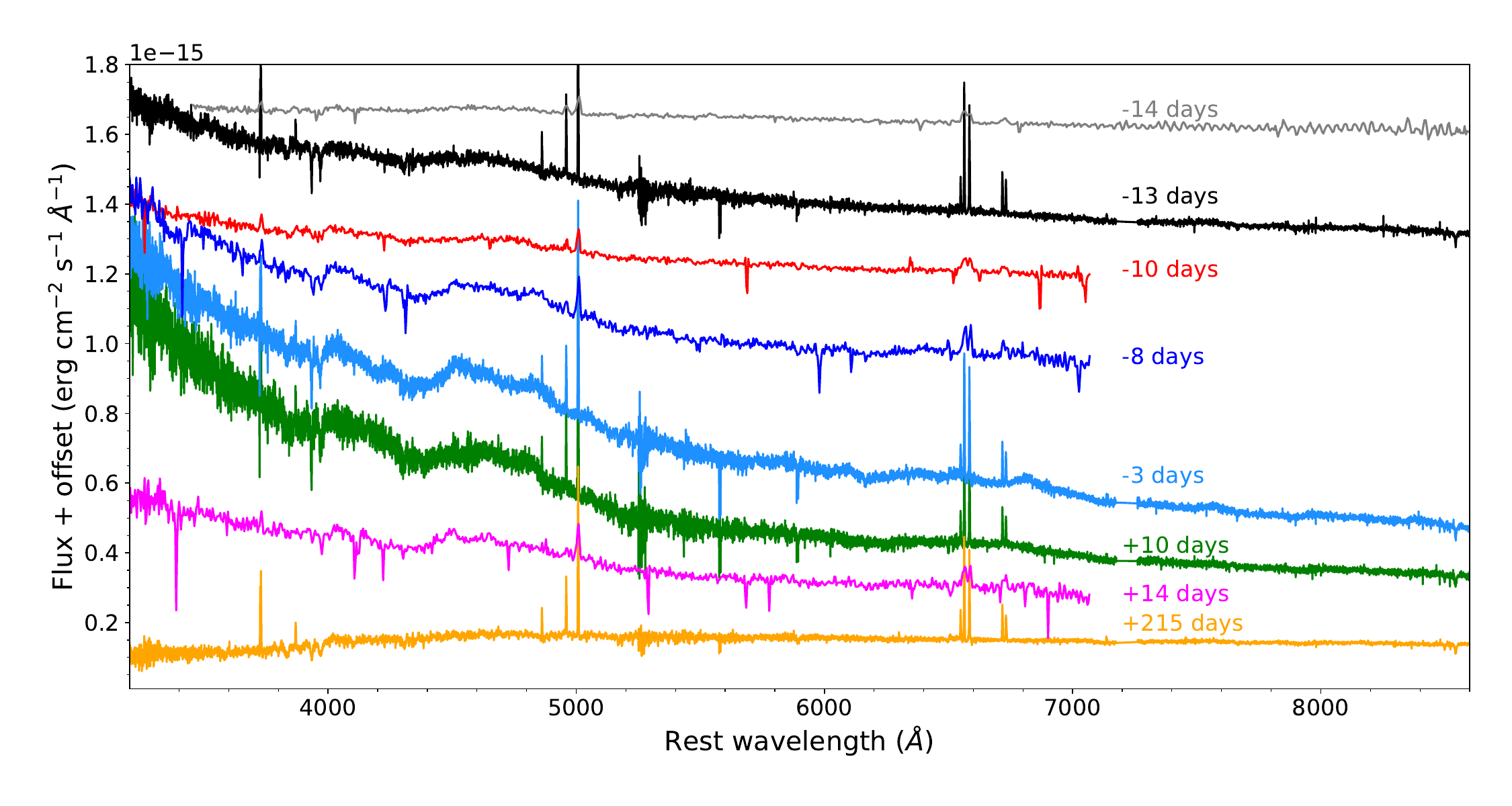}
    \includegraphics[width=1.\linewidth]{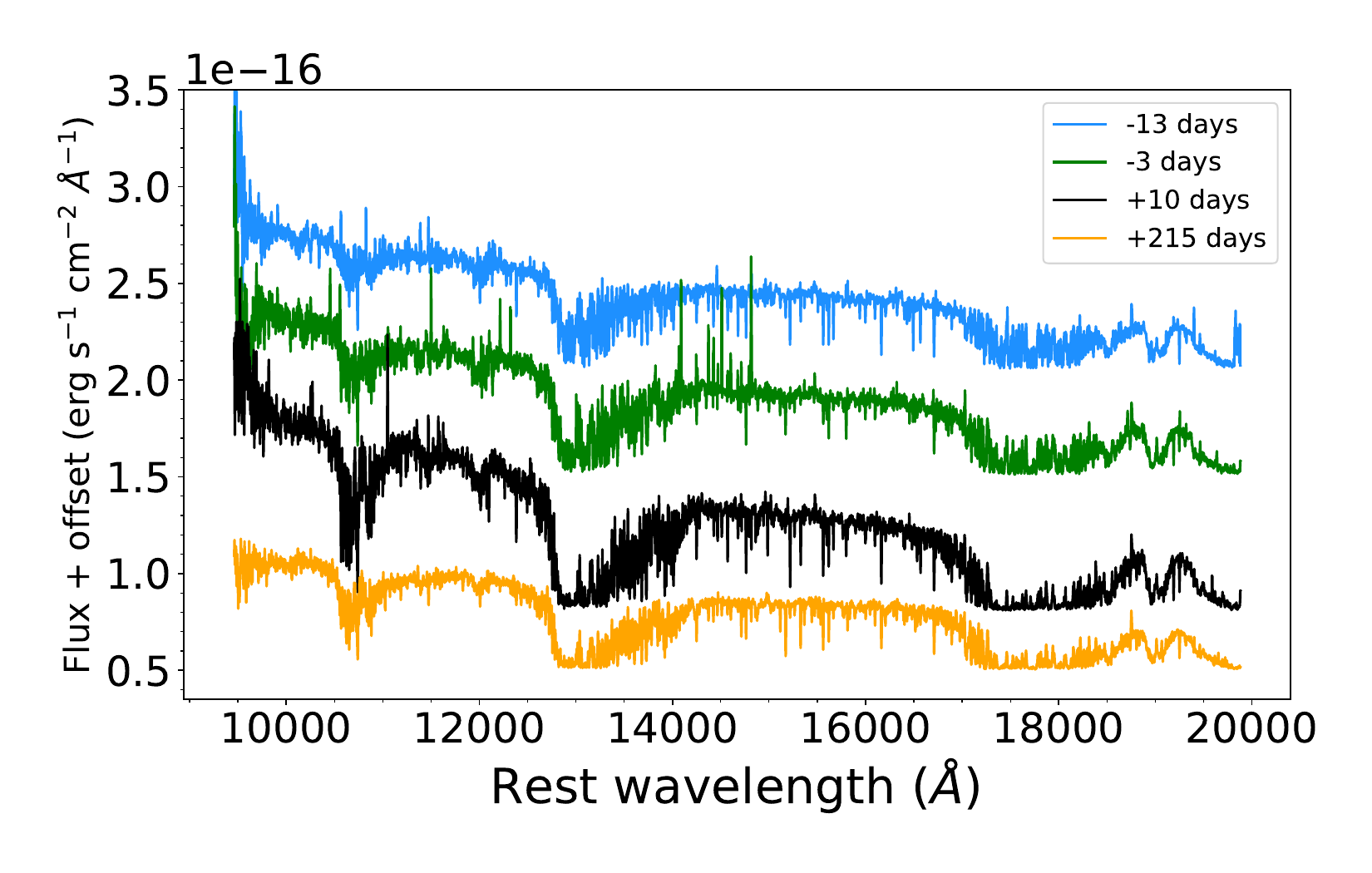}
    \caption{Flux calibrated spectra of AT2020zso (top panel: visible wavelengths, bottom panel: NIR arm of X-shooter). The X-shooter spectra have been resampled to a 0.5 $\AA$ dispersion and offset for clarity.}
    \label{fig:xshooter}
\end{figure*}


\begin{figure*}
    \centering
    \includegraphics[width=0.6\linewidth]{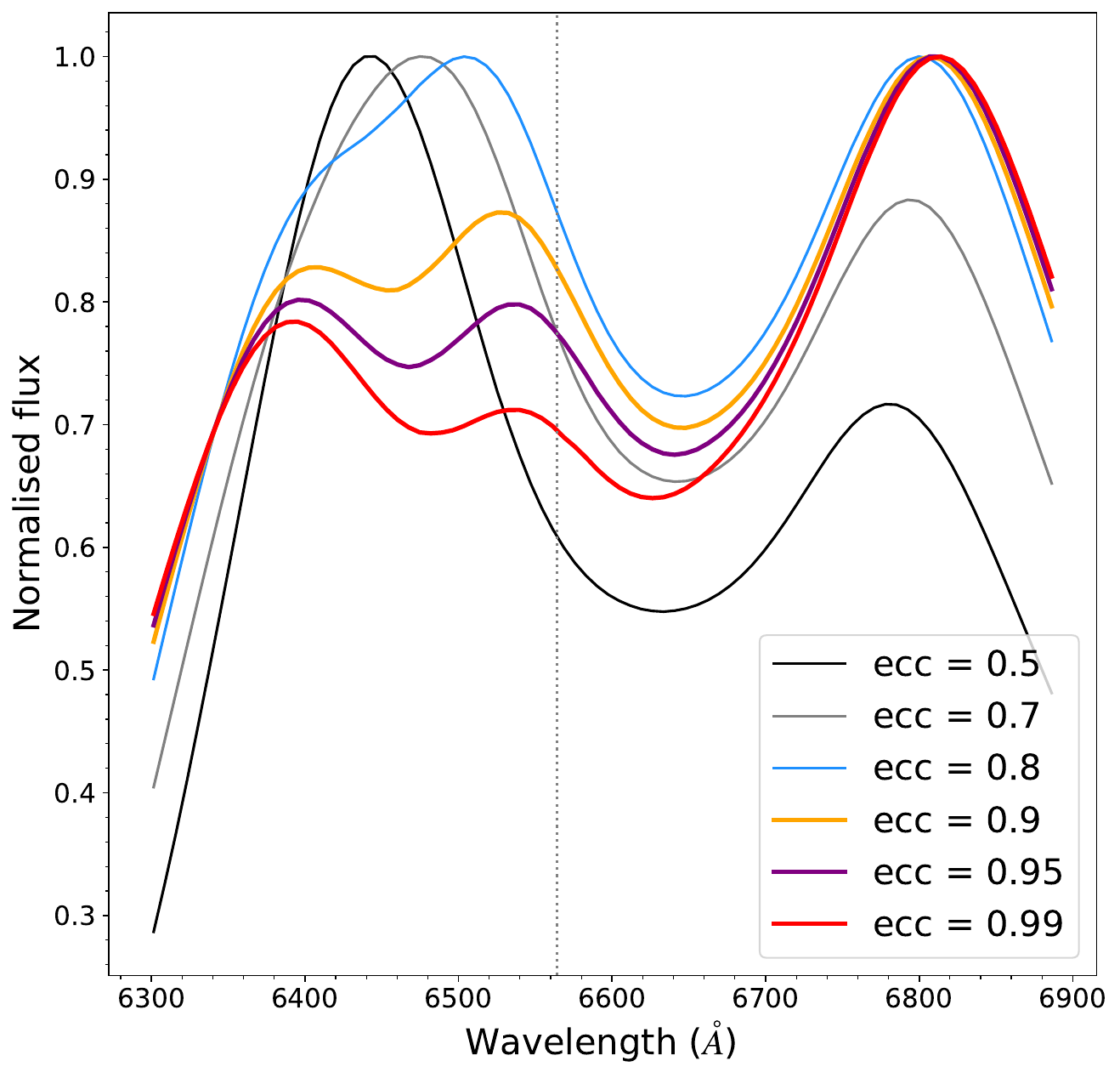}
    \caption{Illustration of the appearance of a large third middle bump with increasing eccentricity in the accretion disk model (all other disk parameters are similar to the ones inferred from the fitting). The dashed vertical line marks the rest wavelength, in this case H$\alpha$.}
    \label{fig:modelevolution}
\end{figure*}

\begin{figure*}
    \centering
    \includegraphics[width=0.6\linewidth]{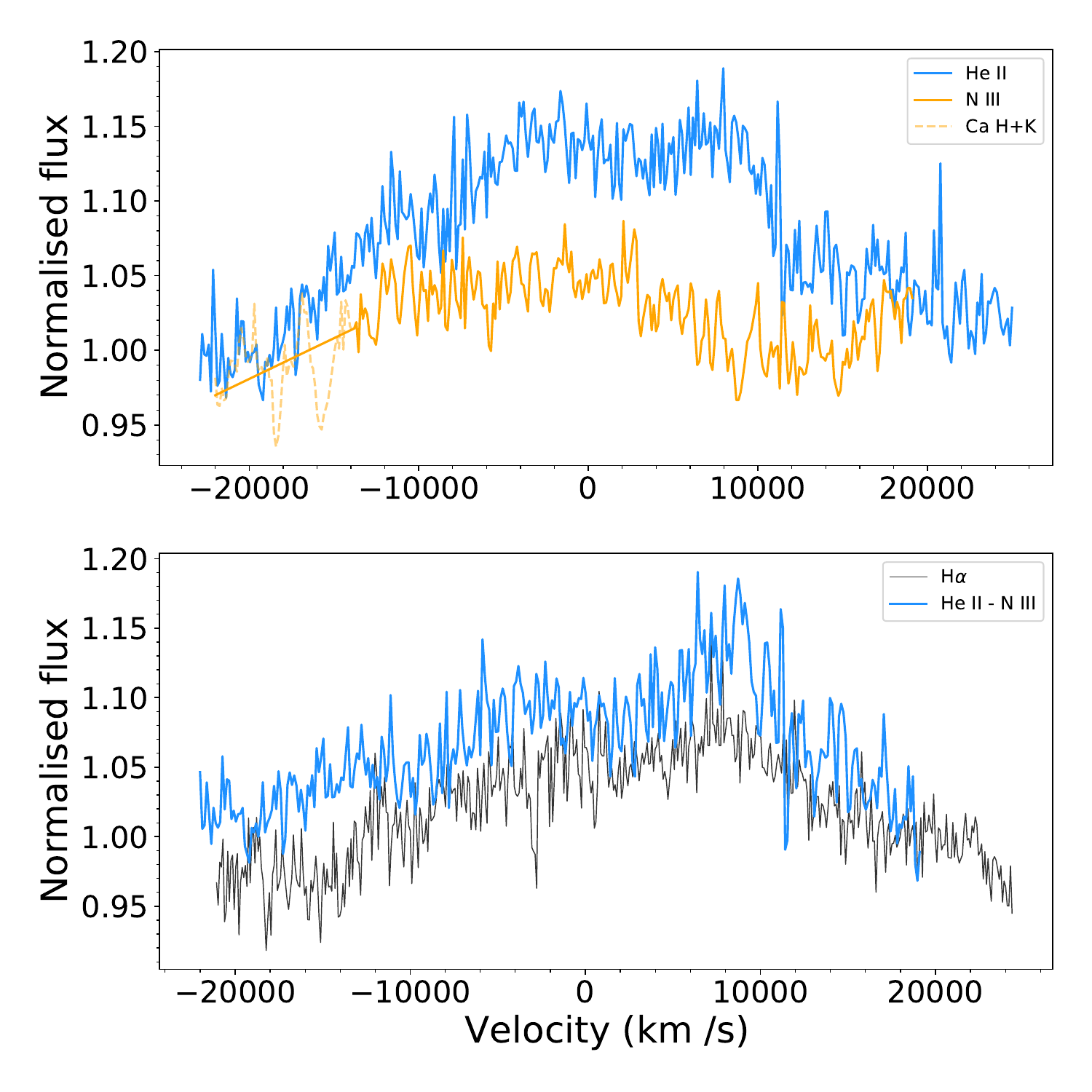}
    \caption{Top panel: He\,\textsc{ii} and N\,\textsc{iii} profiles as observed at --3 days with X-shooter. The Ca H+K doublet is interpolated to avoid biasing the subtraction. Bottom panel: subtracted profile (He\,\textsc{ii} -- N\,\textsc{iii}) in blue, and H$\alpha$ (black) for comparison. }
    \label{fig:he2subtraction_app}
\end{figure*}

\begin{figure*}
    \centering
    \includegraphics[width=0.6\linewidth]{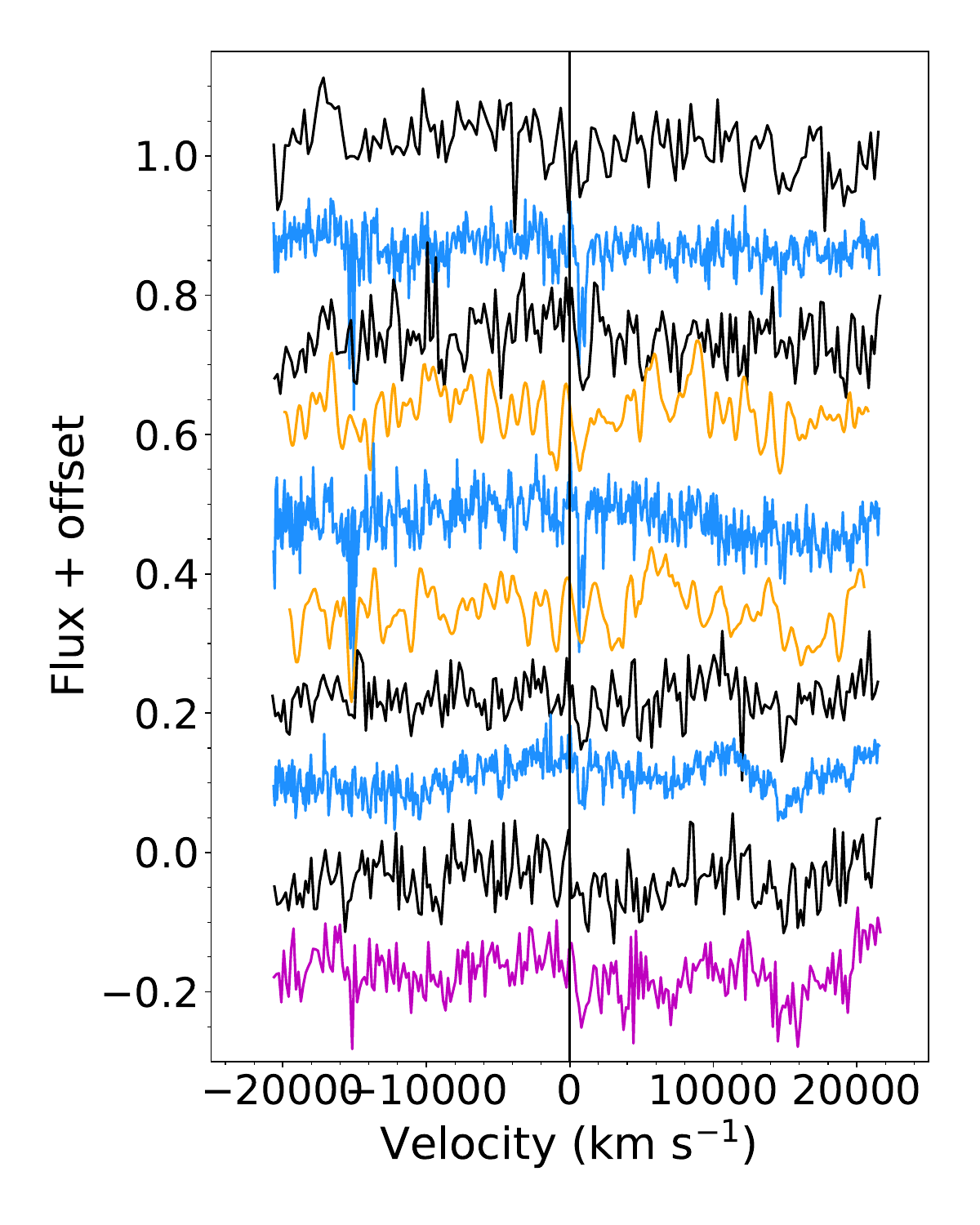}
    \caption{Same as Figure \ref{fig:normspec2} but for the He\,\textsc{i} line at 5876$\AA$. }
    \label{fig:he1evolution}
\end{figure*}

\begin{figure*}
    \centering
    \includegraphics[width=1.\linewidth]{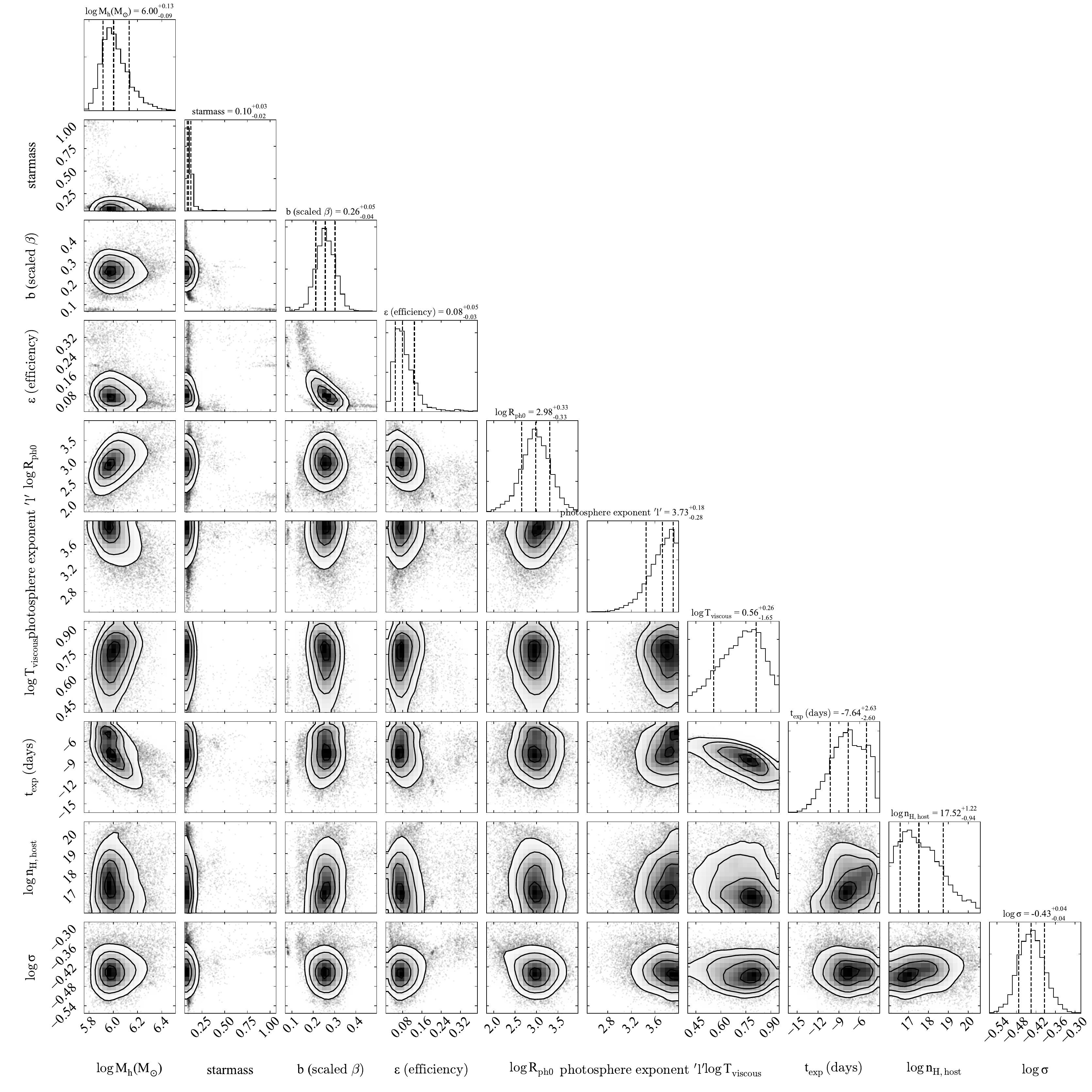}
    \caption{Full posterior distributions of the model parameter values for the lightcurve fitting with MOSFit.}
    \label{fig:posteriors_mosfit}
\end{figure*}

\begin{figure*}
    \centering
    \includegraphics[width=1.\linewidth]{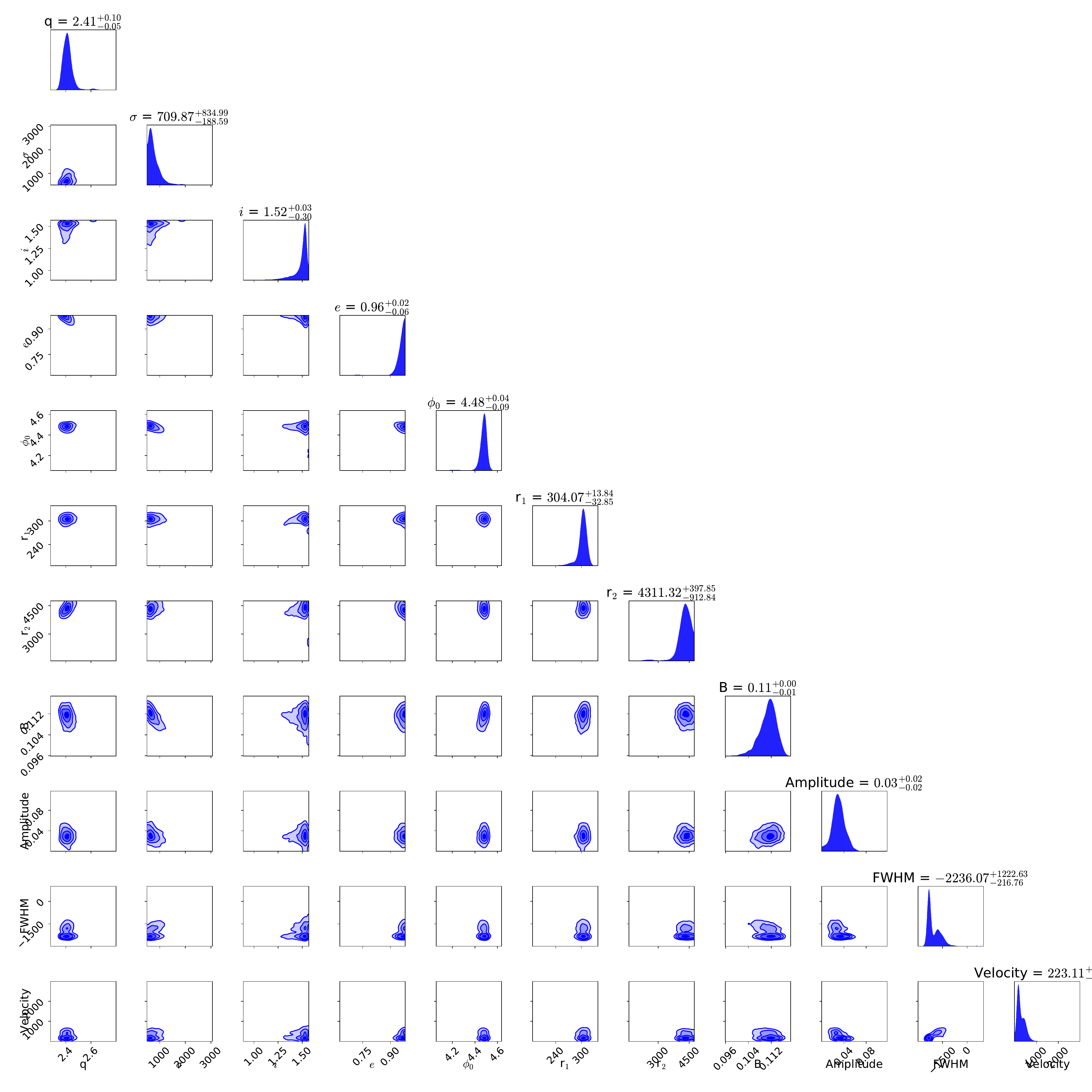}
    \caption{Full posterior distributions of the model parameter values for the H$\alpha$ emission line (epoch --3 days). The inclination and orientation angle are sampled in units of radians. Note that to estimate parameter values and uncertainties, these samples need to be combined with their associated importance weights (and hence the values and uncertainties can differ from those reported in Table \ref{tab:elldisk_result}).}
    \label{fig:posterior}
\end{figure*}

\begin{figure*}
    \centering
    \includegraphics[width=1.\linewidth]{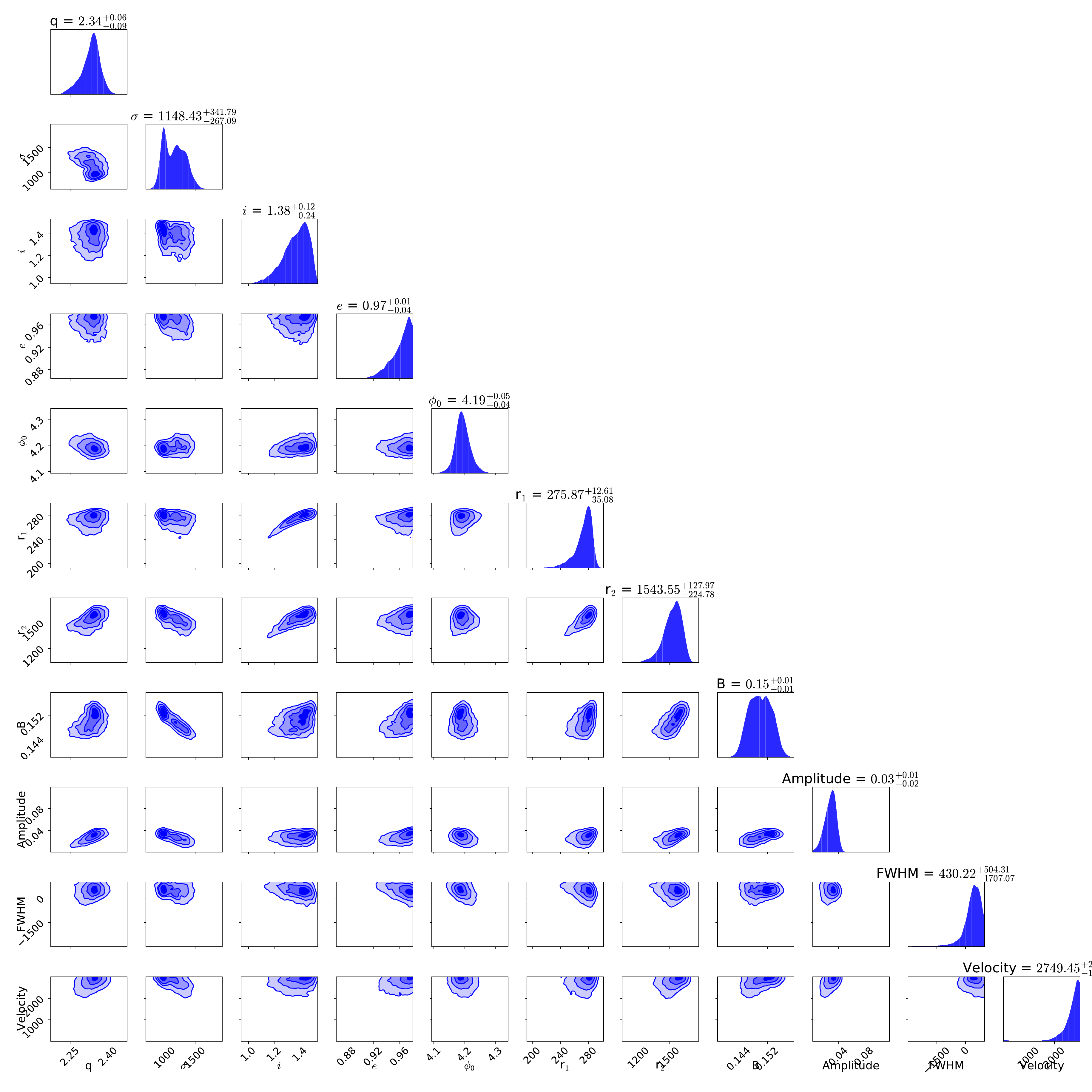}
    \caption{Full posterior distributions of the model parameter values for the H$\alpha$ emission line (epoch +10 days). The inclination and orientation angle are sampled in units of radians. Note that to estimate parameter values and uncertainties, these samples need to be combined with their associated importance weights (and hence the values and uncertainties can differ from those reported in Table \ref{tab:elldisk_result}).}
    \label{fig:posterior}
\end{figure*}

\begin{figure*}
    \centering
    \includegraphics[width=1.\linewidth]{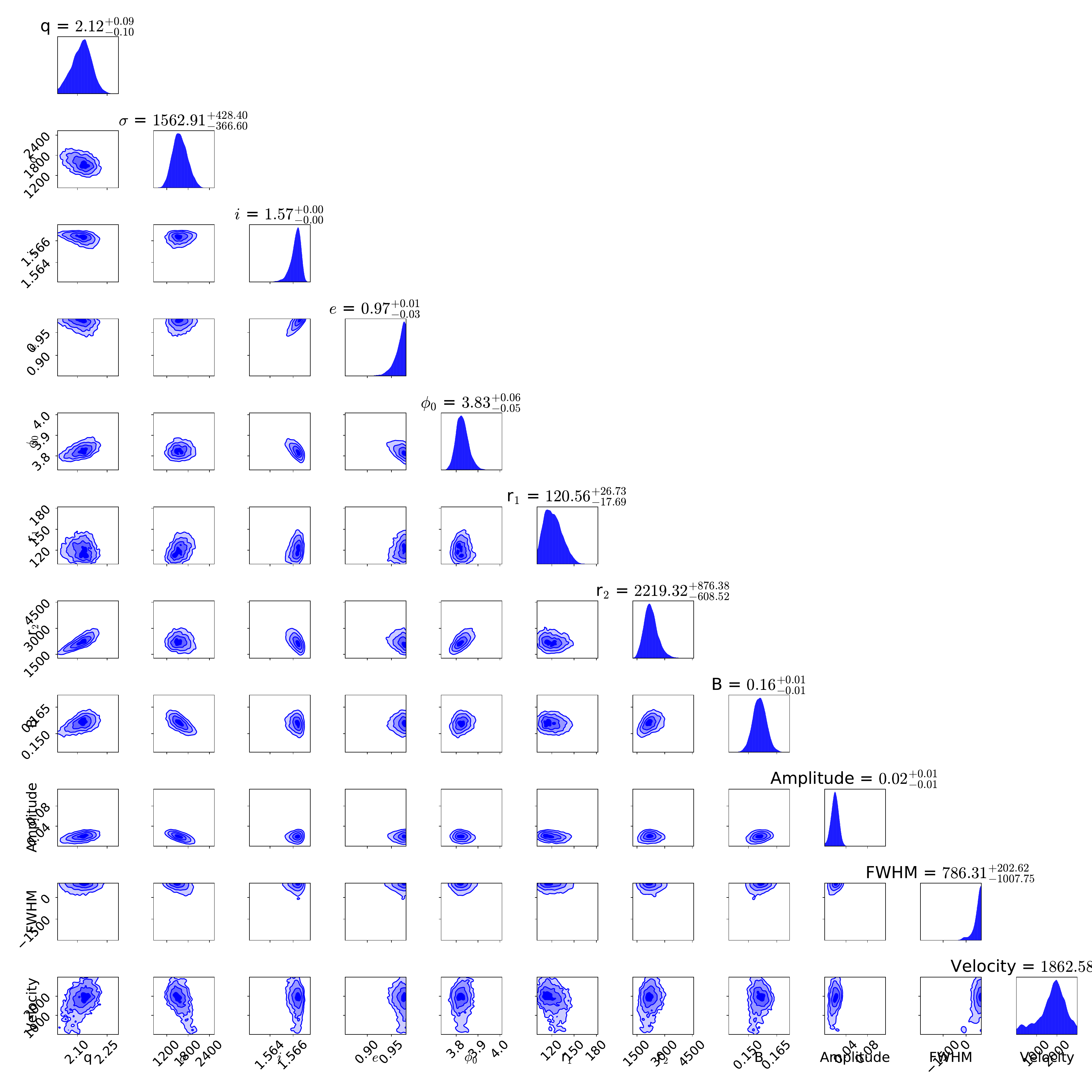}
    \caption{Full posterior distributions of the model parameter values for the (N\,\textsc{iii} subtracted) He\,\textsc{ii} emission line (epoch --3 days). The inclination and orientation angle are sampled in units of radians. Note that to estimate parameter values and uncertainties, these samples need to be combined with their associated importance weights (and hence the values and uncertainties can differ from those reported in Table \ref{tab:elldisk_result}).}
    \label{fig:posterior}
\end{figure*}

\begin{figure*}
    \centering
    \includegraphics[width=1.\linewidth]{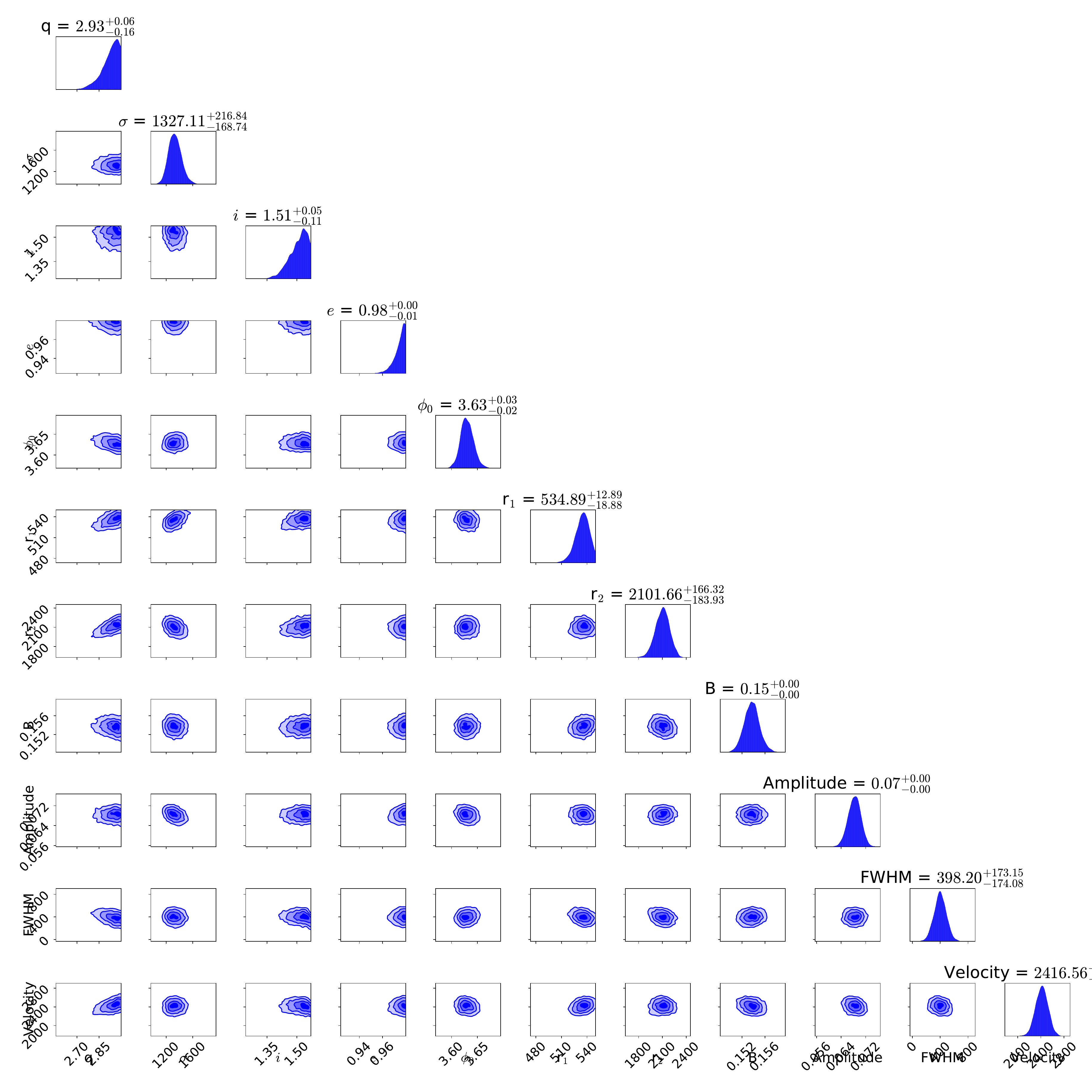}
    \caption{Full posterior distributions of the model parameter values for the (N\,\textsc{iii} subtracted) He\,\textsc{ii} emission line (epoch +10 days). The inclination and orientation angle are sampled in units of radians. Note that to estimate parameter values and uncertainties, these samples need to be combined with their associated importance weights (and hence the values and uncertainties can differ from those reported in Table \ref{tab:elldisk_result}).}
    \label{fig:posterior}
\end{figure*}

\section{Photometry}

\begin{table*}[h!]
 \caption{Photometric observations of AT2020zso.}
 \label{tab:photlog}
 \begin{tabular}{cccc}\hline
 Telescope/filter & MJD & Magnitude & Uncertainty\\\hline
ZTF/g &59165.150 & 20.40 & 0.23\\ 
ZTF/g &59167.188 & 20.15 & 0.25\\ 
ZTF/g &59168.101 & 19.69 & 0.16\\ 
ZTF/g &59170.164 & 19.00 & 0.10\\ 
ZTF/g &59172.117 & 18.56 & 0.08\\ 
ZTF/g &59178.158 & 18.29 & 0.09\\ 
ZTF/g &59182.155 & 18.03 & 0.09\\ 
ZTF/g &59184.149 & 17.98 & 0.07\\ 
ZTF/g &59189.110 & 17.65 & 0.05\\ 
ZTF/g &59192.187 & 17.58 & 0.13\\ 
ZTF/g &59194.181 & 17.78 & 0.10\\ 
ZTF/g &59197.109 & 17.64 & 0.23\\ 
ZTF/g &59199.149 & 17.93 & 0.09\\ 
ZTF/g &59205.097 & 18.53 & 0.08\\ 
ZTF/g &59205.097 & 18.53 & 0.08\\ 
ZTF/r &59165.086 & 20.87 & 0.24\\ 
ZTF/r &59167.193 & 20.43 & 0.19\\ 
ZTF/r &59168.167 & 20.13 & 0.15\\ 
ZTF/r &59170.108 & 19.21 & 0.10\\ 
ZTF/r &59178.198 & 18.37 & 0.08\\ 
ZTF/r &59182.197 & 18.21 & 0.07\\ 
ZTF/r &59187.170 & 17.92 & 0.06\\ 
ZTF/r &59189.170 & 17.83 & 0.07\\ 
ZTF/r &59199.124 & 18.13 & 0.08\\ 
ZTF/r &59205.130 & 18.42 & 0.08\\ 
ZTF/r &59205.130 & 18.42 & 0.08\\ 
ZTF/r &59211.141 & 18.73 & 0.15\\ 
Swift/V & 59173.051 & 18.15 & 0.15\\ 
Swift/V & 59173.179 & 17.52 & 0.13\\ 
Swift/V & 59178.369 & 17.94 & 0.14\\ 
Swift/V & 59180.227 & 18.17 & 0.25\\ 
Swift/V & 59182.023 & 19.17 & 0.31\\ 
Swift/V & 59184.207 & 19.83 & 0.34\\ 
Swift/V & 59191.372 & 17.32 & 0.15\\ 
Swift/V & 59195.758 & 17.33 & 0.14\\ 
Swift/V & 59199.207 & 17.34 & 0.14\\ 
Swift/V & 59204.724 & 17.93 & 0.18\\ 
Swift/V & 59207.181 & 18.42 & 0.25\\ 
Swift/V & 59211.497 & 19.44 & 0.35\\ 
Swift/V & 59215.546 & 18.93 & 0.27\\ 
Swift/V & 59327.786 & 19.77 & 0.28\\ 
Swift/V & 59341.665 & 17.80 & 0.30\\ 
Swift/V & 59355.871 & 20.08 & 0.22\\ 
Swift/B & 59173.047 & 17.83 & 0.10\\ 
Swift/B & 59173.175 & 17.78 & 0.12\\ 
Swift/B & 59178.363 & 17.75 & 0.09\\ 
Swift/B & 59180.224 & 17.59 & 0.14\\ 
Swift/B & 59182.021 & 17.62 & 0.14\\ 
Swift/B & 59184.204 & 17.91 & 0.17\\ 
Swift/B & 59191.368 & 17.44 & 0.11\\ 
Swift/B & 59195.753 & 17.55 & 0.10\\ 
Swift/B & 59199.202 & 17.79 & 0.12\\ 
Swift/B & 59204.719 & 18.39 & 0.15\\ 
Swift/B & 59207.177 & 18.24 & 0.16\\ 
Swift/B & 59211.492 & 18.73 & 0.18\\ 
Swift/B & 59215.541 & 18.58 & 0.18\\ 
Swift/B & 59341.664 & 20.55 & 0.38\\ 
Swift/B & 59355.865 & 20.17 & 0.18\\ 
  \hline
 \end{tabular}
\end{table*}

\setcounter{table}{0}
\begin{table}
 \caption{Photometric observations of AT2020zso, continued.}
 \label{tab:photlog}
 \begin{tabular}{cccc}\hline
 Telescope/filter & MJD & Magnitude & Uncertainty\\\hline

Swift/U & 59173.046 & 17.76 & 0.08\\ 
Swift/U & 59173.174 & 18.09 & 0.11\\ 
Swift/U & 59178.362 & 17.88 & 0.08\\ 
Swift/U & 59180.224 & 17.79 & 0.13\\ 
Swift/U & 59182.021 & 17.77 & 0.13\\ 
Swift/U & 59184.204 & 17.71 & 0.13\\ 
Swift/U & 59191.367 & 17.31 & 0.08\\ 
Swift/U & 59195.752 & 17.62 & 0.09\\ 
Swift/U & 59199.201 & 17.61 & 0.09\\ 
Swift/U & 59204.718 & 18.41 & 0.12\\ 
Swift/U & 59207.176 & 18.34 & 0.14\\ 
Swift/U & 59211.491 & 18.78 & 0.16\\ 
Swift/U & 59215.540 & 19.73 & 0.25\\ 
Swift/U & 59321.217 & 19.94 & 0.25\\ 
Swift/U & 59327.781 & 20.69 & 0.30\\ 
Swift/UVW1 & 59173.044 & 17.72 & 0.07\\ 
Swift/UVW1 & 59173.172 & 17.96 & 0.08\\ 
Swift/UVW1 & 59178.359 & 17.83 & 0.06\\ 
Swift/UVW1 & 59180.223 & 17.86 & 0.10\\ 
Swift/UVW1 & 59182.019 & 17.85 & 0.10\\ 
Swift/UVW1 & 59184.203 & 17.73 & 0.09\\ 
Swift/UVW1 & 59191.365 & 17.58 & 0.07\\ 
Swift/UVW1 & 59195.750 & 17.87 & 0.07\\ 
Swift/UVW1 & 59199.199 & 18.16 & 0.08\\ 
Swift/UVW1 & 59204.715 & 18.46 & 0.09\\ 
Swift/UVW1 & 59207.174 & 18.65 & 0.11\\ 
Swift/UVW1 & 59211.489 & 18.95 & 0.11\\ 
Swift/UVW1 & 59215.538 & 18.86 & 0.11\\ 
Swift/UVW1 & 59321.215 & 20.36 & 0.24\\ 
Swift/UVW1 & 59327.780 & 23.36 & 0.39\\ 
Swift/UVW1 & 59355.862 & 21.17 & 0.21\\ 
Swift/UVW2 & 59173.048 & 17.66 & 0.05\\ 
Swift/UVW2 & 59173.176 & 17.67 & 0.05\\ 
Swift/UVW2 & 59178.364 & 17.67 & 0.04\\ 
Swift/UVW2 & 59180.225 & 17.47 & 0.06\\ 
Swift/UVW2 & 59182.021 & 17.55 & 0.07\\ 
Swift/UVW2 & 59184.205 & 17.38 & 0.06\\ 
Swift/UVW2 & 59191.368 & 17.49 & 0.05\\ 
Swift/UVW2 & 59195.754 & 17.83 & 0.05\\ 
Swift/UVW2 & 59199.203 & 18.09 & 0.06\\ 
Swift/UVW2 & 59204.720 & 18.88 & 0.08\\ 
Swift/UVW2 & 59207.177 & 18.91 & 0.09\\ 
Swift/UVW2 & 59211.493 & 19.21 & 0.09\\ 
Swift/UVW2 & 59215.542 & 19.27 & 0.10\\ 
Swift/UVW2 & 59321.219 & 20.88 & 0.18\\ 
Swift/UVW2 & 59327.783 & 21.48 & 0.22\\ 
Swift/UVW2 & 59341.664 & 20.44 & 0.25\\ 
Swift/UVW2 & 59355.866 & 20.98 & 0.16\\ 
Swift/UVM2 & 59173.052 & 17.91 & 0.08\\ 
Swift/UVM2 & 59173.180 & 17.81 & 0.07\\ 
Swift/UVM2 & 59178.371 & 17.80 & 0.05\\ 
Swift/UVM2 & 59180.227 & 17.74 & 0.07\\ 
Swift/UVM2 & 59182.024 & 17.65 & 0.07\\ 
Swift/UVM2 & 59184.207 & 17.48 & 0.06\\ 
Swift/UVM2 & 59191.373 & 17.49 & 0.05\\ 
Swift/UVM2 & 59195.759 & 17.80 & 0.05\\ 
Swift/UVM2 & 59199.208 & 17.94 & 0.05\\ 
Swift/UVM2 & 59204.725 & 18.72 & 0.07\\ 
Swift/UVM2 & 59207.181 & 18.69 & 0.08\\ 
Swift/UVM2 & 59215.547 & 19.02 & 0.08\\ 
  \hline
 \end{tabular}
\end{table}

\setcounter{table}{0}
\begin{table}
 \caption{Photometric observations of AT2020zso, continued.}
 \label{tab:photlog}
 \begin{tabular}{cccc}\hline
 Telescope/filter & MJD & Magnitude & Uncertainty\\\hline

Swift/UVM2 & 59321.224 & 20.49 & 0.21\\ 
Swift/UVM2 & 59327.787 & 20.73 & 0.24\\ 
Swift/UVM2 & 59341.666 & 21.28 & 0.29\\ 
Swift/UVM2 & 59355.872 & 21.50 & 0.16\\ 
LCO/B & 59171.073 & 18.84 & 0.06\\ 
LCO/B & 59171.076 & 18.86 & 0.05\\ 
LCO/V & 59171.080 & 18.90 & 0.04\\ 
LCO/V & 59171.083 & 18.96 & 0.05\\ 
LCO/g & 59171.110 & 18.69 & 0.02\\ 
LCO/g & 59171.114 & 18.69 & 0.02\\ 
LCO/r & 59171.118 & 18.85 & 0.03\\ 
LCO/r & 59171.121 & 18.89 & 0.04\\ 
LCO/i & 59171.124 & 18.91 & 0.06\\ 
LCO/i & 59171.126 & 18.86 & 0.05\\ 
LCO/B & 59174.059 & 18.48 & 0.09\\ 
LCO/V & 59174.067 & 18.54 & 0.04\\ 
LCO/g & 59177.149 & 18.20 & 0.04\\ 
LCO/g & 59177.153 & 18.41 & 0.06\\ 
LCO/r & 59177.157 & 18.55 & 0.11\\ 
LCO/B & 59178.094 & 18.09 & 0.08\\ 
LCO/B & 59178.097 & 18.26 & 0.08\\ 
LCO/i & 59180.074 & 18.29 & 0.06\\ 
LCO/B & 59180.417 & 18.27 & 0.04\\ 
LCO/B & 59180.421 & 18.27 & 0.04\\ 
LCO/V & 59180.425 & 18.33 & 0.04\\ 
LCO/V & 59180.427 & 18.17 & 0.05\\ 
LCO/g & 59185.420 & 17.94 & 0.04\\ 
LCO/g & 59185.424 & 17.89 & 0.02\\ 
LCO/g & 59188.775 & 17.77 & 0.01\\ 
LCO/g & 59188.779 & 17.66 & 0.01\\ 
LCO/r & 59188.783 & 18.04 & 0.04\\ 
LCO/r & 59188.785 & 17.76 & 0.01\\ 
LCO/i & 59188.788 & 17.74 & 0.02\\ 
LCO/i & 59188.791 & 17.70 & 0.03\\ 
LCO/B & 59191.777 & 17.83 & 0.03\\ 
LCO/B & 59193.778 & 17.92 & 0.03\\ 
LCO/B & 59193.782 & 17.90 & 0.03\\ 
LCO/V & 59193.786 & 17.82 & 0.03\\ 
LCO/V & 59193.788 & 17.88 & 0.02\\ 
LCO/B & 59210.054 & 18.86 & 0.12\\ 
LCO/B & 59210.058 & 18.71 & 0.10\\ 
LCO/V & 59210.062 & 18.77 & 0.13\\ 
LCO/V & 59210.065 & 18.76 & 0.11\\ 
LCO/g & 59219.045 & 19.26 & 0.07\\ 
LCO/g & 59219.049 & 19.23 & 0.05\\ 
LCO/r & 59219.053 & 19.29 & 0.06\\ 
LCO/r & 59219.056 & 19.29 & 0.06\\ 
LCO/i & 59219.058 & 18.99 & 0.07\\ 
LCO/i & 59219.061 & 18.88 & 0.09\\ 
  \hline
 \end{tabular}
\end{table}

\end{document}